\documentclass[twocolumn,tighten]{aastex62}

\pdfoutput=1 
\usepackage{amsmath,amstext,longtable}
\usepackage[T1]{fontenc}
\usepackage{apjfonts} 
\usepackage[figure,figure*]{hypcap}
\received{October 11, 2018}
\accepted{December 3, 2018}
\submitjournal{{\it The Astrophysical Journal Supplement Series}}
\shortauthors{Kirkpatrick et al.}
\shorttitle{Spitzer Parallaxes}

\usepackage{color,amsmath,longtable}

\begin{document}

\title{Preliminary Trigonometric Parallaxes of 184 Late-T and Y Dwarfs and an Analysis of the Field Substellar Mass Function into the "Planetary" Mass Regime}

\correspondingauthor{J.\ Davy Kirkpatrick}
\email{davy@ipac.caltech.edu}

\author[0000-0003-4269-260X]{J.\ Davy Kirkpatrick}
\affiliation{IPAC, Mail Code 100-22, Caltech, 1200 E. California Blvd., Pasadena, CA 91125, USA; davy@ipac.caltech.edu}

\author[0000-0002-0618-5128]{Emily C.\ Martin}
\affiliation{Department of Physics and Astronomy, University of California Los Angeles, 430 Portola Plaza, Box 951547, Los Angeles, CA, 90095-1547, USA}
\affiliation{Department of Astronomy \& Astrophysics, University of California Santa Cruz, 1156 High Street, Santa Cruz, CA 95064, USA}

\author[0000-0002-4424-4766]{Richard L.\ Smart}
\affiliation{Istituto Nazionale di Astrofisica, Osservatorio Astrofisico di Torino, Strada Osservatorio 20, 10025 Pino Torinese, Italy}

\author{Alfred J.\ Cayago}
\affiliation{Department of Statistics, University of California Riverside, 900 University Avenue, Riverside, CA, 92521, USA}

\author{Charles A.\ Beichman}
\affiliation{IPAC, Mail Code 100-22, Caltech, 1200 E. California Blvd., Pasadena, CA 91125, USA}

\author[0000-0001-7519-1700]{Federico Marocco}
\affiliation{IPAC, Mail Code 100-22, Caltech, 1200 E. California Blvd., Pasadena, CA 91125, USA}
\affiliation{Jet Propulsion Laboratory, California Institute of Technology, MS 169-237, 4800 Oak Grove Drive, Pasadena, CA 91109, USA}

\author{Christopher R.\ Gelino}
\affiliation{IPAC, Mail Code 100-22, Caltech, 1200 E. California Blvd., Pasadena, CA 91125, USA}

\author[0000-0001-6251-0573]{Jacqueline K.\ Faherty}
\affiliation{Department of Astrophysics, American Museum of Natural History, Central Park West at 79th Street, New York, NY 10034, USA}

\author[0000-0001-7780-3352]{Michael C.\ Cushing}
\affiliation{The University of Toledo, 2801 West Bancroft Street, Mailstop 111, Toledo, OH 43606, USA}

\author[0000-0002-6294-5937]{Adam C.\ Schneider}
\affiliation{School of Earth and Space Exploration, Arizona State University, Tempe, AZ, 85282, USA}

\author[0000-0001-7875-6391]{Gregory N.\ Mace}
\affiliation{McDonald Observatory and the Department of Astronomy, The University of Texas at Austin, Austin, TX 78712, USA}

\author[0000-0002-7595-0970]{Christopher G.\ Tinney}
\affiliation{School of Physics, University of New South Wales, NSW 2052, Australia}
\affiliation{Australian Centre for Astrobiology, University of New South Wales, NSW 2052, Australia}

\author[0000-0001-5058-1593]{Edward L.\ Wright}
\affiliation{Department of Physics and Astronomy, University of California Los Angeles, 430 Portola Plaza, Box 951547, Los Angeles, CA, 90095-1547, USA}

\author[0000-0001-8014-0270]{Patrick J.\ Lowrance}
\affiliation{IPAC, Mail Code 314-6, Caltech, 1200 E. California Blvd., Pasadena, CA 91125, USA}

\author[0000-0003-4714-1364]{James G.\ Ingalls}
\affiliation{IPAC, Mail Code 314-6, Caltech, 1200 E. California Blvd., Pasadena, CA 91125, USA}

\author{Frederick J.\ Vrba}
\affiliation{US Naval Observatory, Flagstaff Station, P.O. Box 1149, Flagstaff, AZ 86002, USA}

\author[0000-0002-4603-4834]{Jeffrey A.\ Munn}
\affiliation{US Naval Observatory, Flagstaff Station, P.O. Box 1149, Flagstaff, AZ 86002, USA}

\author[0000-0002-2968-2418]{Scott E.\ Dahm}
\affiliation{US Naval Observatory, Flagstaff Station, P.O. Box 1149, Flagstaff, AZ 86002, USA}

\author{Ian S.\ McLean}
\affiliation{Department of Physics and Astronomy, University of California Los Angeles, 430 Portola Plaza, Box 951547, Los Angeles, CA, 90095-1547, USA}

\begin{abstract}

We present preliminary trigonometric parallaxes of 184 late-T and Y dwarfs using observations from {\it Spitzer} (143), USNO (18), NTT (14), and UKIRT (9). To complete the 20-pc census of $\ge$T6 dwarfs, we combine these measurements with previously published trigonometric parallaxes for an additional 44 objects and spectrophotometric distance estimates for another 7. For these 235 objects, we estimate temperatures, sift into five 150K-wide $T_{\rm eff}$ bins covering the range 300-1050K, determine the completeness limit for each, and compute space densities. To anchor the high-mass end of the brown dwarf mass spectrum, we compile a list of early- to mid-L dwarfs within 20 pc. We run simulations using various functional forms of the mass function passed through two different sets of evolutionary code to compute predicted distributions in $T_{\rm eff}$. The best fit of these predictions to our L, T, and Y observations is a simple power-law model with $\alpha \approx 0.6$ (where $dN/dM \propto M^{-\alpha}$), meaning that the slope of the field substellar mass function is in rough agreement with that found for brown dwarfs in nearby star forming regions and young clusters. Furthermore, we find that published versions of the log-normal form do not predict the steady rise seen in the space densities from 1050K to 350K. We also find that the low-mass cutoff to formation, if one exists, is lower than $\sim$5 $M_{Jup}$, which corroborates findings in young, nearby moving groups and implies that extremely low-mass objects have been forming over the lifetime of the Milky Way. 

\end{abstract}

\keywords{stars: luminosity function, mass function -- brown dwarfs -- parallaxes -- stars: distances -- solar neighborhood -- binaries: close}

\section{Introduction}

Understanding the creation mechanisms for brown dwarfs has long been a stumbling block of star formation theory. Simplified arguments predicting the minimum Jeans mass fragment forming from a molecular cloud suggest a value of $\sim$7 M$_{Jup}$ (\citealt{low1976}), although more complicated considerations such as the role of magnetic fields and rotation are thought to drive this value higher. Lower formation masses are possible via secondary mechanisms. For example, sites replete with high-mass stars can create lower-mass brown dwarfs through the ablation, via O star winds, of protostellar embryos that would otherwise have formed bona fide stars (e.g., \citealt{whitworth2004}). Also, protostars in rich clusters with many high mass members may, via dynamical interactions, be stripped from their repository of accreting material, thus artificially stunting their growth (e.g., \citealt{reipurth2001}). Star forming regions lacking higher mass stars will form objects via neither of these processes but still create low-mass objects down to at least $\sim$3 M$_{Jup}$, as discoveries of low-mass brown dwarfs in nearby, sparse moving groups suggest (\citealt{liu2013, gagne2015, faherty2016, best2017}). Providing direct measurements on the frequency of low-mass formation and its low-mass limit are thus key parameters needed to inform modified predictions.

Field brown dwarfs -- the well-mixed, low-mass by-products of star formation -- can be used to derive the field substellar mass function. This function averages over any site-to-site differences and allows us to study the global efficiency of substellar formation integrated over the history of the Galactic disk. Does this field function suggest that low-mass star formation can be characterized by a simple form (\citealt{chabrier2001}) or only by a more complicated one (\citealt{kroupa2013})? What is the low-mass cutoff (\citealt{andersen2008})? Have these field objects, scattered from their stellar nurseries, been joined by other low-mass objects that escaped their birthplaces within a proto-planetary disk (\citealt{sumi2011, mroz2017})? Is the formation efficiency of low-mass objects higher today than it was in the distant past (\citealt{bate2005})?

Our ability to measure the low-mass cutoff is critically dependent on measuring the space density of objects with temperatures below 500K (Figure 12 of \citealt{burgasser2004b}), which corresponds to late-T and Y dwarf spectral types believed to span a mass range of $\sim$3-20 M$_{Jup}$ (\citealt{cushing2011, luhman2014-0855}). The functional form of the mass function is also most easily discerned at these same temperatures (Figure 5 of \citealt{burgasser2004b}). Both of these measurements can be accomplished by establishing a volume-limited sample of nearby brown dwarfs. 

The recent spate of cold, field brown dwarf discoveries by the NASA {\it Wide-field Infrared Survey Explorer} ({\it WISE}) now makes these measurements possible. The all-sky nature of {\it WISE} enables the creation of a sample of the closest, brightest brown dwarfs that exist, and these are the easiest brown dwarfs to characterize in detail. In this paper, we aim to use these discoveries to deduce the functional form and low-mass cutoff of the substellar mass function. 

In Section~\ref{establishing_data_set} we discuss the empirical sample needed to address our goals, and in Section~\ref{target_selection} we build a preliminary version of the sample. In subsequent sections we describe astrometric measurements from {\it Spitzer} (Sections~\ref{spitzer_observations} and \ref{reductions}), the U.S.\ Naval Observatory (Section~\ref{usno_section}), and other telescopes (Section~\ref{other_telescopes}) needed to further characterize the sample. In Section~\ref{mean_trends} we examine the measured trends in spectral type, color, and absolute magnitude to identify hidden binaries. In Section~\ref{underlying_mf} we build predictions of the space density as a function of $T_{\rm eff}$ using two different suites of evolutionary models and several different forms of the underlying mass function; these are compared to our empirical results to deduce the most likely functional form and to place constraints on the mass of the low-mass cutoff. We discuss these results in context with theoretical predictions in Section~\ref{discussion}. In Section~\ref{conclusions} we summarize our conclusions and discuss future plans for making the sample even more robust.

\section{Establishing the Empirical Data Set\label{establishing_data_set}}

To reach the above goals, we define an empirical sample and a list of follow-up observations necessary to adequately characterize it. We consider the following four points:

(1) Accurate distances to these objects must be measured so that space densities can be calculated. These objects will have large parallactic signatures, but they are extremely faint in the optical and near-infrared wavelengths where ground-based astrometric monitoring takes place. For the faintest objects ($J >$ 21 mag), the Earth's atmosphere and thermal environment preclude parallax measurements entirely. {\it Gaia} is unable to measure parallaxes for cold brown dwarfs because it observes only shortward of 1.05 $\mu$m, a wavelength regime in which these sources emit very little flux. Observations are more easily done at wavelengths near 5 $\mu$m where these objects are brightest. The InfraRed Array Camera (IRAC; \citealt{fazio2004}) onboard the {\it Spitzer Space Telescope} is therefore a natural choice, and accurate parallaxes can be measured using only a modest amount of {\it Spitzer} observing time.

(2) Objects must be characterized to the extent that unresolved binarity can be deduced. Unresolved binaries can lead to overestimates of the space density when spectrophotometric distance estimates are used, or underestimates if objects truly in the volume are not recognized as double. This is an inherent problem with any magnitude-limited sample. Although high-resolution imaging has been acquired for some of the nearest brown dwarfs (\citealt{gelino2011, liu2012, opitz2016}), very tight binaries or those with unfavorable orientations as seen from the Earth will go undetected. Measuring the absolute magnitude directly for each object is an excellent means of detecting unresolved doubles with small magnitude differences. 

(3) The number of sources as a function of $T_{\rm eff}$ (or other observable quantity) must be known with sufficient resolution in mass to determine the low-mass cutoff to a few M$_{Jup}$. As Table 9 of \cite{kirkpatrick2012} shows, a volume-limited sample with a distance limit of 20 pc provides between one dozen and four dozen objects in each half-subclass spectral type bin from T6 through early Y, although the magnitude limits of {\it WISE} restrict the T9 and later bins to somewhat smaller distances. This binning provides sufficient resolution to determine the low-mass cut-off to the precision required. Including brown dwarfs as early as T6 allows us to properly measure the shape of the density distribution below $\sim$1100K so that we can also place constraints on the overall functional form of the mass function, as Figure 14 of \cite{kirkpatrick2012} illustrates.

(4) A sufficient sample size must be considered to robustly measure biases due to metallicity effects. At late-T and Y types, a higher percentage of old objects may be expected relative to field stars because the only young objects possible at these types are those that are exceedingly low in mass, although the percentage is critically dependent upon the low-mass cutoff of formation. (See, e.g., Figure 8 of \citealt{burgasser2004b}). Several late-T subdwarfs, believed to be older objects with lower-than-average metal content, are already recognized in the nearby sample. These can serve as metallicity calibrators with which to calibrate the absolute magnitude vs.\ spectral type relations as a function of metallicity and to provide the first evidence on the efficiency of low-mass formation at lower metallicities. These objects are relatively rare, and to measure the spread in age for each mass bin, sufficient statistics (i.e., a large sample size over a sizable volume) is needed.

In conclusion, a volume-limited sample out to 20 pc that covers spectral types of T6 and later would fulfill our research goals. In the next section, we describe how we tabulated sources belonging to the sample itself.

\section{Target Selection for the 20-pc Sample of $\ge$T6 Dwarfs\label{target_selection}}

Using the compilation of known T and Y dwarfs at Dwarf Archives\footnote{See {\url http://www.DwarfArchives.org}.} along with a listing of more recent additions that one of us (CRG) has compiled since the last Dwarf Archives update, we cataloged all published objects of type T6 and later ($>$350 total), regardless of distance. We then used spectral types and $H$- and/or W2-band magnitudes to compute spectrophotometric distance estimates (\citealt{kirkpatrick2012}) for those not already having measured parallaxes. Objects were retained in our list if their spectrophotometric distance estimates placed them within 22 pc (to account for distance uncertainties) or if they had published parallaxes good to 10\% accuracy that placed them, within the measurement uncertainties, inside the 20-pc volume. This list of 235 objects is presented in Table~\ref{20pc_sample_preliminary}. This table gives the discovery designation and reference in columns 1 and 2, the AllWISE designation in column 3, the measured infrared spectral type and its reference in columns 4 and 5, and the measured parallax and its reference in columns 6 and 7. For previously published parallax values in column 6, measurements of absolute parallax are commented with "(abs)" and those of relative parallax with "(rel)". In some cases, it is not clear whether the published value is an absolute or relative value, so those are left uncommented.

Of these 235 dwarfs in Table~\ref{20pc_sample_preliminary}, 142 are being astrometrically monitored by our {\it Spitzer} program discussed in Section~\ref{spitzer_observations}, 41 have high-accuracy parallax measurements from our ground-based programs discussed in Sections~\ref{usno_section} and \ref{other_telescopes}, and 49 have high-accuracy\footnote{We list the published parallax value with the smallest quoted errors, but only if it meets our goal of $<$10\% measurement uncertainty.} parallax measurements from the literature, although four of these are in common with objects in our own parallax programs.
Only seven objects lack astrometric monitoring, and these were not added to our {\it Spitzer} programs because they either do not have any prior {\it Spitzer} measurements -- and thus may not have had a large time baseline to decouple proper motion from parallax had we observed them only in Cycle 13 -- or were published after the Cycle 13 deadline; all of these, however, are being astrometrically monitored by us in our {\it Spitzer} Cycle 14 program. For these, the parallax values in column 6 are given in italics and are spectrophotometric estimates only. With one exception, these estimates take the apparent $H$ and/or W2 magnitudes in Table~\ref{20pc_sample_photometry} together with the spectral types in column 4 to compute distance estimates via the equations (those that include the WISE 1828+2650 data point) from section 4.3 of \cite{kirkpatrick2012}. The sole exception is the parallax estimate of ULAS 0745+2332 (whose photometry was not extracted by the {\it WISE} pipeline), which is calculated from the minimum and maximum distance estimates provided by \cite{burningham2008}.

\startlongtable

\onecolumngrid
\tablecomments{References:
(1) this paper,
(2) \citealt{kirkpatrick2012},
(3) \citealt{burgasser2004},
(4) \citealt{kirkpatrick2011},
(5) \citealt{albert2011},
(6) \citealt{tinney2018},
(7) \citealt{pinfield2014b},
(8) \citealt{mace2013},
(9) \citealt{schneider2015},
(10) \citealt{cushing2011},
(11) \citealt{mainzer2011},
(12) \citealt{kirkpatrick2013-0647},
(13) \citealt{pinfield2014a},
(14) \citealt{luhman2014-solar_comp},
(15) \citealt{kirkpatrick2014},
(16) \citealt{luhman2014-0855},
(17) \citealt{tinney2012},
(18) \citealt{cushing2014},
(19) \citealt{burningham2013},
(20) \citealt{thompson2013},
(21) \citealt{tinney2014},
(22) Martin et al., submitted,
(23) \citealt{lodieu2012},
(24) \citealt{looper2007},
(25) \citealt{mace2013b},
(26) \citealt{scholz2010b},
(27) \citealt{warren2007},
(28) \citealt{mugrauer2006},
(29) \citealt{tinney2005},
(30) \citealt{delorme2008},
(31) \citealt{burningham2010},
(32) \citealt{burgasser2002},
(33) \citealt{scholz2011},
(34) \citealt{burgasser2003b},
(35) \citealt{bihain2013},
(36) \citealt{nakajima1995},
(37) \citealt{lucas2010},
(38) \citealt{artigau2010},
(39) \citealt{lodieu2007},
(40) \citealt{leggett2012},
(41) \citealt{luhman2012},
(42) \citealt{burningham2008},
(43) \citealt{burgasser1999},
(44) \citealt{wright2013},
(45) \citealt{goldman2010},
(46) \citealt{burningham2011},
(47) \citealt{tsvetanov2000},
(48) \citealt{cardoso2015},
(49) \citealt{scholz2010a},
(50) \citealt{pinfield2012},
(51) \citealt{burgasser2000},
(52) \citealt{burgasser2003a},
(53) \citealt{chiu2006},
(54) \citealt{murray2011},
(55) \citealt{strauss1999},
(56) \citealt{gelino2011},
(57) \citealt{knapp2004},
(58) \citealt{biller2006},
(59) \citealt{burningham2009},
(60) \citealt{scholz2003},
(61) \citealt{burgasser2006},
(62) \citealt{burgasser2010},
(63) \citealt{kasper2007}.,
(64) \citealt{pinfield2008},
(65) \citealt{luhman2011},
(66) \citealt{faherty2012},
(67) \citealt{dupuy2012},
(68) \citealt{vrba2004},
(69) \citealt{subasavage2009},
(70) \citealt{marocco2010},
(71) \citealt{burgasser2008},
(72) \citealt{gaia2018},
(73) \citealt{cushing2016},
(74) \citealt{smart2013},
(75) \citealt{vanaltena1995},
(76) \citealt{tinney2003},
(77) \citealt{manjavacas2013},
(78) \citealt{henry2006},
(79) \citealt{harrington1980}.
}
\tablenotetext{a}{0039+2115: Also known as HD 3651B.}
\tablenotetext{b}{0309$-$5016: Type estimated from methane imaging.}
\tablenotetext{c}{0323$-$5907: The ch1$-$ch2 color of 1.244$\pm$0.033 mag (Table~\ref{20pc_sample_photometry}) suggests a type of T6 based on Figure 11 of \cite{kirkpatrick2011}.}
\tablenotetext{d}{0628$-$8057: Type estimated from methane imaging.}
\tablenotetext{e}{0807$-$6618: Given the fact that the absolute ch2 magnitude of this object (15.43$\pm$0.09 mag) and ch1$-$ch2 color (2.81$\pm$0.16 mag) are most like the Y1 dwarfs, this object has been assigned a temporary spectral type of Y1.}
\tablenotetext{f}{0855$-$0714: Given the fact that the absolute $H$ and ch2 magnitudes of this object (27.04$\pm$0.24 and 17.13$\pm$0.02 mag, respectively) are {\it much} fainter, and the $H-$W2 and ch1$-$ch2 colors (10.13$\pm$0.24 and 3.55$\pm$0.07 mag, respectively) {\it much} redder, than that of the other Y dwarfs typed as late as $\ge$Y2, this object has been assigned a temporary spectral type of $\ge$Y4.}
\tablenotetext{g}{0950+0117: This is a common-proper-motion companion to LHS 6176.}
\tablenotetext{h}{1118+3125: Also known as the distant companion to $\xi$ UMa (Gl 423).}
\tablenotetext{i}{1231+0837: Object earlier in type than T6, but nonetheless included here because we obtained a parallax with {\it Spitzer}.}
\tablenotetext{j}{1300+1221: Also known as Ross 458C.}
\tablenotetext{k}{1333$-$1607: The Table~\ref{20pc_sample_photometry} colors of ch1$-$ch2 = 1.811$\pm$0.050 mag and $H-$ch2 = 3.369$\pm$0.132 mag suggest, based on Figures 11 and 14 of \cite{kirkpatrick2011}, a type of T7.5.}
\tablenotetext{l}{1416+1348: The parallax of the sdL primary is quoted here for the (sd)T companion.}
\tablenotetext{m}{1423+0114: Also known as BD+01 2920B.}
\tablenotetext{n}{1504+0538: Also known as HIP 73786B.}
\tablenotetext{p}{1541$-$2250: Source not extracted in the AllWISE Source Catalog, so this designation is the one from the WISE All-Sky Source Catalog.}
\tablenotetext{q}{1758+4633: Also known as GJ 4040B}
\tablenotetext{r}{2017$-$3421: Type estimated from methane imaging.}
\tablenotetext{s}{2146$-$0010: Also known as Wolf 940B and GJ 1263B.}
\tablenotetext{t}{2204$-$5646: Also known as $\epsilon$ Indi Bb.}
\tablenotetext{u}{2211$-$4758: Type estimated from methane imaging.}
\tablenotetext{v}{2302$-$7134: Type estimated from methane imaging. Object earlier in type than T6, but nonetheless included here because we obtained a parallax with {\it Spitzer}.}
\end{deluxetable*}

\twocolumngrid

Photometry for these objects is given in Table~\ref{20pc_sample_photometry}. This table shows an abbreviated name from Table~\ref{20pc_sample_preliminary} in column 1, the $H$-band apparent magnitude and its reference in columns 2 and 3, and the {\it WISE} W1 through W3 magnitudes in columns 4 through 6. It should be noted that ground-based near-infrared photometry in both the $J$ and $H$ bands has been taken for many of these objects in one of two systems: either the Mauna Kea Observatories filter set (MKO; \citealt{tokunaga2002}) or the filter set established by the Two Micron All Sky Survey (2MASS; \citealt{skrutskie2006}). Because the $J$-band filter profiles between the two systems are very different, and the $H$-band profiles are nearly the same, we list in Table~\ref{20pc_sample_photometry} only the $H$-band magnitudes since those can be intercompared regardless of the system used. Additional discussion on this point can be found in section 3.1 of \cite{kirkpatrick2011}. Discussion of the measurements of {\it Spitzer } magnitudes in columns 7-8 and the Astronomical Observation Requests (AORs) from which they were measured can be found in section~\ref{reductions}. 

\startlongtable

\onecolumngrid
\tablecomments{For resolved photometry of binaries, see Table 5 of \cite{leggett2015}. For VHS $H$-band magnitudes, the hAperMag3 was chosen, per the recommendations given at {\url http://horus.roe.ac.uk/vsa/dboverview.html}. The AperMag3 values from ULAS and UGPS were also the ones used.
References for $H$-band magnitudes:
(1) this paper,
(2) \citealt{leggett2015},
(3) \citealt{warren2007},
(4) \citealt{chiu2006},
(5) \citealt{liu2007},
(6) \citealt{leggett2010},
(7) \citealt{delorme2008},
(8) \citealt{kirkpatrick2011},
(9) UKIDSS - \citealt{lawrence2007},
(10) \citealt{knapp2004},
(11) \citealt{liu2011},
(12) \citealt{pinfield2014b},
(13) \citealt{leggett2013},
(14) \citealt{leggett2017},
(15) \citealt{schneider2015},
(16) \citealt{luhman2016},
(17) \citealt{leggett2002},
(18) \citealt{lucas2010},
(19) \citealt{pinfield2008},
(20) \citealt{lodieu2007},
(21) \citealt{leggett2009},
(22) \citealt{burningham2013},
(23) \citealt{wright2013},
(24) \citealt{goldman2010},
(25) \citealt{burningham2008},
(26) \citealt{tsvetanov2000},
(27) \citealt{burningham2010},
(28) \citealt{pinfield2012},
(29) \citealt{geballe2001},
(30) \citealt{scholz2010b},
(31) \citealt{strauss1999},
(32) \citealt{mccaughrean2004},
(33) \citealt{kasper2007},
(34) \citealt{cushing2014},
(35) \citealt{kirkpatrick2012},
(36) \citealt{burningham2009},
(37) \citealt{pinfield2014a},
(38) 2MASS Survey Point Source Reject Table - \citealt{skrutskie2006},
(39) \citealt{albert2011},
(40) \citealt{mace2013},
(41) Martin et al., submitted,
(42) \citealt{dupuy2015},
(43) VHS - \citealt{mcmahon2013},
(44) ULAS - \citealt{warren2007b},
(45) UGPS - \citealt{lucas2008},
(46) 2MASS All-Sky Point Source Catalog - \citealt{skrutskie2006}.
}
\tablenotetext{a}{0039+2115: {\it Spitzer} photometry is from \cite{luhman2007}.}
\tablenotetext{b}{0521+3640: AllWISE W1 photometry highly contaminated by the halo of a much brighter star.}
\tablenotetext{c}{0807$-$6618: W2 photometry is from a preliminary CatWISE detection (\citealt{meisner2018}). {\it Spitzer} photometry is from \cite{luhman2012}.}
\tablenotetext{d}{0855$-$0714: W1 magnitude is measured by ELW from a moving coadd using eleven epochs of {\it WISE} and {\it NEOWISE} data between early 2010 and mid 2018. The epochs themselves are comprised of 177 individual exposures. The software includes one moving source (WISE 0855$-$0714 itself) and a grid of 275 fixed background sources that model the fixed celestial pattern. The two known interfering sources in the 2010 epochs (see \citealt{wright2014}) are measured, free of contamination by WISE 0855$-$0714, and are allowed for when using the 2010 data.  The fit, which uses a scaled PSF for each source (moving or fixed) and a flat background for each frame, is in principle immune to confusion noise. The resulting W1 magnitude is much dimmer than the measurement made by \cite{wright2014}, which used only 1 epoch of {\it NEOWISE} Reactivation data. This new W1 measurement is a 2.1$\sigma$ detection (flux = 1.896$\pm$0.905 $\mu$Jy; most likely value is 1.809 $\mu$Jy).}
\tablenotetext{e}{1541$-$2250: WISE photometry is from the All-Sky Catalog.}
\end{deluxetable*}

\twocolumngrid

\section{Spitzer Observations\label{spitzer_observations}}

A list of 142 objects from Table~\ref{20pc_sample_preliminary} was astrometrically monitored with {\it Spitzer}/IRAC. Most of the {\it Spitzer} data came from programs 70062, 80109, 90007, 11059, and 13012 (Kirkpatrick, PI). The earliest of these programs, 70062 and 80109, were primarily aimed at photometric characterization of {\it WISE}-selected brown dwarf candidates using the 3.6 $\mu$m (hereafter, ch1) and 4.5 $\mu$m (hereafter, ch2) bandpasses of IRAC. In these two programs, a cycling dither pattern of medium scale and a five-position dither were used with frame times of 30 s in both channels. However, candidates with the reddest {\it WISE} W1$-$W2 colors were also observed at subsequent epochs only in ch2 to provide early astrometric monitoring to supplement {\it Spitzer} astrometric programs planned for later. Additional photometric characterization of brown dwarf candidates selected later in the {\it WISE} mission was obtained in program 11059. In this program, ch1 and ch2 observations were obtained using a random dither pattern of medium scale and nine dither positions, again with frame times of 30 s. Our main astrometric monitoring campaigns were programs 90007 and 13012, which are described further below.

We chose ch2 for our astrometric follow-up for a variety of reasons, which are listed in \cite{martin2018} but reiterated here. During {\it Spitzer} cryogenic operations, ch1 was more sensitive than ch2, but after cryogen depletion the deep image noise was found to be 12\% worse in ch1 and 10\% better in ch2, making the channels comparable in sensitivity for average field stars (ch1$-$ch2 $\approx$ 0 mag; \citealt{carey2010}). Another change during warm operations was the behavior of latent images from bright objects. Whereas latents in ch2 decay rapidly (typically within ten minutes), ch1 latents decay on timescales of hours. Moreover, the ch2 intra-pixel sensitivity variation (aka, the pixel phase effect) is about half that of ch1. Given these points, the fact that the point response function (PRF) is less undersampled in ch2 than in ch1, and the fact that our cold brown dwarfs are also much brighter in ch2 than in ch1 (1.0 $<$ ch1$-$ch2 $<$ 3.0 mag), we choose to do our imaging in ch2.

Programs 90007 and 13012 were designed exclusively for astrometric monitoring, with program 90007 providing data to satisfy the minimum astrometric requirement and program 13012 extending the time baseline to improve the astrometric uncertainties. The Lutz-Kelker bias sets the fundamental requirement for the precision we targeted. The Lutz-Kelker effect, a systematic error inherent in the measurements of trigonometric parallax for a volume-limited set of stars, is induced as follows. Near the maximum distance, $d_{max}$, the volume of stars just inside, ($d_{max} - {\Delta}d$), is smaller than that just outside, ($d_{max} + {\Delta}d$), meaning that there are more stars able to scatter into the volume than can scatter outside. This means that the true average parallax is smaller than the average parallax that is measured. Figure 1 of \cite{lutz1973}
demonstrates that the astrometric error needed to correct this effect must be $\le$15\% or else the effect is uncorrectable. For our distance limit of 20 pc (a parallax of 50 mas), this sets the error floor at 7.5 mas. Ideally, we would like to have errors even smaller; for example, an error of 10\% (5 mas) reduces the bias on the absolute magnitude determination from 0.28 mag to 0.11 mag (see Table 1 of \citealt{lutz1973}).

Our AORs for program 90007 were designed so targets were acquired with a signal-to-noise ratio (S/N) of $\ge$100 per epoch. Theoretically, the best astrometric precision we can expect from a single measurement (see equation 1 of \citealt{monet2010}) is FWHM/(2 $\times$ S/N) = 1$\farcs$7/(2 $\times$ 100) = 8.5 mas, where the full width at half maximum (FWHM) is the ch2 value from Table 2.1 of the IRAC Instrument Handbook\footnote{See {\url https://irsa.ipac.caltech.edu/data/SPITZER/docs/irac/}.}. However, our ability to fully correct for astrometric distortion keeps us from reaching this theoretical floor. We conservatively estimated that we could achieve $\sim$15 mas precision after distortion is taken into account. Because we had already acquired earlier {\it Spitzer} epochs that would allow us to measure the proper motion and disentangle it from the parallax, we assumed that the error in the parallax would be roughly equal to the per-epoch-precision divided by the square root of the number of epochs measured in program 90007. For a per-epoch-precision of 15 mas and a final parallactic error of $<$10\%, at least nine measurement epochs for each source were required. Hence, in program 90007, the per-epoch goal of S/N $>$ 100 per target was accomplished with a total ch2 integration time of 270 s acquired using a random dither pattern of medium scale, nine dither positions, and individual frame times of 30 s. This same procedure was used for program 13012.

Timing constraints were placed on the observations to optimize the measurement of parallax. The {\it Spitzer} viewing zone falls between 82$\fdg$5 deg and 120$\degr$ solar elongation, which encompasses the maximum parallax factor in ecliptic longitude (at 90$\degr$) critical to our program. For targets having two {\it Spitzer} visibility windows per year, one Astronomical Observation Request (AOR) was placed within 3.5 days of each maximum parallax factor. The other two or three AORs were evenly spaced through the rest of the visibility period. For targets near, but not in, the Continuous Viewing Zone, five or six observations were obtained at roughly evenly spaced intervals within each window, with an AOR landing within $\pm$3.5 days of each available maximum parallax factor. For targets located within the Continuous Viewing Zone, ten to twelve observations were used each year, again with an AOR falling within 3.5 days of each available maximum parallax factor. Slight adjustments to these constraints had to be made to fit within the calendar dates of each cycle, and other post facto adjustments were included when the observed data were corrupted by (rarely occurring) coronal mass ejection events from the Sun. These timing constraints were determined using a program written by one of us (ELW) that calculates the time of maximum parallax factor for each target using the predicted {\it Spitzer} ephemerides available from JPL Horizons\footnote{See {\url https://ssd.jpl.nasa.gov/?horizons}.}. 







Although program 13012 was granted a full two years of observation and is ongoing, this paper uses only those data taken by 2017 Oct 20 (UT). This date was chosen so that each target will have been observed through at least two maximum parallax factors in {\it Spitzer} Cycle 13. Future Cycle 13 observations along with data from our new Cycle 14 program (discussed further in Section~\ref{conclusions}) to be completed in 2019 Nov, will be included in a subsequent paper presenting our final parallactic measurements. A listing of the first and last observation date, total time baseline, and number of AORs per program for each target is given in Table~\ref{spitzer_targets}.

\startlongtable
\begin{deluxetable*}{lcccc}
\tabletypesize{\scriptsize}
\tablenum{3}
\tablecaption{Objects on the IRAC ch2 {\it Spitzer} Parallax Program\label{spitzer_targets}}
\tablehead{
\colhead{Object} &                          
\colhead{First Obs.\ Date} &
\colhead{Last Obs.\ Date} &     
\colhead{Baseline}  &
\colhead{Program \# (and \# of Epochs)} \\
\colhead{Name} &                          
\colhead{(UT)} &
\colhead{(UT)} &     
\colhead{(yr)}  &
\colhead{with ch2 Coverage} \\ 
\colhead{(1)} &                          
\colhead{(2)} &  
\colhead{(3)} &     
\colhead{(4)} &
\colhead{(5)}    
}
\startdata
WISE  0005+3737    & 2012 Sep 06 & 2017 Apr 22 &  4.6 & 80109(1), 90007(12), 13012(6)   \\
WISE  0015$-$4615  & 2010 Dec 17 & 2017 Sep 10 &  6.7 & 70062(2), 90007(12), 13012(5)   \\
WISE  0032$-$4946  & 2012 Jul 28 & 2017 Sep 12 &  5.1 & 80109(1), 90007(12), 13012(6)   \\
2MASS 0034+0523    & 2012 Feb 15 & 2017 Apr 14 &  5.2 & 80109(2), 90007(12), 13012(6)   \\
WISE  0038+2758    & 2012 Mar 22 & 2017 Apr 22 &  5.1 & 80109(2), 90007(14), 13012(6)   \\
WISE  0049+2151    & 2012 Mar 22 & 2017 Apr 24 &  5.1 & 80109(2), 90007(12), 13012(6)   \\
WISE  0123+4142    & 2011 Mar 25 & 2017 May 13 &  6.1 & 70062(1), 13012(6)   \\
CFBDS 0133+0231    & 2012 Sep 13 & 2017 Apr 24 &  4.6 & 80109(1), 90007(12), 13012(6)   \\
WISE  0146+4234AB  & 2011 Apr 05 & 2017 May 15 &  6.1 & 70062(1), 80109(2), 90007(12), 13012(6)   \\
WISE  0221+3842    & 2011 Mar 31 & 2017 May 17 &  6.1 & 70062(1), 13012(6)   \\
WISE  0226$-$0211AB& 2010 Sep 18 & 2017 May 08 &  6.6 & 70062(2), 80109(1), 13012(6)   \\
WISE  0233+3030    & 2012 Mar 22 & 2017 May 18 &  5.2 & 80109(2), 13012(6)   \\
WISE  0241$-$3653  & 2011 Jan 19 & 2017 Apr 23 &  6.3 & 70062(1), 80109(2), 90007(12), 13012(6)   \\
WISE  0247+3725    & 2011 Apr 12 & 2017 May 23 &  6.1 & 70062(1), 80109(1), 90007(12), 13012(6)   \\
WISE  0302$-$5817  & 2015 Feb 10 & 2017 Oct 02 &  2.6 & 11059(1), 13012(8)   \\
WISE  0304$-$2705  & 2014 Oct 20 & 2017 May 09 &  2.6 & 10135(1), 13012(6)   \\
WISE  0309$-$5016  & 2010 Dec 22 & 2017 May 05 &  6.4 & 70062(1), 80109(2), 13012(6)   \\
WISE  0313+7807    & 2010 Dec 21 & 2017 Jun 16 &  6.5 & 70062(2), 80109(2), 13012(6)   \\
WISE  0316+4307    & 2011 Nov 19 & 2017 Jun 02 &  5.5 & 80109(1), 90007(12), 13012(6)   \\
WISE  0323$-$5907  & 2010 Dec 18 & 2017 Apr 23 &  6.3 & 70062(2), 13012(6)   \\
WISE  0323$-$6025  & 2010 Sep 18 & 2017 Apr 23 &  6.6 & 70062(2), 80109(1), 90007(10), 13012(5)   \\
WISE  0325$-$5044  & 2011 Nov 19 & 2017 May 09 &  5.5 & 80109(1), 90007(10), 13012(5)   \\
WISE  0325$-$3854  & 2011 Jan 30 & 2017 May 03 &  6.3 & 70062(1), 80109(1), 13012(6)   \\
WISE  0325+0831    & 2011 Apr 12 & 2017 May 23 &  6.1 & 70062(1), 90007(12), 13012(6)   \\
WISE  0335+4310    & 2011 Apr 19 & 2017 Jun 05 &  6.1 & 70062(1), 80109(1), 90007(12), 13012(6)   \\
WISE  0336$-$0143  & 2011 Apr 12 & 2017 May 29 &  6.1 & 70062(1), 80109(1), 90007(12), 13012(6)   \\
WISE  0350$-$5658  & 2010 Sep 18 & 2017 May 13 &  6.6 & 70062(2), 80109(2), 90007(10), 13012(6)   \\
WISE  0359$-$5401  & 2010 Sep 18 & 2017 May 19 &  6.7 & 70062(2), 80109(2), 90007(12), 13012(5)   \\
WISE  0404$-$6420  & 2011 Nov 20 & 2017 May 05 &  5.5 & 80109(1), 90007(10), 13012(5)   \\
WISE  0410+1502    & 2010 Oct 21 & 2017 Jun 05 &  6.6 & 70062(2), 80109(2), 90007(12), 13012(6)   \\
WISE  0413$-$4750  & 2011 Nov 19 & 2017 May 28 &  5.5 & 80109(1), 13012(6)   \\
WISE  0430+4633    & 2012 Apr 19 & 2017 Jun 16 &  5.2 & 80109(1), 13012(6)   \\
WISE  0458+6434AB  & 2010 Oct 29 & 2017 Jun 22 &  6.6 & 70062(2), 80109(1), 13012(6)   \\
WISE  0500$-$1223  & 2010 Oct 18 & 2017 Jun 19 &  6.7 & 70062(2), 80109(1), 13012(6)   \\
WISE  0512$-$3004  & 2010 Dec 19 & 2017 Jun 20 &  6.5 & 70062(2), 80109(1), 90007(12), 13012(6)   \\
WISE  0535$-$7500  & 2010 Oct 17 & 2017 Sep 12 &  6.9 & 70062(2), 80109(2), 90007(10), 13012(6)   \\
WISE  0540+4832    & 2012 Apr 23 & 2017 Jun 02 &  5.1 & 80109(1), 90007(12), 13012(4)   \\
WISE  0614+0951    & 2010 Dec 17 & 2017 Jul 05 &  6.5 & 70062(2), 80109(2), 90007(12), 13012(6)   \\
WISE  0645$-$0302  & 2012 Jan 07 & 2017 Jul 20 &  5.5 & 80109(1), 13012(6)   \\
WISE  0647$-$6232  & 2010 Sep 19 & 2017 Sep 10 &  7.0 & 70062(2), 80109(1), 90007(10), 13012(6)   \\
WISE  0713$-$5854  & 2012 Mar 20 & 2017 Oct 13 &  5.6 & 80109(1), 13012(5)   \\
WISE  0713$-$2917  & 2012 Jan 02 & 2017 Aug 01 &  5.6 & 80109(1), 90007(12), 13012(6)   \\
WISE  0723+3403    & 2010 Nov 25 & 2017 Jul 24 &  6.7 & 70062(2), 80109(1), 13012(6)   \\
WISE  0734$-$7157  & 2011 Apr 19 & 2017 Aug 19 &  6.3 & 70062(1), 80109(1), 90007(10), 13012(5)   \\
WISE  0744+5628    & 2011 Jun 02 & 2017 Jul 18 &  6.1 & 70062(1), 80109(2), 90007(12), 13012(6)   \\
WISE  0759$-$4904  & 2011 Jan 02 & 2017 Sep 15 &  6.7 & 70062(2), 80109(1), 90007(10), 13012(6)   \\
WISE  0812+4021    & 2010 Dec 09 & 2017 Aug 02 &  6.6 & 70062(2), 80109(1), 90007(12), 13012(6)   \\
WISE  0825+2805    & 2012 Jan 07 & 2017 Aug 04 &  5.6 & 80109(2), 90007(12), 13012(6)   \\
WISE  0833+0052    & 2013 Jan 03 & 2017 Aug 13 &  4.6 & 80077(1), 13012(6)   \\
WISE  0836$-$1859  & 2011 Jan 01 & 2017 Aug 24 &  6.6 & 70062(2), 80109(1), 13012(6)   \\
WISE  0855$-$0714  & 2013 Jun 21 & 2017 Aug 24 &  4.2 & 90095(3), 10168(4), 13012(6)   \\
WISE  0857+5604    & 2010 Dec 18 & 2017 Aug 03 &  6.6 & 70062(2), 80109(1), 90007(12), 13012(6)   \\
WISE  0906+4735    & 2010 Dec 23 & 2017 Aug 04 &  6.6 & 70062(2), 80109(2), 13012(6)   \\
WISE  0914$-$3459  & 2012 Jun 19 & 2017 Sep 10 &  5.2 & 80109(1), 13012(6)   \\
WISE  0940$-$2208  & 2012 Feb 03 & 2017 Sep 10 &  5.6 & 80109(1), 13012(6)   \\
WISE  0943+3607    & 2010 Dec 23 & 2017 Aug 19 &  6.7 & 70062(2), 80109(1), 90007(12), 13012(6)   \\
WISE  0952+1955    & 2011 Jan 01 & 2017 Aug 26 &  6.7 & 70062(2), 80109(2), 13012(6)   \\
WISE  1018$-$2445  & 2011 Jan 26 & 2017 Sep 19 &  6.6 & 70062(1), 80109(2), 13012(6)   \\
WISE  1025+0307    & 2012 Jul 07 & 2017 Sep 13 &  5.2 & 80109(1), 90007(11), 13012(6)   \\
CFBDS 1028+5654    & 2010 Dec 20 & 2017 Aug 19 &  6.7 & 70021(1), 90007(12), 13012(6)   \\
WISE  1039$-$1600  & 2011 Jan 26 & 2017 Sep 19 &  6.6 & 70062(1), 80109(1), 90007(12), 13012(6)   \\
ULAS  1043+1048    & 2012 Feb 01 & 2017 Sep 09 &  5.6 & 70058(1), 13012(6)   \\
WISE  1050+5056    & 2012 Jun 12 & 2017 Aug 26 &  5.2 & 80109(1), 13012(6)   \\
WISE  1051$-$2138  & 2011 Mar 13 & 2017 Sep 30 &  6.6 & 70062(1), 90007(11), 13012(6)   \\
WISE  1052$-$1942  & 2011 Jan 31 & 2017 Sep 30 &  6.7 & 70062(1), 13012(6)   \\
WISE  1055$-$1652  & 2012 Jul 16 & 2017 Sep 30 &  5.2 & 80109(1), 90007(9), 13012(6)   \\
WISE  1124$-$0421  & 2012 Jul 19 & 2017 Sep 30 &  5.2 & 80109(1), 90007(11), 13012(6)   \\
WISE  1139$-$3324  & 2012 Aug 22 & 2017 Oct 13 &  5.1 & 80109(1), 13012(7)   \\
WISE  1141$-$3326  & 2012 Aug 22 & 2017 Oct 13 &  5.1 & 80109(1), 13012(7)   \\
WISE  1143+4431    & 2012 Mar 04 & 2017 Sep 10 &  5.5 & 80109(1), 13012(6)   \\
WISE  1150+6302    & 2010 Dec 23 & 2017 Aug 23 &  6.7 & 70062(2), 80109(1), 90007(12), 13012(6)   \\
ULAS  1152+1134    & 2012 Mar 18 & 2017 Sep 28 &  5.5 & 80109(1), 13012(5)   \\
WISE  1206+8401    & 2010 Dec 09 & 2017 Aug 03 &  6.6 & 70062(2), 80109(1), 90007(12), 13012(6)   \\
WISE  1217+1626AB  & 2011 Mar 12 & 2017 Sep 30 &  6.6 & 70062(1), 80109(2), 13012(6)   \\
WISE  1220+5407    & 2015 Feb 10 & 2017 Sep 10 &  2.6 & 11059(1), 13012(6)   \\
WISE  1221$-$3136  & 2012 Aug 12 & 2017 May 19 &  4.8 & 80109(1), 90007(12), 13012(5)   \\
WISE  1225$-$1013  & 2012 Mar 18 & 2017 Oct 12 &  5.6 & 80109(1), 90007(12), 13012(6)   \\
2MASS 1231+0847    & 2012 Sep 01 & 2017 Oct 09 &  5.1 & 80109(1), 13012(6)   \\
WISE  1243+8445    & 2011 Nov 21 & 2017 Jul 29 &  5.7 & 80109(1), 13012(6)   \\
WISE  1254$-$0728  & 2012 Mar 30 & 2017 Oct 20 &  5.6 & 80109(1), 13012(8)   \\
WISE  1257+4008    & 2011 Jan 29 & 2017 Sep 25 &  6.7 & 70062(1), 13012(6)   \\
VHS   1258$-$4412  & 2012 Apr 21 & 2017 Jun 06 &  5.1 & 80109(1), 13012(6)   \\
WISE  1301$-$0302  & 2012 Aug 22 & 2017 Oct 20 &  5.2 & 80109(1), 13012(8)   \\
WISE  1318$-$1758  & 2011 Apr 12 & 2017 May 27 &  6.1 & 70062(1), 80109(2), 90007(12), 13012(5)   \\
WISE  1333$-$1607  & 2011 Apr 12 & 2017 May 31 &  6.1 & 70062(1), 13012(5)   \\
WISE  1405+5534    & 2011 Jan 22 & 2017 Oct 10 &  6.7 & 70062(1), 80109(2), 90007(10), 13012(6)   \\
VHS   1433$-$0837  & 2011 Apr 19 & 2017 Jun 07 &  6.1 & 70062(1), 80109(2), 13012(6)   \\
WISE  1436$-$1814  & 2011 Apr 19 & 2017 Jun 10 &  6.1 & 70062(1), 80109(2), 90007(12), 13012(6)   \\
WISE  1448$-$2534  & 2012 Apr 29 & 2017 Jun 17 &  5.1 & 80109(1), 90007(12), 13012(6)   \\
WISE  1501$-$4004  & 2012 Apr 21 & 2017 Jun 28 &  5.2 & 80109(1), 13012(6)   \\
WISE  1517+0529    & 2012 Apr 04 & 2017 Jun 19 &  5.2 & 80109(1), 13012(6)   \\
WISE  1519+7009    & 2010 Dec 09 & 2017 Sep 25 &  6.8 & 70062(2), 80109(1), 90007(10), 13012(6)   \\
WISE  1523+3125    & 2011 Apr 16 & 2017 Jun 28 &  6.2 & 70062(1), 80109(2), 13012(6)   \\
WISE  1541$-$2250  & 2011 Apr 13 & 2017 Jun 26 &  6.2 & 70062(1), 80109(2), 90007(12), 13012(6)   \\
WISE  1542+2230    & 2011 Apr 18 & 2017 Jun 26 &  6.2 & 70062(1), 80109(2), 13012(6)   \\
WISE  1612$-$3420  & 2010 Sep 18 & 2017 Jul 05 &  6.8 & 70062(2), 80109(1), 13012(6)   \\ 
WISE  1614+1739    & 2010 Sep 20 & 2017 Jul 07 &  6.8 & 70062(2), 80109(2), 13012(6)   \\
2MASS 1615+1340    & 2012 Apr 28 & 2017 Jul 07 &  5.2 & 80109(2), 90007(12), 13012(6)   \\
WISE  1622$-$0959  & 2010 Sep 18 & 2017 Jul 05 &  6.8 & 70062(2), 80109(2), 13012(6)   \\
WISE  1639$-$6847  & 2012 May 12 & 2017 Aug 01 &  4.2\tablenotemark{a} & 
                                                        80109(1), 90007(12), 11059(1), 13012(6)   \\
WISE  1653+4444    & 2010 Sep 20 & 2017 May 31 &  6.7 & 70062(2), 80109(2), 90007(10), 13012(4)   \\
WISE  1711+3500AB  & 2010 Sep 18 & 2017 Aug 24 &  6.9 & 70062(2), 80109(1), 90007(12), 13012(6)   \\
WISE  1717+6128    & 2010 Sep 19 & 2017 Jul 15 &  6.8 & 70062(2), 80109(1), 13012(6)   \\
WISE  1721+1117    & 2012 May 12 & 2017 Jul 24 &  5.2 & 80109(1), 13012(6)   \\
WISE  1735$-$8209  & 2011 Jun 17 & 2017 Sep 07 &  6.2 & 70062(1), 13012(6)   \\
WISE  1738+2732    & 2010 Sep 18 & 2017 Aug 14 &  6.9 & 70062(2), 80109(2), 90007(12), 13012(6)   \\
WISE  1804+3117    & 2010 Sep 26 & 2017 Sep 04 &  6.9 & 70062(2), 80109(2), 13012(6)   \\
WISE  1812+2007    & 2011 Jun 11 & 2017 Aug 15 &  6.2 & 70062(1), 80109(1), 13012(6)   \\
WISE  1813+2835    & 2012 May 19 & 2017 Sep 01 &  5.3 & 80109(1), 90007(12), 13012(6)   \\
WISE  1828+2650    & 2010 Jul 10 & 2017 Aug 30 &  7.1 & 551(1), 70062(1), 80109(2), 90007(12), 13012(6)   \\
WISE  1928+2356    & 2012 Jun 21 & 2017 Sep 14 &  5.2 & 80109(1), 90007(12), 13012(6)   \\
WISE  1955$-$2540  & 2011 Jun 26 & 2017 Aug 24 &  6.2 & 70062(1), 90007(11), 13012(6)   \\
WISE  1959$-$3338  & 2011 Jun 07 & 2017 Aug 24 &  6.2 & 70062(2), 80109(2), 90007(12), 13012(6)   \\
WISE  2000+3629    & 2012 Jul 18 & 2017 Oct 17 &  5.2 & 80109(1), 90007(12), 13012(7)   \\
WISE  2005+5424    & 2010 Dec 31 & 2017 Sep 10 &  6.7 & 70062(2), 80109(1), 13012(7)   \\
WISE  2015+6646    & 2010 Dec 31 & 2017 May 19 &  6.4 & 70062(2), 90007(10), 13012(5)   \\
WISE  2017$-$3421  & 2011 Jun 09 & 2017 Sep 01 &  6.2 & 70062(1), 13012(6)   \\
WISE  2019$-$1148  & 2011 Jun 15 & 2017 Sep 01 &  6.2 & 70062(1), 80109(2), 13012(6)   \\
WISE  2056+1459    & 2010 Dec 10 & 2017 Oct 01 &  6.8 & 70062(2), 80109(2), 90007(12), 13012(6)   \\
WISE  2147$-$1029  & 2010 Dec 22 & 2017 Sep 25 &  6.8 & 70062(2), 13012(6)   \\
WISE  2157+2659    & 2010 Dec 23 & 2017 Sep 20 &  6.7 & 70062(2), 80109(2), 90007(12), 13012(6)   \\
WISE  2159$-$4808  & 2011 Jun 22 & 2017 Sep 21 &  6.2 & 70062(2), 80109(1), 90007(12), 13012(6)   \\
WISE  2203+4619    & 2014 Sep 22 & 2017 Oct 01 &  3.0 & 10135(1), 13012(7)   \\
WISE  2209+2711    & 2010 Dec 31 & 2017 Sep 23 &  6.7 & 70062(2), 80109(1), 90007(12), 13012(6)   \\
WISE  2209$-$2734  & 2010 Dec 10 & 2017 Sep 23 &  6.8 & 70062(2), 80109(2), 90007(11), 13012(6)   \\
WISE  2211$-$4758  & 2011 Nov 30 & 2017 Sep 19 &  5.8 & 80109(1), 13012(6)   \\
WISE  2212$-$6931  & 2011 Jun 15 & 2017 Sep 22 &  6.3 & 70062(1), 80109(1), 90007(12), 13012(6)   \\
WISE  2220$-$3628  & 2012 Jan 23 & 2017 Sep 24 &  5.7 & 80109(2), 90007(12), 13012(6)   \\
WISE  2232$-$5730  & 2010 Nov 15 & 2017 Sep 24 &  6.9 & 70062(2), 13012(6)   \\
WISE  2237+7228    & 2012 Apr 15 & 2017 Jun 02 &  5.1 & 80109(2), 90007(10), 13012(6)   \\
WISE  2255$-$3118  & 2010 Dec 17 & 2017 Oct 06 &  6.8 & 70062(2), 80109(2), 13012(6)   \\
WISE  2301+0216    & 2012 Jan 13 & 2017 Oct 16 &  5.8 & 80109(2), 90007(12), 13012(6)   \\
WISE  2302$-$7134  & 2011 Jun 17 & 2017 Oct 07 &  6.3 & 70062(1), 13012(6)   \\
WISE  2313$-$8037  & 2011 Jun 17 & 2017 Sep 22 &  6.3 & 70062(1), 80109(2), 90007(10), 13012(6)   \\
WISE  2319$-$1844  & 2010 Dec 31 & 2017 Oct 12 &  6.8 & 70062(2), 80109(1), 90007(12), 13012(6)   \\
ULAS  2321+1354    & 2009 Aug 19 & 2017 Oct 13 &  8.2 & 60093(1), 70021(1), 80109(2), 90007(12), 13012(7)   \\
ULAS  2326+0201    & 2012 Jan 31 & 2017 Oct 09 &  5.7 & 80077(1), 13012(6)   \\
WISE  2332$-$4325  & 2010 Dec 18 & 2017 Oct 12 &  6.8 & 70062(2), 80109(1), 90007(12), 13012(6)   \\
WISE  2343$-$7418  & 2011 Jun 17 & 2017 Oct 07 &  6.3 & 70062(1), 80109(2), 90007(10), 13012(6)   \\
WISE  2344+1034    & 2011 Jan 18 & 2017 Oct 04 &  6.7 & 70062(1), 80109(2), 13012(6)   \\
WISE  2354+0240    & 2012 Sep 03 & 2017 Oct 04 &  5.1 & 80109(1), 13012(6)   \\
WISE  2357+1227    & 2012 Sep 03 & 2017 Oct 07 &  5.1 & 80109(1), 13012(6)   \\
\enddata
\tablenotetext{a}{1639$-$6847: In the observation from program 80109 and the first observation from program 90007, the object is blended with a background source. This reduces the usable time baseline for astrometric measurements from 5.2 to 4.2 yr because the first clean image was taken 2013 May 19 (UT).}
\end{deluxetable*}


For a few targets, the disentangling of proper motion from parallax benefited from using ch2 observations taken prior to our own: 

\begin{enumerate}

\item WISE 0304$-$2705 and WISE 2203+4619: We supplemented our data with an earlier observation from program 10135 (D.J.\ Pinfield, PI) that used a sixteen-point spiral dither pattern with medium spacing, two frames at each dither position, and exposure times of 30 s per frame. 

\item WISE 0833+0052 and ULAS 2326+0201: We used an earlier observation from program 80077 (S.K.\ Leggett, PI) that used a twelve-point Reuleaux dither pattern with medium spacing, four frames at each dither position, and exposure times of 30 s per frame. 

\item WISE 0855$-$0714: We used earlier observations from programs 90095 and 10168 (K.L.\ Luhman, PI). Program 90095 acquired three separate epochs on the target object, using a 5-point Gaussian dither with small spacing and exposure times of 30 s per frame. Program 10168 acquired four separate epochs using a 9-point random dither pattern of small spacing and exposure times of 30 s per frame. 

\item CFBDS 1028+5654: An earlier observation from program 70021 (K.L.\ Luhman, PI) was used to extend our time baseline. This program acquired data using a 5-point Gaussian dither with small spacing, one frame per dither position, and exposure times of 30 s per frame. 

\item ULAS 1043+1048: An observation from program 70058 (S.K.\ Leggett, PI) was used to extend the time baseline for this object. This program took data using a sixteen-point spiral dither pattern with medium spacing, six frames at each dither position, and exposure times of 30 s per frame. 

\item WISE 1828+2650: Data from program 551 (A.K.\ Mainzer, PI) were the first taken as part of the WISE brown dwarf team's follow-up of WISE color-seleted objects. This program, which targeted only WISE 1828+2650 at a single epoch, used a twelve-point Reuleaux dither pattern with medium spacing, one frame per dither position, and exposure times of 12 s per frame. 

\item ULAS 2321+1354: Program 60093 (S.K.\ Leggett, PI) provided a supplemental data point for this object. Data were taken using a five-point Gaussian dither pattern with medium spacing, one frame per dither, and exposure times of 30 s per frame.

\end{enumerate}

\section{Spitzer Reductions\label{reductions}}

\subsection{Photometry in ch1 and ch2}

With the exception of Gl 27B (\citealt{luhman2007}) and WD0806-661B (\citealt{luhman2012}), the {\it Spitzer} photometry for all sources was measured by CRG and JDK as follows. For each object, we performed aperture photometry on an AOR that included both ch1 and ch2 data so that measurements were contemporaneous. These aperture measurements were done either on custom mosaics of the corrected basic calibrated data (CBCD) frames or on the post-basic calibrated data (pBCD) mosaics archived at the {\it Spitzer} Heritage Archive at IRSA\footnote{See \url{http://irsa.ipac.caltech.edu}.} using the MOsaicker and Point Source EXtractor with point source extraction package (MOPEX/APEX), also available at IRSA. As mentioned in \cite{kirkpatrick2011}, for objects as bright as our targets, we have seen negligible difference in the photometry between the two methods. Our scripts used a 4-pixel aperture ({\it aperture1} in the MOPEX output files) with a 24-to-40-pixel sky annulus. Resultant raw fluxes were then multiplied by the aperture corrections recommended in Table 4.7 of the IRAC Instrument Handbook, also available at IRSA -- 1.208 for ch1 and 1.221 for ch2 -- to obtain the flux in units of $\mu$Jy, and then converted to magnitudes using the flux zero points in the Handbook's Table 4.1 (280.9$\pm$4.1 Jy in  ch1 and 179.7$\pm$2.6 Jy in ch2), propagating the uncertainty in zero point and flux into the final measurement error. This resulting photometry, along with the AOR number from which the photometry was extracted, is given in columns 7-9 of Table~\ref{20pc_sample_photometry}.

\subsection{Astrometry in ch2}

Our basic astrometric reduction methodology is discussed in detail in \cite{martin2018}. Whereas ECM wrote Python code for the reductions presented in that paper, JDK wrote independent IDL code and other wrappers for the reductions presented here, and those included one fundamental change to improve astrometric repeatability. We outline the basic steps below.

\subsubsection{Identifying Gaia Astrometric Anchor Points\label{mosaicking_issue}}

For each AOR, we created a mosaic from the individual CBCD frames and extracted astrometry and photometry of detected sources using the MOPEX/APEX software. We examined the resulting mosaics at all epochs to ensure that the target object was near the center of the field, as expected. (This step eliminated three AORs for WISE 2301+0216 -- 46523392, 46523136, and 46522880 -- from further use, due to a typographical error in the requested pointings for those epochs.) In addition to these per-epoch mosaics, we also created a supermosaic and associated source list by combining all available CBCD files for each object over the full time baseline. We then used the {\it tmatch2} routine in the Starlink Tables Infrastructure Library Tool Set (STILTS; \citealt{taylor2006}) and a 3-arcsec radius to match sources from each per-epoch mosaic source list to the supermosaic source list. Parameters were set so that we retained for each supermosaic source only the closest match in the per-epoch source list ($find=best1$) and listed only those supermosaic sources having a match ($join=1and2$). 

We then used the Interactive Data Language (IDL) to perform a preliminary re-registration on each per-epoch mosaic to place it on the same astrometric grid as the supermosaic. Only those objects with S/N $\gtrapprox$ 30 on the per-epoch mosaics were used. We found that a simple translation in RA and Dec provided excellent results. The per-frame offsets were computed by performing a robust mean of the offsets and including a 3$\sigma$ clipping of outliers. 
We uploaded this list of high-S/N re-registration stars into IRSA and matched to the {\it Gaia} Data Release 1 (DR1) catalog using a 1 arcsec search radius. Most of these stars have {\it Gaia} matches, since they are bright at both {\it Gaia} $G$ band and IRAC ch2 band, but we dropped any {\it Gaia} matches for which the {\it Gaia} DR1 astrometric uncertainties were $\gtrapprox$100 milliarcsec.

Once the {\it Gaia} astrometric anchor stars had been identified in each field, more precise refinements to the astrometry could be performed, since the {\it Gaia} DR1 positions for these well detected objects typically have uncertainties of under a milliarcsec. Originally, we wrote code like that described in \cite{martin2018}, to do these refinements on the source lists derived from the per-epoch mosaics, but we found that this was degrading our final astrometric precision. After some experimentation, we concluded that the mosaicking step in MOPEX was responsible. As described below, the actual astrometric measurements for this paper use source positions as measured on the individual {\it frames}, not the per-epoch mosaics. 

\subsubsection{Astrometric Refinements to Frame Source Positions\label{astrometric_refinements}}

For each AOR, we ran scripts in MOPEX/APEX to measure the positions and photometry of our target and its associated re-registration stars on each individual frame. These scripts used the CBCD, uncertainty (CBUNC), and mask (BIMSK) files available in the {\it Spitzer} Heritage Archive. Data were pulled from the {\it Spitzer} Heritage Archive after all image headers had been updated by the {\it Spitzer} Science Center to incorporate the new fifth-order correction to the array distortion\footnote{See also {\url https://irachpp.spitzer.caltech.edu/page/dist\_correct}} (\citealt{lowrance2014,lowrance2016}). Point response function (PRF) fitting was done using a set of warm PRFs built by one of us (JGI), adapted from the cold mission PRFs to mimic the intrapixel gain variations measured with warm IRAC.  Because the optical path did not change from the cold to warm {\it Spitzer} missions, the optical point {\it spread} function (PSF) did not change, only the overall detector sensitivity.  In other words, while the dynamic range of the PRF has changed, its spatial structure has not. As usual, the PRFs we used are tabulated images of point sources at various sub-pixel offsets and therefore strongly mitigate the astrometric bias caused by intrapixel distortion (\citealt{ingalls2012, ingalls2014}), which is a consequence of the fact that IRAC data are undersampled and the intrinsic sensitivity of an IRAC detector is not uniform across a pixel.  While this analysis was underway, warm mission PRFs developed by \cite{hora2012} were released on the IRSA website\footnote{See {\url http://irsa.ipac.caltech.edu/data/SPITZER/docs/irac/calibrationfiles/}.}, but, as expected, the astrometric differences between the official PRFs and the ones used here were found to be negligible.

The resulting, native astrometry of the re-registration stars on each frame was then compared to the {\it Gaia} DR1 positions of these same sources. Specifically, we used the ($x$, $y$) positions computed by APEX and translated these into (RA, Dec) using the World Coordinate System (WCS) information in the frame FITS header\footnote{Note that FITS and APEX use the convention that the first pixel is located at (1,1), whereas IDL convention places this at (0,0).}. Na\"ively, one would assume that the ($x$, $y$) positions reported by APEX would be true array coordinates, but through experimentation we determined that these positions are actually rubber-sheeted versions; i.e., ones with the fifth-order distortion correction already included. Hence, we were able to convert to (RA, Dec) using the WCS specifications only.

With the ($x$, $y$)-derived sky positions in hand for both the re-registration stars and the target, we converted all to tangent plane coordinates ($\xi$, $\eta$), where we used the mean position of our target across all epochs as the tangent point itself; i.e., ($\xi$, $\eta$) = (0,0). We then computed the ($\Delta\xi$, $\Delta\eta$) values between the ($x$, $y$)-derived sky positions for the re-registration stars and their {\it Gaia} reported positions, and computed a robust mean of the differences using 3$\sigma$ outlier clipping and a simple ($\xi$, $\eta$) translation. These mean differences were then used to place the ($\xi$, $\eta$) coordinates of the target object and re-registration stars onto the {\it Gaia} reference frame. 

For most of our targets, we were able to find a sufficient number of {\it Gaia} DR1 stars above a single-frame S/N value of 100 to perform an adequate re-registration. For some targets in sparser fields, however, as few as three re-registration stars were found above S/N = 100. 
(For the two targets closest to the Galactic Center, WISE 1928+2356 and WISE 2000+3629, we had the opposite problem of too many potential re-registration stars, so we reset the selection threshold for re-registration stars to S/N $>$ 500.) For any target having ten or more re-registration stars with S/N $\ge$ 100, we found that the inclusion of re-registration stars with 30 $<$ S/N $<$ 100 did not generally decrease the reduced $\chi^2$ values in the final proper motion and parallax fits. For targets having fewer than ten re-registration stars with S/N $\ge$ 100, however, these reduced $\chi^2$ values were generally much better. For these targets, a single-frame S/N floor of $\sim$30 was used for selection of the re-registration stars. With this adjustment, no fewer than 5 re-registation stars were used for each field.

\begin{figure}
\figurenum{1}
\includegraphics[scale=0.4,angle=0]{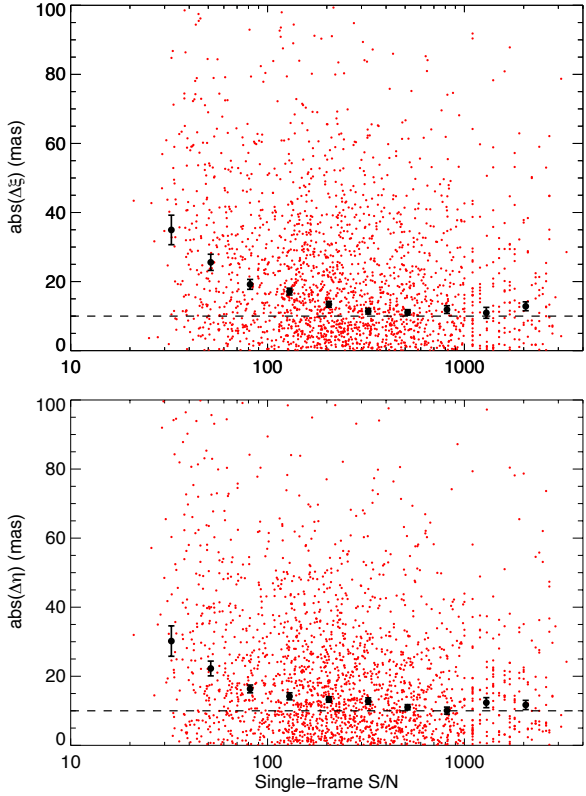}
\caption{Astrometric repeatability as a function of the single-frame signal-to-noise ratio. Plotted in red are the per-axis differences in position ($\xi$ in the top panel and $\eta$ in the bottom) between our measurements and those of {\it Gaia} DR1. These are the mean position in the AOR determined from the (generally) nine independent measurements in each individual frame. Shown in the figure are sources from one representative AOR from each of our 142 targets, excepting nineteen with $|{\beta}| < 15^\circ$ that have been removed since their higher source densities would otherwise dominate the plot. The black points are robust means -- determined with a 1$\sigma$ clipping to remove outliers -- that sample ten S/N bins over the range 25 $<$ S/N $<$ 2500 using logarithmic spacing. As shown by the dashed black lines, an asymptotic floor of $\sim$10 mas is reached at very high S/N values.
\label{astrometric_repeatability}}
\end{figure}

For the target object, the robust mean of these per-frame adjusted positions (again using 3$\sigma$ clipping) was used as the measured position of the object for this AOR. For the astrometric uncertainty on each AOR position, we use a measurement of the positional repeatability for stars lying at a S/N value similar to the target. Figure~\ref{astrometric_repeatability} illustrates this repeatability. The plot demonstrates that our target T and Y dwarfs, which lie in the single-frame S/N range between 70 and 200, have per-axis astrometric precisions of $\sim$15 mas. There are some AORs for which, due to small number statistics, this measured repeatability falls below 10 mas. For these, we set a floor of 10 mas on the per-axis positional uncertainty of the target, since the broader analysis in Figure~\ref{astrometric_repeatability} shows this is the best we actually achieve. For each AOR, the mean ($\xi$, $\eta$) position of the target was then converted back to (RA, Dec). The FITS header of the middle frame in each AOR was then consulted for the UT date and barycentric (X,Y,Z) position of the {\it Spitzer} Space Telescope at the time of observation. 

\subsubsection{Fitting for Proper Motion and Parallax}

For each target, our list of measured astrometry and its uncertainties along with the observation times and locations of the observing platform were fed into a least-squares fitting code to determine the proper motion, parallax, and position at a fiducial time. We supply two non-linear equations (see \citealt{smart1977}, \citealt{green1985}, \citealt{beichman2014}) to the code:
\begin{equation}
\begin{split}
&\alpha_{residual} = \Big( \alpha(t) - \alpha^\prime\Big) cos\,\delta^\prime \\
&-\pi_{trig}\Big( X(t) sin\,\alpha^\prime - Y(t) cos\,\alpha^\prime \Big) \Big) /3600 
\end{split}
\end{equation}
\begin{equation}
\begin{split}
&\delta_{residual} = \delta(t) - \delta^\prime \\
&-\Big( \pi_{trig}\Big( X(t) cos\,\alpha^\prime sin \,\delta^\prime+ Y(t) sin\,\alpha^\prime sin\,\delta^\prime -Z(t) cos\,\delta^\prime \Big) /3600
\end{split}
\end{equation}
where
\begin{equation}
\alpha^\prime= \alpha_0+(\mu_\alpha(t-t_0)/cos\,\delta^\prime)/3600
\end{equation}
\begin{equation}
\delta^\prime= \delta_0+(\mu_\delta(t-t_0))/3600
\end{equation}

Here, $\alpha(t)$ and $\delta(t)$ are the RA and Dec positions (and their uncertainties) of the target in degrees; $X(t)$, $Y(t)$, and $Z(t)$ are the locations of the spacecraft in astronomical units; $t_0$ is a fiducial time in years;  $\alpha_0$ and $\delta_0$ are the RA and Dec in degrees at time $t_0$; $\mu_\alpha$ and $\mu_\delta$ are the RA and Dec components of the proper motion in arcsec yr$^{-1}$; and $\pi_{trig}$ is the trigonometric parallax in arcsec. We employed the IDL module {\tt mpfitfun} (\citealt{markwardt2009}), which uses the Levenberg-Marquardt least-squares algorithm to attempt to drive both $\alpha_{residual}$ and $\delta_{residual}$ simultaneously to zero, given the observational data and their uncertainties. 

The fiducial time was chosen to be at a point in the middle of our observational data set, epoch 2014.0, since setting this to be within the timeframe of the observations minimizes the associated uncertainties. Choosing a much later or earlier time, say 2000.0, would result in larger positional uncertainties since the uncertainty from the proper motion compounds with the difference in the time interval between the observations and the chosen fiducial time.

Table~\ref{spitzer_results} gives the best fitting solutions for each of our 142 targets. The abbreviated object name is in column 1 followed by the position of the object at epoch 2014.0 along with the positional uncertainties in column 2-5. The value of parallax obtained by the best-fit solution above is a measurement relative to the background stars and requires an adjustment to the absolute reference frame as described further below. The resulting value of the absolute parallax and the correction needed to convert from relative to absolute units are given in columns 6 and 7. The best-fit proper motion per axis and its uncertainties are listed in columns 8 and 9. The $\chi^2$ value of the best fit, the number of degrees of freedom, and the reduced $\chi^2$ value are listed in columns 10-12. The number of {\it Spitzer} epochs used in the solution is shown in column 13. Finally, the number of re-registration stars used along with a flag value for the S/N floor used for registration star selection ("H" for the high S/N cut and "L" for the low S/N cut) is given in column 14.

\clearpage

\startlongtable
\begin{deluxetable*}{lrrrrrrrrrrrrr}
\tabletypesize{\footnotesize}
\tablenum{4}
\tablecaption{Parallax and Motion Fits for Objects on the {\it Spitzer} Parallax Program\label{spitzer_results}}
\tablehead{
\colhead{Object} &                          
\colhead{J2000 RA} &
\colhead{RA} &
\colhead{J2000 Dec} &
\colhead{Dec} &
\colhead{$\pi_{abs}$} &
\colhead{$corr_{\pi}$} &
\colhead{$\mu_{RA}$} &
\colhead{$\mu_{Dec}$} &
\colhead{$\chi^2$} &                          
\colhead{Dof} &
\colhead{Red.} &
\colhead{\# of} &
\colhead{\# of} \\
\colhead{Name} &
\colhead{Ep.\ 2014.0} &
\colhead{Unc.} &
\colhead{Ep.\ 2014.0} &
\colhead{Unc.} &
\colhead{(mas)} &
\colhead{(mas)} &
\colhead{(mas yr$^{-1}$)} &
\colhead{(mas yr$^{-1}$)} &
\colhead{} &                          
\colhead{} &
\colhead{$\chi^2$} &
\colhead{Ep.} &
\colhead{Reg.} \\
\colhead{} &
\colhead{(deg)} &
\colhead{(mas)} &
\colhead{(deg)} &
\colhead{(mas)} &
\colhead{} &
\colhead{} &
\colhead{} &
\colhead{} &
\colhead{} &                          
\colhead{} &
\colhead{} &
\colhead{} &
\colhead{Stars\tablenotemark{a}} \\
\colhead{(1)} &                          
\colhead{(2)} &  
\colhead{(3)} &  
\colhead{(4)} &
\colhead{(5)} &
\colhead{(6)} &
\colhead{(7)} &
\colhead{(8)} &
\colhead{(9)} &                          
\colhead{(10)} &
\colhead{(11)} &
\colhead{(12)} &
\colhead{(13)} &
\colhead{(14)} \\
}
\startdata
 WISE 0005+3737    &   1.324120&  2.6&  37.622066&  2.5& 127.0$\pm$2.4& 1.3$\pm$0.4&   998.9$\pm$1.4&  -267.0$\pm$1.4&  19.925&  33&  0.604& 19&  22H \\
 WISE 0015$-$4615  &   3.775288&  2.7& -46.255526&  2.6&  71.8$\pm$2.7& 1.5$\pm$0.5&   409.7$\pm$1.5&  -682.9$\pm$1.3&  30.137&  33&  0.913& 19&  15L \\
 WISE 0032$-$4946  &   8.128906&  3.3& -49.781951&  3.0&  63.6$\pm$2.9& 1.4$\pm$0.3&  -379.3$\pm$1.7&  -858.1$\pm$1.7&  28.256&  33&  0.856& 19&  13L \\
 2MASS 0034+0523   &   8.717410&  3.3&   5.385402&  3.7& 120.1$\pm$3.0& 1.6$\pm$0.4&   672.6$\pm$1.7&   184.7$\pm$1.9&  17.909&  35&  0.512& 20&  16L \\
 WISE 0038+2758    &   9.621136&  2.5&  27.981175&  2.5&  89.7$\pm$2.5& 1.5$\pm$0.6&   -15.4$\pm$1.6&    96.3$\pm$1.5&  37.107&  39&  0.951& 22&  12H \\
 WISE 0049+2151    &  12.439742&  2.5&  21.855396&  2.5& 139.9$\pm$2.5& 1.5$\pm$0.5&  -483.2$\pm$1.4&   -46.5$\pm$1.4&  18.928&  35&  0.541& 20&  15L \\
 WISE 0123+4142    &  20.889465&  8.1&  41.701367&  5.6&  44.0$\pm$4.1& 1.3$\pm$0.3&   601.7$\pm$2.6&    91.5$\pm$1.9&   5.439&   9&  0.604&  7&  10H \\
 CFBDS 0133+0231   &  23.260890&  3.7&   2.524552&  2.7&  56.3$\pm$3.4& 2.2$\pm$1.2&   598.0$\pm$2.0&  -113.9$\pm$1.5&  44.910&  33&  1.361& 19&   8L \\
\enddata
\tablenotetext{a}{The letter following the number of re-registration stars indicates whether we used the high-S/N ("H") or low-S/N ("L") limit for their selection. See Section~\ref{astrometric_refinements} for details.}
\end{deluxetable*}

We are measuring parallaxes relative to re-registration stars in each field, so we need to account for the fact that those re-registration stars themselves each have a small parallax that is partially damping the parallactic signal of our target. In order to correct for this, we need estimates of the mean distance to the re-registration stars. To do this, we tabulated the stars' $J$-band magnitudes measured from the 2MASS All-Sky Point Source Catalog, supplemented in some cases by the 2MASS Survey Point Source Reject Table. The distance to each object is then estimated by comparing that magnitude to a model prediction of Galactic structure for that region of sky, as provided by \cite{mendez1996}. For fields where 2MASS $J$ magnitudes were available for all re-registration stars, we found that the correction was very small and varied from 0.4 to 2.2 mas. For some fields, not all of our re-registration stars were detected by 2MASS, so these stars were assigned a floor value for $J = 17.0$. To test whether this is a reasonable assumption, we chose WISE 0032$-$4946 because five of its thirteen re-registration stars have no 2MASS mags. Setting all five of these to have $J = 17.0$ mag gives a correction of 1.4$\pm$0.3 mas. If we instead set all five to an absurdly faint value of $J = 19.0$ mag, we find a correction of 1.1$\pm$0.6 mas. Because these two assumptions give corrections that are essentially identical within the error bars and the former is likely more realistic, we assumed a floor of $J = 17.0$ mag (or $J = 16.5$ mag in fields of higher source density) for all 2MASS undetected sources. 

\begin{figure*}
\figurenum{2}
\plotone{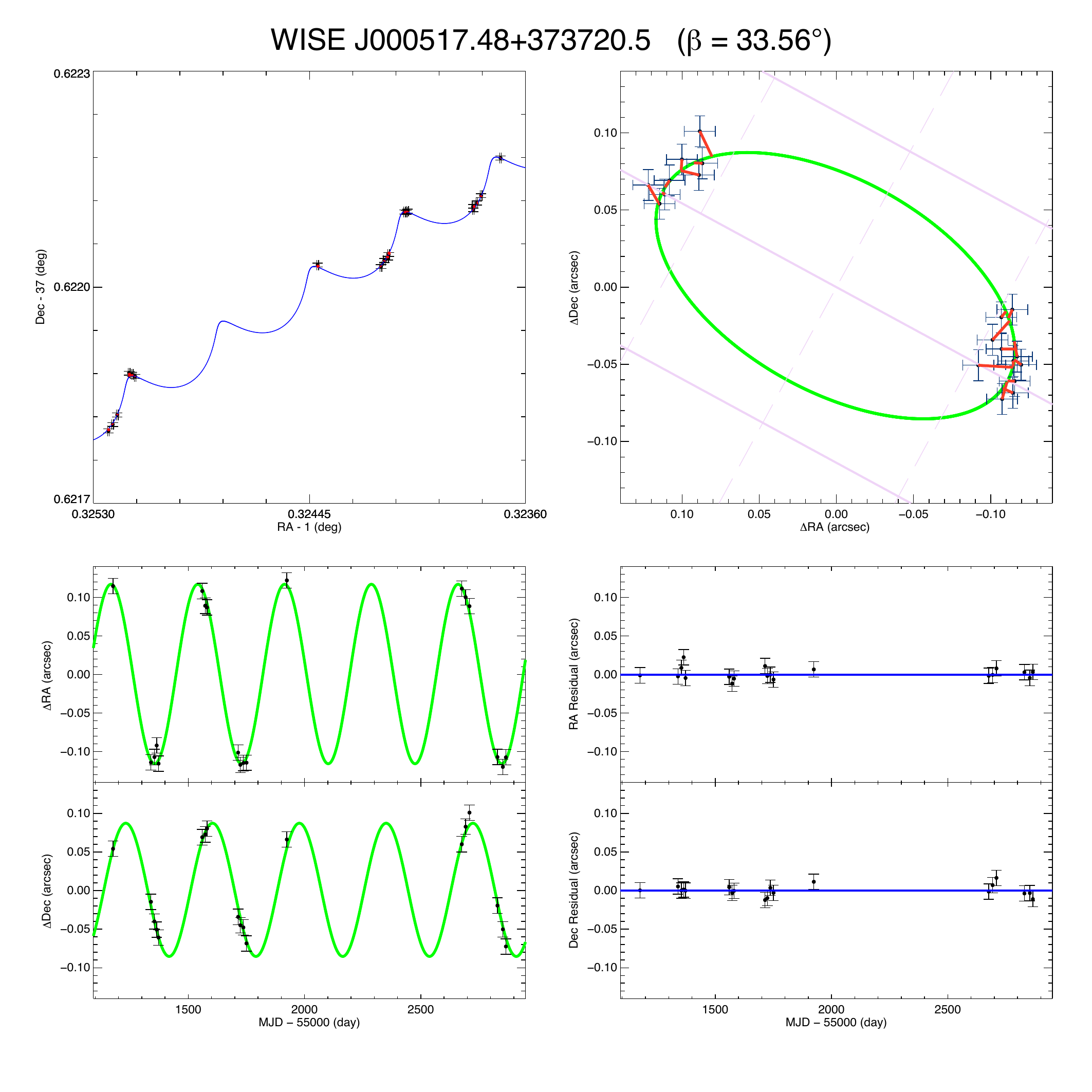}
\caption{Plots of our astrometric measurements and their best fits, divided 
into four panels. (Upper left) The measured astrometry and its
uncertainty at each epoch (black points with error bars) plotted 
in RA and Dec with the best fit model shown as the blue curve. Red 
lines connect each observation to its corresponding time point 
along the best-fit curve. (Upper right) A square patch of sky 
centered at the mean equatorial position of the target. The green 
curve is the parallactic fit, which is just the blue curve in the 
previous panel with the proper motion vector removed. In the 
background is the ecliptic coordinate grid, with lines of constant 
$\beta$ shown in solid pale purple and lines of constant $\lambda$ 
shown in dashed pale purple. Grid lines are shown at 0$\farcs$1 
spacing. (Lower left) The change in RA and Dec as a function of time
with the proper motion component removed. The parallactic fit is again 
shown in green. (Lower right) The overall RA and Dec residuals 
between the observations and the best fit model as a function of 
time. The complete figure set contains these plots for all 142
objects in the {\it Spitzer} parallax program.
\label{plots_0005p3737}}
\end{figure*}

Plots of our astrometric measurements and their best fits are shown in Figure~\ref{plots_0005p3737}. 
\figsetstart
\figsetnum{2}
\figsettitle{Parallax and proper motion fits to the 142 objects in the Spitzer parallax program}

\figsetgrpstart
\figsetgrpnum{2.1}
\figsetgrptitle{Parallax and proper motion fit for WISE 0005+3737}
\figsetplot{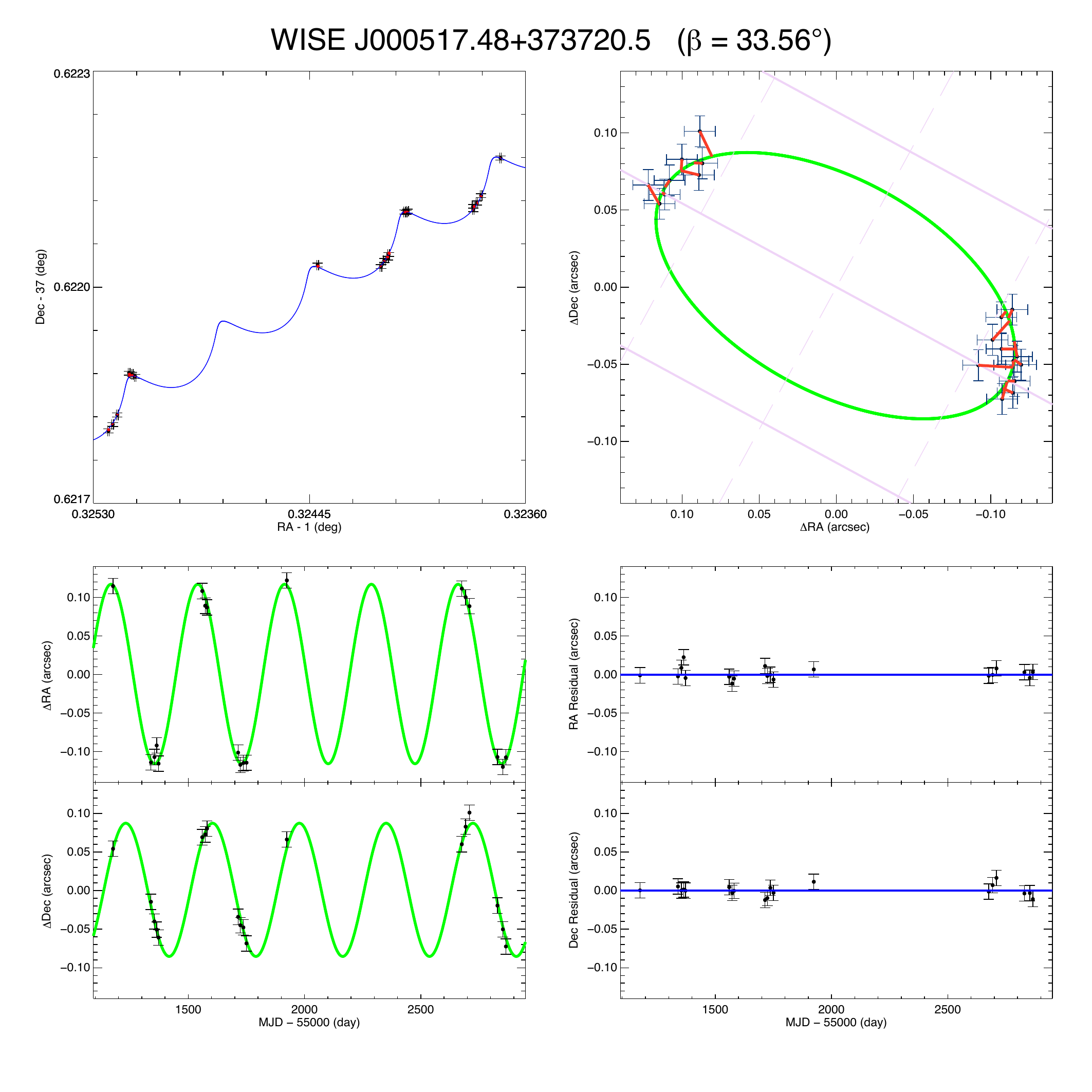}
\figsetgrpnote{Plots of our astrometric measurements and their best fits, divided 
into four panels. (Upper left) The measured astrometry and its
uncertainty at each epoch (black points with error bars) plotted 
in RA and Dec with the best fit model shown as the blue curve. Red 
lines connect each observation to its corresponding time point 
along the best-fit curve. (Upper right) A square patch of sky 
centered at the mean equatorial position of the target. The green 
curve is the parallactic fit, which is just the blue curve in the 
previous panel with the proper motion vector removed. In the 
background is the ecliptic coordinate grid, with lines of constant 
$\beta$ shown in solid pale purple and lines of constant $\lambda$ 
shown in dashed pale purple. Grid lines are shown at 0$\farcs$1 
spacing. (Lower left) The change in RA and Dec as a function of time
with the proper motion component removed. The parallactic fit is again 
shown in green. (Lower right) The overall RA and Dec residuals 
between the observations and the best fit model as a function of 
time.}
\figsetgrpend

\figsetend

We note from Table~\ref{spitzer_results} that most of our objects have values of reduced $\chi^2$ near 1.0, confirming that our methodology for measuring the astrometric uncertainties per epoch is sound. This enables us to identify targets for which either our centroiding is poor because a marginally resolved companion creates a profile not well fit by our PRF, or our single-object solution is a poor assumption due to the presence of an unseen companion. Objects with fits having reduced $\chi^2$ values $>$2  have been placed on continued monitoring with {\it Spitzer} through the end of Cycle 14 to see if any cyclical variations in the residuals, indicative of an unseen astrometric companion, can be found. These and other targets of interest are discussed in Section~\ref{notes_on_individual_objects}.

\begin{figure*}
\figurenum{3}
\includegraphics[scale=0.60,angle=0]{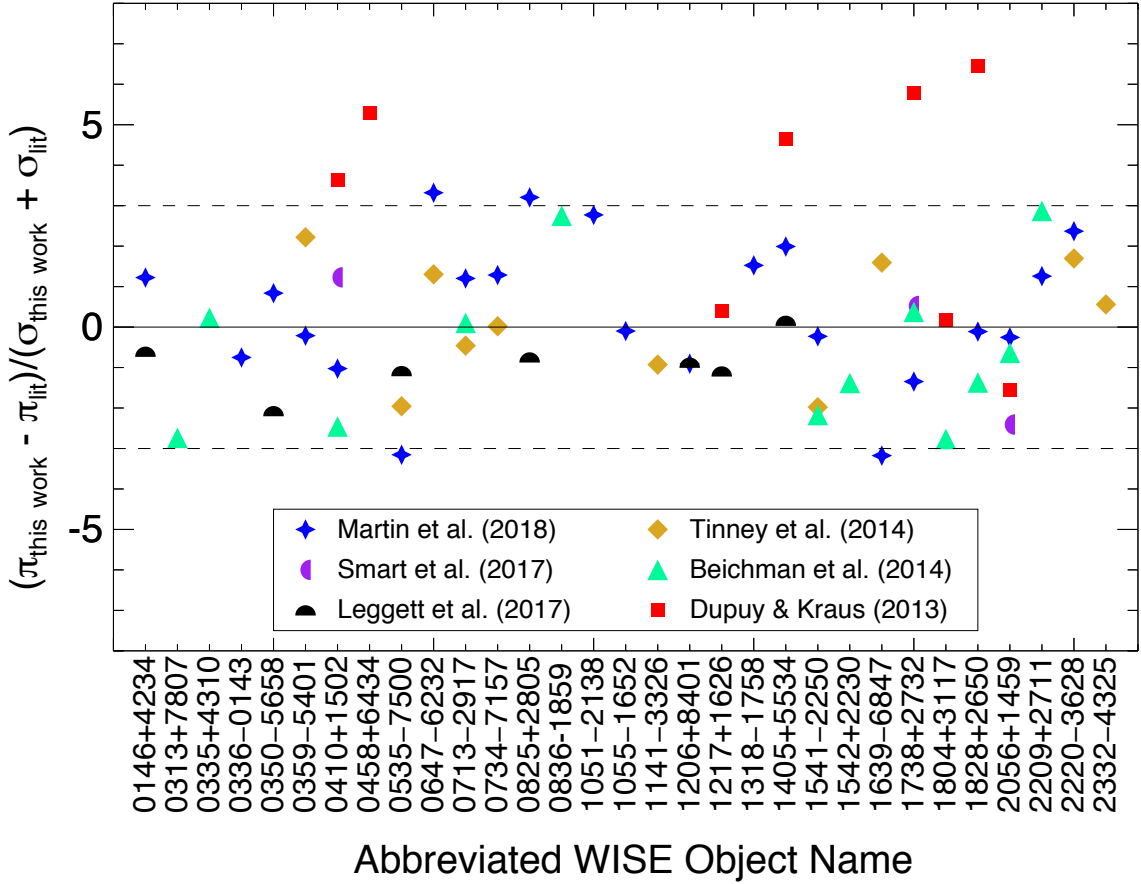}
\caption{A comparison of the parallax determinations for thirty-one objects in common between this paper and those of other published works. Our results show excellent agreement with other values, with the exception of measurements from \cite{dupuy2013}. See text for discussion.
\label{pi_comparison}}
\end{figure*}

By design, most of the T dwarfs in our parallax program were selected for observation because they were not being targeted by other ground- or space-based programs. On the other hand, {\it all} known Y dwarfs were targeted regardless of whether they had parallaxes measured elsewhere. As a result, only 31 of the 142 objects for which we have {\it Spitzer}-measured parallaxes have previously measured values. Figure~\ref{pi_comparison} compares our values to those in the literature for these 31 objects. Three comparison data points fall outside of the boundaries of this plot, but those are attributed to poor measurements in the literature: WISE 0146+4234 and WISE 2220$-$3628 from \cite{beichman2014}, which are discrepant from our values by $\sim$9$\sigma$, and WISE 1541$-$2250 from \cite{dupuy2013}, which is discrepant from our value by $\sim$20$\sigma$. These literature values are also discrepant with other published values; see analysis of WISE 2220$-$3628 in \cite{martin2018}, WISE 1541$-$2250 in \cite{martin2018} and \cite{bedin2018}, and WISE 0146+4234 in \cite{leggett2017}.

We find excellent agreement between our values and ground-based Magellan/FourStar measurements by \cite{tinney2014}, between our values and ground-based UKIRT/WFCAM measurements by \cite{smart2017}, and between our values and {\it Spitzer}/IRAC ch2 measurements by \cite{leggett2017}, some data for which had been published earlier in \cite{luhman2016} and which uses a subset of the {\it Spitzer}/IRAC ch2 imaging taken by us. With the exception of the two objects noted above, we also find excellent agreement with a hybrid approach from \cite{beichman2014} that uses astrometry compiled from Keck/NIRC2, {\it HST}/WFC3, {\it WISE}, and {\it Spitzer}/IRAC. 

We can also compare the parallax measurements for objects in common between this paper and \cite{martin2018}. The latter paper uses a subset of the same {\it Spitzer}/IRAC ch2 data used in the current paper, but there are a few differences to note: (1) the current paper uses measurements taken from individual frames as opposed to the \cite{martin2018} method of using measurements from the epochal coadds, (2) the current paper uses a different set of analysis code following the MOPEX frame extraction step, even though the guiding logic is very similar, and (3) the current paper benefits from the longer time baseline afforded by the additional observations in Cycle 13. Four of the twenty-two objects in common between the two papers have differences exceeding three times the sum of the individual uncertainties. Two of these objects, WISE 0535$-$7500 and WISE 0647$-$6323, are located near the Continuous Viewing Zone of {\it Spitzer} and thus have frames taken at many different roll angles, where the aforementioned issue with the MOPEX mosaicking step (Section~\ref{mosaicking_issue}) is known to have problems. A comparison of the parallax-only plots for these two objects (Figure~\ref{plots_0005p3737}, upper right panels) show individual measurements with much smaller uncertainties and a more convincing overall fit than do similar plots in \cite{martin2018} (their Figure 5). For the other two objects -- WISE 0825+2805 and WISE 1639$-$6847 -- the improved astrometry provided by individual frame measurements along with the added time baseline (an additional three years) appears to have vastly improved the fits and as a consequence moved them significantly from the parallax values determined by \cite{martin2018}.

Just as both \cite{smart2017} and \cite{martin2018} found, we see a systematic offset between our values and the parallaxes measured by \cite{dupuy2013}, who used {\it Spitzer}/IRAC ch1 imaging. \cite{martin2018} advanced three hypotheses to attempt to explain the discrepancy: (1) chromatic distortion in the ch1 data, (2) a fundamental error in the \cite{dupuy2013} fitting analysis, or (3) an insufficient time baseline with which to beat down random errors and to disentangle proper motion and parallax. Despite testing each of these hypotheses, \cite{martin2018} were unable to come to a firm conclusion. 

\clearpage

\section{USNO Astrometry\label{usno_section}}

A number of early {\it WISE} discoveries from Table~\ref{20pc_sample_preliminary} were astrometrically monitored at the US Naval Observatory (USNO) in Flagstaff. These objects were measured on the 1.55 m Strand Astrometric Reflector using the ASTROCAM infrared imager (\citealt{fischer2003}), which was commissioned in Sep 2000. Sadly, the original ASTROCAM was destroyed in a cryogenic explosion ultimately caused by a local forest fire in Jun 2006. In May 2011, a repaired version of ASTROCAM was commissioned, using the same basic setup as the old instrument. Observation and reduction procedures for these new targets are nearly identical to those described in \cite{vrba2004}, which presented preliminary parallaxes and proper motions of 40 L and T dwarfs using the first 1.7 years of data from the original instrument. The main difference is that these new reductions use a combined X+Y solution, rather than just using the X solution as was done in \cite{vrba2004}, along with minor quality control software improvements to ensure that all useful frames are included in the solutions. Final parallax and proper motion results for the original 40 objects plus an additional 19 L and T dwarfs using the full, original ASTROCAM database will be given in Vrba et al.\ (in prep.).

The eighteen objects in Table~\ref{usno_results} were placed on the program over a number of years after recommissioning and the results shown are for data obtained through early fall of 2017. All of these objects and an additional 103 sources continue on the USNO ASTROCAM astrometry program. All results in Table~\ref{usno_results} were obtained through a $J$-band filter. The abbreviated object name is given in column 1, the relative parallax and its uncertainty in column 2, the proper motion along with its uncertainty and position angle (east of north) in columns 3 and 4, the total time baseline of the observations in column 5, the number of independent nights of observation in column 6, and the number of registration stars in column 7. While several factors contribute to the quality of in-frame relative astrometry (e.g., the distribution of registration stars), objects having smaller measurement errors are generally those that have been observed over a longer time baseline. As a case in point, WISE 2226+0440 has the poorest measurements because it was added much later into the program. (In fact, it is the only object in Table~\ref{usno_results} not already listed in Table 3 of \citealt{vrba2015}.) Given the typical magnitudes of the registration stars, the correction from relative to absolute parallax is $\sim$1.6$\pm$0.3 mas (\citealt{vrba2004}). For subsequent analyses we have added this correction, and its uncertainty in quadrature, to the relative parallax to convert to near-absolute units.

\startlongtable
\begin{deluxetable*}{lrrrrrr}
\tabletypesize{\footnotesize}
\tablenum{5}
\tablecaption{Parallax and Motion Fits for Objects on the USNO Parallax Program\label{usno_results}}
\tablehead{
\colhead{Object} &                          
\colhead{$\pi_{rel}$} &
\colhead{$\mu_{rel}$} &
\colhead{$\theta$} &
\colhead{Baseline} &                          
\colhead{\# of} &
\colhead{\# of} \\
\colhead{Name} &
\colhead{(mas)} &
\colhead{(mas yr$^{-1}$)} &
\colhead{(deg)} &
\colhead{(yr)} &                          
\colhead{Ep.} &
\colhead{Reg.} \\
\colhead{} &
\colhead{} &
\colhead{} &
\colhead{} &
\colhead{} &                          
\colhead{} &
\colhead{Stars} \\
\colhead{(1)} &                          
\colhead{(2)} &  
\colhead{(3)} &  
\colhead{(4)} &
\colhead{(5)} &
\colhead{(6)} &
\colhead{(7)}  \\
}
\startdata
WISE 0254+0223  & 144.49$\pm$1.45&      2572.2$\pm$0.1&     85.05$\pm$0.03&        5.30&       107&          13\\       
WISE 0513+0608  &  69.17$\pm$1.49&       433.0$\pm$1.0&    180.69$\pm$0.07&        5.41&       133&          16\\      
WISE 0614+3912  &  52.09$\pm$1.70&       529.3$\pm$1.2&    200.55$\pm$0.07&        5.30&        71&          20\\      
WISE 0625+5646  &  47.94$\pm$2.02&        52.7$\pm$1.4&    251.86$\pm$0.76&        5.24&        74&          15\\    
WISE 1019+6529  &  41.25$\pm$1.75&       150.6$\pm$1.1&    323.89$\pm$0.21&        5.92&        93&           7\\    
WISE 1122+2550  &  64.74$\pm$2.26&      1028.6$\pm$1.2&    252.34$\pm$0.04&        5.98&        88&           8\\    
WISE 1320+6034  &  59.00$\pm$2.54&       561.3$\pm$1.4&    264.80$\pm$0.08&        6.05&        50&          10\\    
WISE 1322$-$2340&  75.86$\pm$4.22&       524.1$\pm$1.9&    318.35$\pm$0.11&        5.92&        56&           9\\    
WISE 1457+5815  &  53.43$\pm$2.30&       502.0$\pm$1.1&    262.75$\pm$0.07&        6.08&       107&          12\\    
WISE 1506+7027  & 191.91$\pm$0.51&      1587.3$\pm$0.3&    311.19$\pm$0.03&        6.08&       100&           9\\    
WISE 1627+3255  &  52.77$\pm$1.85&       351.6$\pm$0.8&    193.55$\pm$0.07&        6.10&        66&          11\\    
WISE 1741+2553  & 212.65$\pm$2.77&      1556.9$\pm$1.3&    198.94$\pm$0.04&        5.10&        56&          14\\    
WISE 1852+3537  &  70.41$\pm$1.88&       381.5$\pm$1.1&    138.82$\pm$0.09&        5.29&        63&          15\\    
WISE 1906+4508  &  62.45$\pm$1.60&       351.0$\pm$0.1&    183.73$\pm$0.08&        5.35&        53&          16\\    
WISE 2213+0911  &  52.88$\pm$2.45&       128.0$\pm$1.2&    256.46$\pm$0.27&        5.37&       133&           9\\    
WISE 2226+0440  &  52.75$\pm$5.87&       543.2$\pm$5.7&    211.42$\pm$0.30&        2.23&        38&           9\\    
WISE 2340$-$0745&  46.19$\pm$3.11&       293.1$\pm$1.4&    148.58$\pm$0.14&        5.32&        61&           9\\    
WISE 2348$-$1028&  56.77$\pm$3.46&       642.5$\pm$1.5&     75.96$\pm$0.08&        5.24&        62&           7\\    
\enddata
\end{deluxetable*}

\clearpage

\section{NTT and UKIRT Astrometry\label{other_telescopes}}

Twenty-three additional late-T dwarfs were monitored astrometrically at either the 3.5m New Technology Telescope (NTT) or at the 3.8m United Kingdom Infrared Telescope (UKIRT); one of these objects was observed at both facilities. Table~\ref{ntt_ukirt_results} lists the results. An abbreviated name of the object from Table~\ref{20pc_sample_preliminary} is given in column 1, and the source of the measurement as either NTT or UKIRT is given in column 2. The J2000 equinox RA and Dec along with the epoch of the position are given in columns 3-5. The measured value of the absolute parallax and the correction from relative parallax is given in columns 6-7, and the measured proper motions per axis are given in columns 8-9. Columns 10-12 list the total time baseline, the number of registration stars, and the total number of observations.

Fourteen objects were targeted as part of the NTT Parallaxes of Southern Extremely Cool objects (NPARSEC) project. NPARSEC is a European Southern Observatory long term program (186.C-0756; R.\ Smart, PI) of 96 nights on the NTT's infrared spectrograph and imaging camera Son Of ISAAC (SOFI; \citealt{moorwood1998}). The main observational program ran from 01 Oct 2010 through 15 Sep 2013, although various ad hoc requests were made to extend the temporal baseline through 2018. Target selection, observing methodology, and reduction procedures are explained in detail in \cite{smart2013}. 

Ten objects were targeted as part of a UKIRT program primarily designed to follow-up T dwarfs from the UKIRT Infrared Deep Sky Survey (UKIDSS; \citealt{lawrence2007})  This program, described in \cite{marocco2010}, used service
observations on the UKIRT Wide Field Camera (WFCAM; \citealt{casali2007}), which is a large-field infrared imager. This program started as a director's discretionary request in 2007 and continued under various proposals and target lists until 2016.  As the UKIDSS discovery image was often used in the parallax determination, some targets also have observations starting as early as 2005. Scheduling, observing methodlogy, and reduction procedures are described in \cite{smart2010} and \cite{marocco2010}. 

The results in Table~\ref{ntt_ukirt_results} are regarded as preliminary only. Final reductions on both data sets will be completed in the near future, once observations from the NTT program conclude.

\startlongtable
\begin{deluxetable*}{lrrrcrcrrccc}
\tabletypesize{\footnotesize}
\tablenum{6}
\tablecaption{Parallax and Motion Fits for Objects on the NTT and UKIRT Parallax Programs\label{ntt_ukirt_results}}
\tablehead{
\colhead{Object} & 
\colhead{Program} &
\colhead{J2000 RA} &
\colhead{J2000 Dec} &
\colhead{Epoch} &
\colhead{$\pi_{abs}$} &
\colhead{Abs.\ Corr.} &
\colhead{$\mu_{RA}$} &
\colhead{$\mu_{Dec}$} &
\colhead{Baseline} &                          
\colhead{\# of} &
\colhead{\# of} \\
\colhead{Name} &
\colhead{} &
\colhead{(deg)} &
\colhead{(deg)} &
\colhead{} &
\colhead{(mas)} &
\colhead{(mas)} &
\colhead{(mas yr$^{-1}$)} &
\colhead{(mas yr$^{-1}$)} &
\colhead{(yr)} &                          
\colhead{Ref.\ Stars} &
\colhead{Obs.} \\
\colhead{(1)} &                          
\colhead{(2)} &  
\colhead{(3)} &  
\colhead{(4)} &
\colhead{(5)} &
\colhead{(6)} &
\colhead{(7)} &
\colhead{(8)} &
\colhead{(9)} &                          
\colhead{(10)} &
\colhead{(11)} &
\colhead{(12)} \\
}
\startdata
WISE 0040+0900    & UKIRT   &  10.103778&	 9.0151823&	2010.50&	69.8$\pm$1.5&	1.0&	-52.3$\pm$0.9&	 -53.0$\pm$2.0&	6.0&	117&	11\\
ULAS 0150+1359    & UKIRT   &  27.601590&	 13.990027&	2011.01&	39.8$\pm$3.2&	1.1&	 25.0$\pm$1.9&	  24.8$\pm$1.4&	7.4&	177&	31\\
WISE 0223$-$2932  & NPARSEC &  35.844782& $-$29.550163&	2016.79&    80.7$\pm$2.6&	0.6&    780.2$\pm$1.1&	-535.1$\pm$1.5&	5.8&	50&	    13\\
2MASS 0348$-$6022 & NPARSEC &  57.030457& $-$60.376464&	2010.75&   120.1$\pm$1.8&	0.6&   -279.7$\pm$0.6&  -768.5$\pm$0.7&	6.4&	87&	    16\\
UGPS 0521+3640    & UKIRT   &  80.364579&	 36.678113&	2012.65&   122.2$\pm$1.6&	0.3&	569.0$\pm$0.9& -1511.0$\pm$1.0&	8.3&	311&	27\\
WISE 0528$-$3308  & NPARSEC &  82.185418& $-$33.140095&	2011.93&    49.4$\pm$3.9&	0.7&	  1.0$\pm$1.2&   -19.5$\pm$1.6&	6.2&    71&	    16\\
WISE 0542$-$1628  & NPARSEC &  85.629964& $-$16.474330&	2015.89&    61.5$\pm$2.7&  	0.7&   -217.7$\pm$0.9&	 294.6$\pm$0.9&	6.2&	131&	23\\
WISE 0612$-$3036AB& NPARSEC &  93.057745& $-$30.603509&	2011.12&    41.1$\pm$1.6&	0.6&   -120.3$\pm$0.7&  -258.5$\pm$1.7&	6.2&	274&	18\\
\enddata
\end{deluxetable*}


\startlongtable
\begin{longrotatetable}
\begin{deluxetable*}{lcccccccccccc}
\tabletypesize{\scriptsize}
\tablenum{7}
\tablecaption{Summary of Spectral Types, Parallaxes, Proper Motions, Absolute Magnitudes, and Tangential Velocities\label{absmag_vtan_table}}
\tablehead{
\colhead{Object} &
\colhead{Spec.} &
\colhead{Spec.} &
\colhead{$\pi_{abs}$} &
\colhead{$\mu_{total}$} &
\colhead{$\pi, \mu_{total}$} &
\colhead{$M_H$} &
\colhead{$M_{W1}$} &     
\colhead{$M_{W2}$} &     
\colhead{$M_{W3}$} &     
\colhead{$M_{ch1}$} &     
\colhead{$M_{ch2}$} &
\colhead{$v_{tan}$} \\
\colhead{Name} & 
\colhead{Type} &
\colhead{Code} &
\colhead{(mas)} &
\colhead{(mas/yr)} &
\colhead{Ref.} &
\colhead{(mag)} &
\colhead{(mag)} &     
\colhead{(mag)} &     
\colhead{(mag)} &     
\colhead{(mag)} &     
\colhead{(mag)} &
\colhead{(km/s)} \\
\colhead{(1)} &                          
\colhead{(2)} &  
\colhead{(3)} &     
\colhead{(4)} &
\colhead{(5)} &                          
\colhead{(6)} &
\colhead{(7)} &
\colhead{(8)} &
\colhead{(9)} &                          
\colhead{(10)} &  
\colhead{(11)} &     
\colhead{(12)} &
\colhead{(13)}                          
}
\startdata
 WISE  0005+3737&   T9     & 9     &   127.0$\pm$2.4&   1034.0$\pm$ 2.0& Table~\ref{spitzer_results}  & 18.50$\pm$0.05&  17.28$\pm$0.10&  13.81$\pm$0.05&  12.30$\pm$0.24&  15.95$\pm$0.05&  13.80$\pm$0.04&   38.6$\pm$0.7\\
 WISE  0015-4615&   T8     & 8     &    71.8$\pm$2.7&    796.4$\pm$ 2.0& Table~\ref{spitzer_results}  & 17.19$\pm$0.11&  16.24$\pm$0.13&  13.50$\pm$0.09&  \nodata       &  15.38$\pm$0.09&  13.51$\pm$0.08&   52.6$\pm$2.0\\
 WISE  0032-4946&   T8.5   & 8.5   &    63.6$\pm$2.9&    938.2$\pm$ 2.4& Table~\ref{spitzer_results}  & 17.89$\pm$0.17&  16.69$\pm$0.19&  14.09$\pm$0.12&  11.22$\pm$0.36&  15.95$\pm$0.11&  13.95$\pm$0.10&   69.9$\pm$3.2\\
 ULAS  0034-0052&   T8.5   & 8.5   &    68.7$\pm$1.4&    359.2$\pm$ 0.8&                             1& 17.67$\pm$0.06&  16.19$\pm$0.14&  13.73$\pm$0.08&  \nodata       &  15.42$\pm$0.06&  13.68$\pm$0.05&   24.8$\pm$0.5\\
 2MASS 0034+0523&   T6.5   & 6.5   &   120.1$\pm$3.0&    697.5$\pm$ 2.5& Table~\ref{spitzer_results}  & 15.95$\pm$0.06&  15.49$\pm$0.07&  12.95$\pm$0.06&  12.18$\pm$0.32&  14.49$\pm$0.06&  12.98$\pm$0.06&   27.5$\pm$0.7\\
 WISE  0038+2758&   T9     & 9     &    89.7$\pm$2.5&     97.5$\pm$ 2.2& Table~\ref{spitzer_results}  & 18.68$\pm$0.07&  17.21$\pm$0.16&  14.13$\pm$0.07&  12.14$\pm$0.34&  16.22$\pm$0.07&  14.17$\pm$0.06&    5.2$\pm$0.2\\
          Gl 27B&   T8     & 8     &    89.8$\pm$0.1&    591.9$\pm$ 0.1&                             2& 16.49$\pm$0.03&  \nodata       &  \nodata       &  \nodata       &  15.15$\pm$0.04&  13.39$\pm$0.02&   31.2$\pm$0.1\\
 WISE  0040+0900&   T7     & 7     &    69.8$\pm$1.5&     74.5$\pm$ 2.2& Table~\ref{ntt_ukirt_results}& 15.78$\pm$0.05&  15.20$\pm$0.08&  13.05$\pm$0.07&  \nodata       &  14.29$\pm$0.05&  12.98$\pm$0.05&    5.1$\pm$0.2\\
\enddata
\onecolumngrid
\tablecomments{References for parallax and total motions: 
(1) \citealt{dupuy2012},
(2) \citealt{gaia2018},
(3) \citealt{tinney2014},
(4) \citealt{vrba2004},
(5) \citealt{gaia2016},
(6) \citealt{leggett2012},
(7) \citealt{faherty2012},
(8) \citealt{subasavage2009},
(9) \citealt{artigau2010},
(10) \citealt{marocco2010},
(11) \citealt{burgasser2008},
(12) \citealt{zacharias2012},
(13) \citealt{tinney2003},
(14) \citealt{manjavacas2013},
(15) \citealt{henry2006},
(16) \citealt{harrington1980},
(17) \citealt{smart2013},
(18) \citealt{vanaltena1995}.
}
\end{deluxetable*}
\end{longrotatetable}


\section{Analysis of the Mean Trends\label{mean_trends}}

In the following subsections, we amass the color, parallax, and proper motion data from Tables~\ref{20pc_sample_preliminary}, \ref{20pc_sample_photometry}, \ref{spitzer_results}, \ref{usno_results}, and \ref{ntt_ukirt_results} to study trends in spectral type, absolute magnitude, color, and tangential velocity for late-T to early-Y dwarfs. We also discuss objects that may be particularly interesting because of their unusual placement relative to the means trends or their high reduced $\chi^2$ values in the {\it Spitzer} parallax and proper motion fits.

Table~\ref{absmag_vtan_table} summarizes the adopted spectral types (column 2), the measured parallaxes and proper motions (columns 4 and 6), and the derived absolute magnitudes (columns 8-13) and tangential velocities (column 14) for all of the T6 and later objects for which we have newly measured parallaxes or for which quality parallaxes have been published by other groups. Full designations for the abbreviated names listed in column 1 can be found in Table~\ref{20pc_sample_preliminary}, and the spectral code in column 3 gives a conversion from the spectral type to a scale running from 6 at T6.0 to 14 at Y4. Objects lacking spectral types are not shown on subsequent plots having spectral type along one axis, with the exception of WD 0806B and WISE 0855$-$0714, whose types are assumed to be Y1 and $\ge$Y4 (see footnotes to Table~\ref{20pc_sample_preliminary}).

Column 5 indicates whether the parallax value in column 4 is an absolute or a relative value. The only references with relative parallaxes are the USNO results from Table~\ref{usno_results} and those from \cite{tinney2003} and \cite{tinney2014}. As stated in section~\ref{usno_section}, the correction from relative to absolute parallax is expected to be $\sim$1.6$\pm$0.3 mas for the USNO results. The near-infrared results from \cite{tinney2003} were done on a telescope of larger aperture (3.5m NTT) than that of the USNO program (1.55m Strand), so smaller corrections can be assumed based on the fact that stars in the reference frames are generally fainter and more distant. The corrections shown in Table~\ref{ntt_ukirt_results} from both the 3.5m NTT and 3.8m UKIRT indicate that a correction of $\sim$0.9$\pm$0.3 mas is also appropriate here. The near-infrared results from \cite{tinney2014} were obtained on the larger aperture Magellan Baade Telescope (6.5m), so even smaller corrections of $\sim$0.6$\pm$0.3 mas can be assumed. These mean relative-to-absolute parallax corrections have been applied to all parallaxes for these three sources prior to the calculation of the absolute magnitudes and tangential velocities in Table~\ref{absmag_vtan_table}. These corrections have also been applied in the analyses of subsequent sections of this paper.

For the $J$, $H$, $K$ photometry, we list only $H$-band because it is the only one of these three bands that is invariant between the two main near-infrared classification systems, 2MASS and MKO-NIR (see section 4.3 of \citealt{kirkpatrick2012}). Photometry shortward of $J$-band is scarce due to a combination of non-universal filters (such as $z$ and $Y$) and the fact that these brown dwarfs are much fainter, and thus harder to detect, at even shorter wavelengths (e.g., at $R$ or $I$). Longward of $K$-band, we list the {\it WISE} W1, W2, and W3 photometry, although the last of these bands is generally not sensitive enough to detect these brown dwarfs well. In some cases, the object is sufficiently faint in the W1 band that it is not detected well there, either. We also list photometry in the {\it Spitzer}/IRAC ch1 and ch2 bands.

\subsection{Identifying Outliers from the Mean Trends}

As we have large numbers of objects with accurate parallaxes, motions, and photometry, we can now examine trends across the late-T and early-Y spectral types, particularly in relation to the 20-pc sample we are building for mass function analysis. The goal of this section is to examine these trends in an effort to identify outliers that may need special handling in section~\ref{mass_function}. In some cases, outliers may indicate unresolved binarity, and this is an issue because we want to have the most accurate accounting possible of the number of objects within our sample volume. In other cases, outliers may indicate unusual spectroscopic features which may confuse our translation of an observational parameter such as spectral type or color into a physical parameter such as effective temperature. In this subsection, we identify plots useful for identifying unusual objects, and in section~\ref{notes_on_individual_objects} we discuss each of these objects in detail so that they will be properly handled during the mass function computation. In the following figures, only those objects with solid measurements in both axes are plotted -- i.e., objects having absolute magnitude limits or color limits are removed.

\begin{figure*}[t!]
\figurenum{4}
\includegraphics[scale=0.65,angle=0]{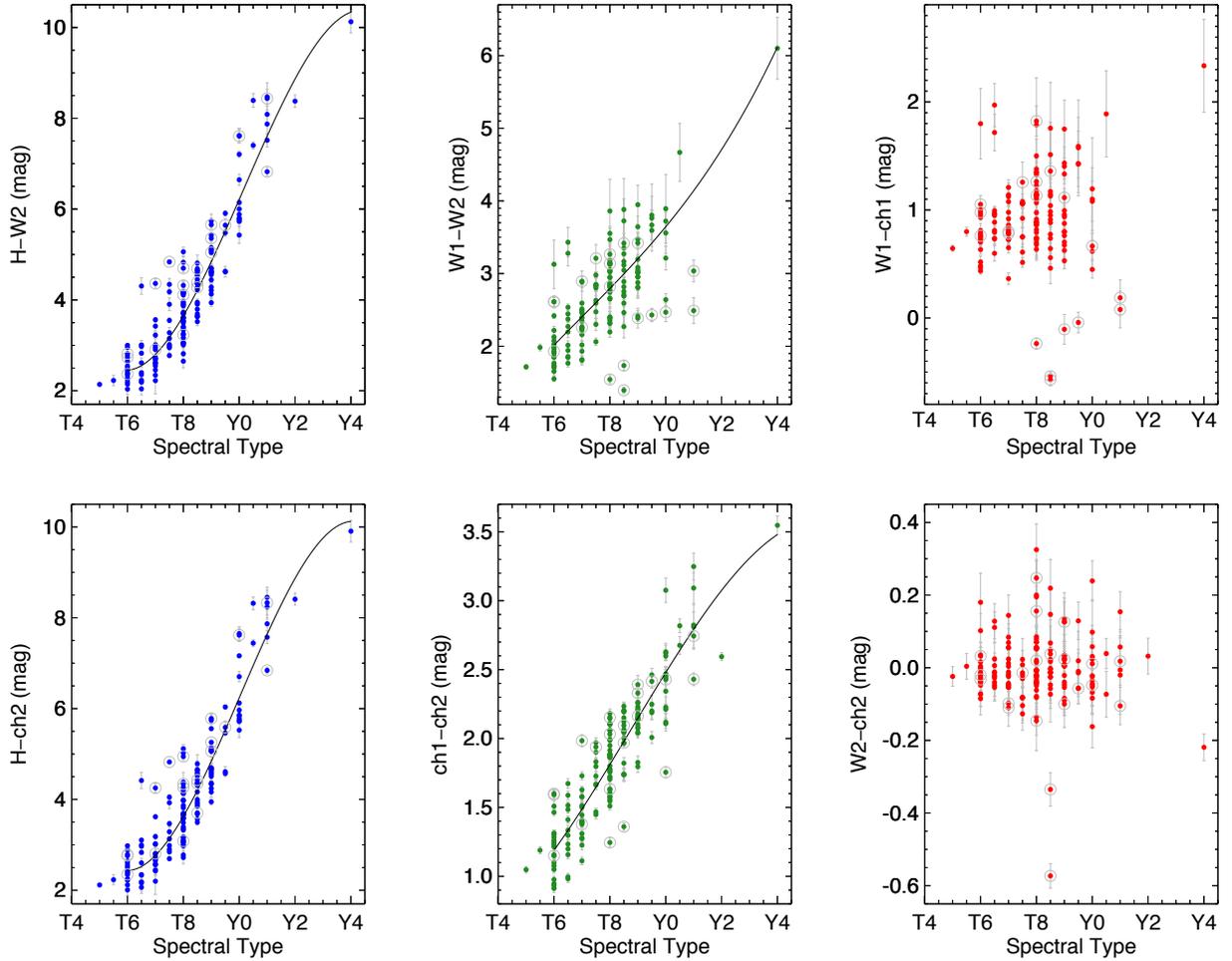}
\caption{Trends of color with spectral type for all objects from Table~\ref{absmag_vtan_table} that fall within 20 pc of the Sun ($\pi_{abs} \ge 50$ mas). Shown as blue points in the two leftmost panels are trends with $H-$W2 and $H-$ch2 color. The two middle panels show trends of W1$-$W2 and ch1$-$ch2 color (green points). The two rightmost panels show trends of W1$-$ch1 and W2$-$ch2 color (red points). Polynomial fits to the relations in the four leftmost panels are shown by the black curves, and the functional forms are presented in Table~\ref{equations}. Points ignored by the fitting (see Section~\ref{fits_to_trends}) are circled in gray.
\label{trend_colors_vs_type}}
\end{figure*}

Shown in the leftmost panels of Figure~\ref{trend_colors_vs_type} (blue points) are trends of $H-$W2 and $H-$ch2 color with spectral type. These colors are very similar because the W2 and ch2 bands sample a similar range in wavelength near the peak flux in cold brown dwarfs (see Figure 2 of \citealt{mainzer2011}). A number of outliers stand out as having unusually red colors for their types between T6.5 and T8.

The middle panels of Figure~\ref{trend_colors_vs_type} (green points) show the trends of W1$-$W2 and ch1$-$ch2 color with spectral type. Notable here are a handful of objects with significantly blue colors, especially on the ch1$-$ch2 plot.

The rightmost panels in Figure~\ref{trend_colors_vs_type} (red points) show the trends of W1$-$ch1 and W2$-$ch2 color with spectral type. Given the similarity of the W2 and ch2 filters, one would expect the W2$-$ch2 color to fall near zero, which it does. Objects falling significantly far from zero color are probably contaminated by a background source in one of the two bands.
The W1 and ch1 filters, on the other hand, sample slightly different wavelengths, with the W1 band centered squarely on the fundamental methane absorption feature near 3.3 $\mu$m and the ch1 band lying somewhat redward of this and sampling more of the rise in the brown dwarf spectrum near 4.0 $\mu$m (Figure 2 of \citealt{mainzer2011}). As expected, objects are generally fainter in the W1 band than in the ch1 band, but the plot shows a few exceptions that again are likely to be objects with contaminated photometry in one of the bands. 

\begin{figure*}
\figurenum{5}
\includegraphics[scale=0.65,angle=0]{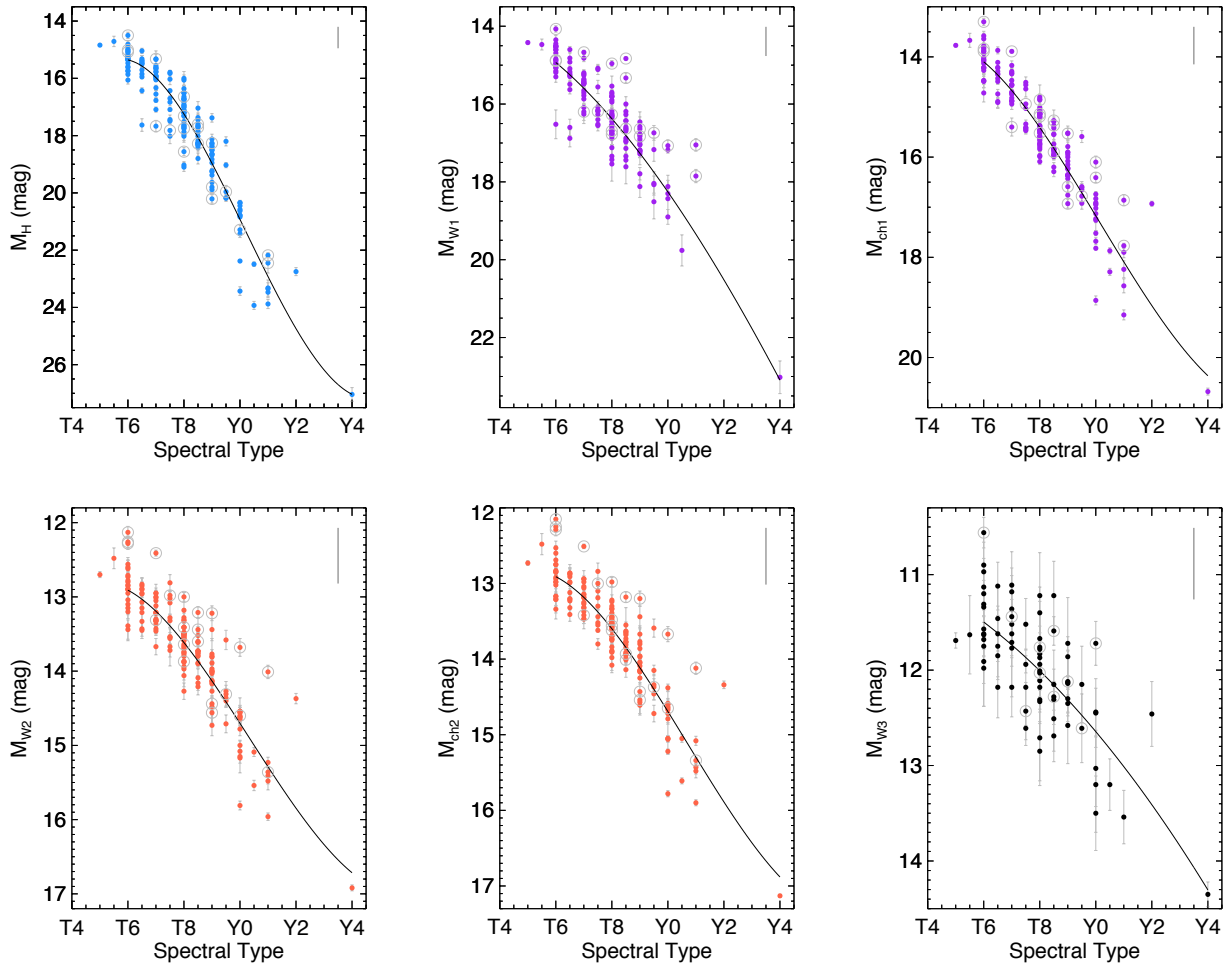}
\caption{Trends of absolute magnitude with spectral type for all objects from Table~\ref{absmag_vtan_table} that fall within 20 pc of the Sun ($\pi_{abs} \ge 50$ mas). The top row shows trends with absolute $H$-band magnitude (light blue) and the similar absolute W1 and ch1 magnitudes (purple). The bottom row shows trends with the similar absolute W2 and ch2 magnitudes (orange red) and the absolute W3 magnitude (black). Polynomial fits to the relations are shown by the black curves, and the functional forms are presented in Table~\ref{equations}. Points ignored by the fitting (see Section~\ref{fits_to_trends}) are circled in gray. The light gray bar at the upper right of each panel shows the size of the 0.75-mag offset expected for an equal-magnitude binary.
\label{trend_abs_mags_vs_type}}
\end{figure*}

Shown in Figure~\ref{trend_abs_mags_vs_type} are trends of absolute magnitude with spectral type. Despite the small uncertainties on these measurements, there is a surprisingly large scatter of absolute magnitude in all bands at each spectral subclass. At W2 and ch2, there is typically a magnitude of scatter at each late-T subclass, and this increases to two magnitudes at $H$, W1, and ch1. (The uncertainties at W3 are large enough that interpretation is more difficult.) At Y0, this scatter increases by roughly another magnitude in all bands. Despite this dispersion, we nonetheless can identify some objects that fall significantly further off the trends than the others.

On the absolute W1 and ch1 plots of Figure~\ref{trend_abs_mags_vs_type}, we consider a T dwarf to be unusual if it is a clear outlier on {\it both} plots, since the T dwarfs in our sample generally have solid photometry in both bands. 
For the absolute W2 and ch2 plots, we consider an object to be unusual only if it appears as a significant outlier at both bands, and we can apply this strategy to T {\it and} Y dwarfs because both are easily detected at these wavelengths. 

\begin{figure*}
\figurenum{6}
\includegraphics[scale=0.65,angle=0]{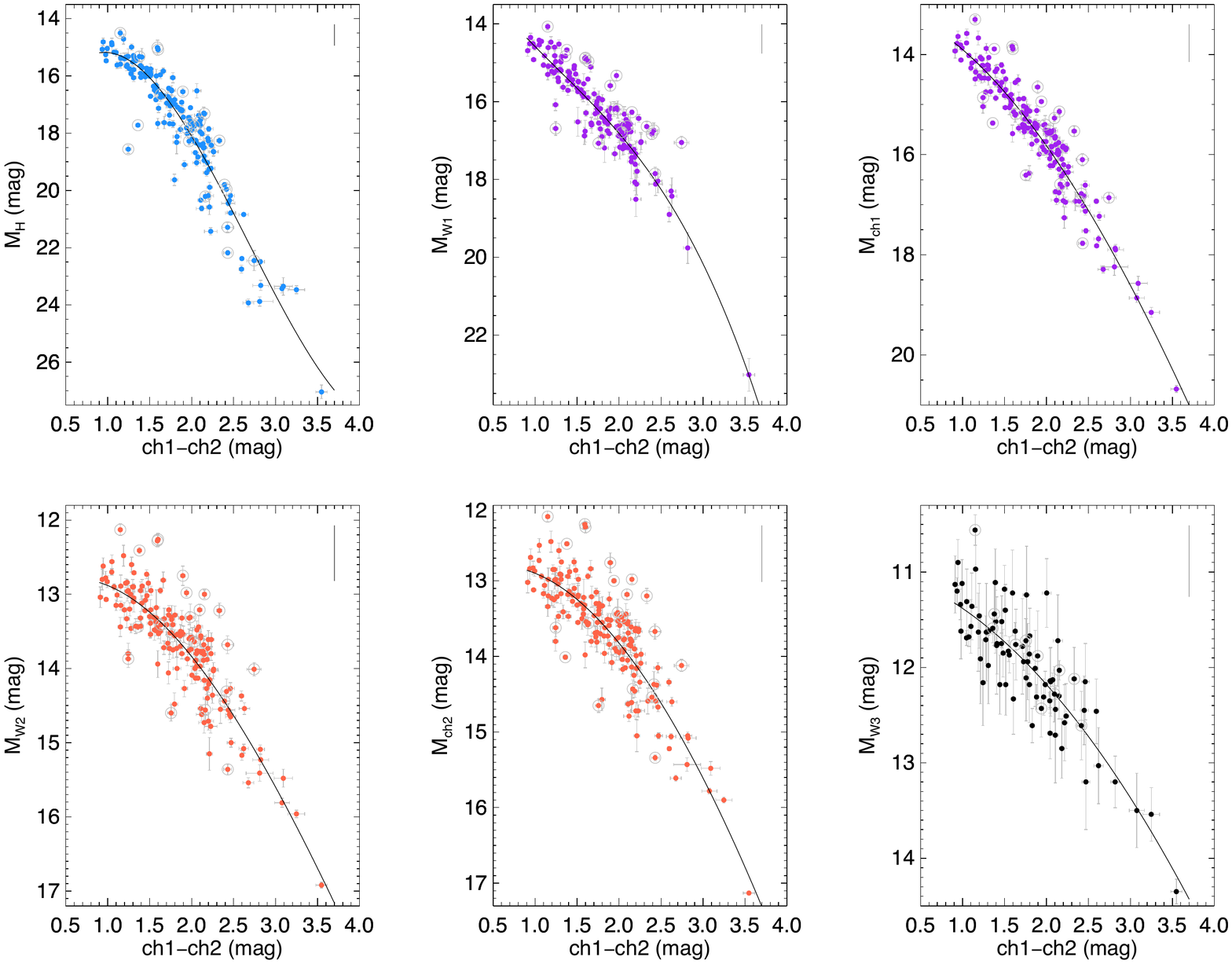}
\caption{Trends of absolute magnitude with ch1$-$ch2 color for all objects from Table~\ref{absmag_vtan_table} that fall within 20 pc of the Sun ($\pi_{abs} \ge 50$ mas). The top row shows trends with absolute $H$-band magnitude (light blue) and the similar absolute W1 and ch1 magnitudes (purple). The bottom row shows trends with the similar absolute W2 and ch2 magnitudes (orange red) and the absolute W3 magnitude (black). Polynomial fits to the relations are shown by the black curves, and the functional forms are presented in Table~\ref{equations}. Points ignored by the fitting (see Section~\ref{fits_to_trends}) are circled in gray. The light gray bar at the upper right of each panel shows the size of the 0.75-mag offset expected for an equal-magnitude binary.
\label{trend_abs_mags_vs_ch1ch2}}
\end{figure*}

Figure~\ref{trend_abs_mags_vs_ch1ch2} shows trends of absolute magnitudes with ch1$-$ch2 color. There is a linear trend of decreasing absolute magnitude with increasing ch1$-$ch2 color, but with a notable inflection point in most of the plots near ch1$-$ch2 $\approx$ 2.2 mag, corresponding to the transition between types T and Y (see Figure~\ref{trend_colors_vs_type}). 

\begin{figure*}
\figurenum{7}
\includegraphics[scale=0.65,angle=0]{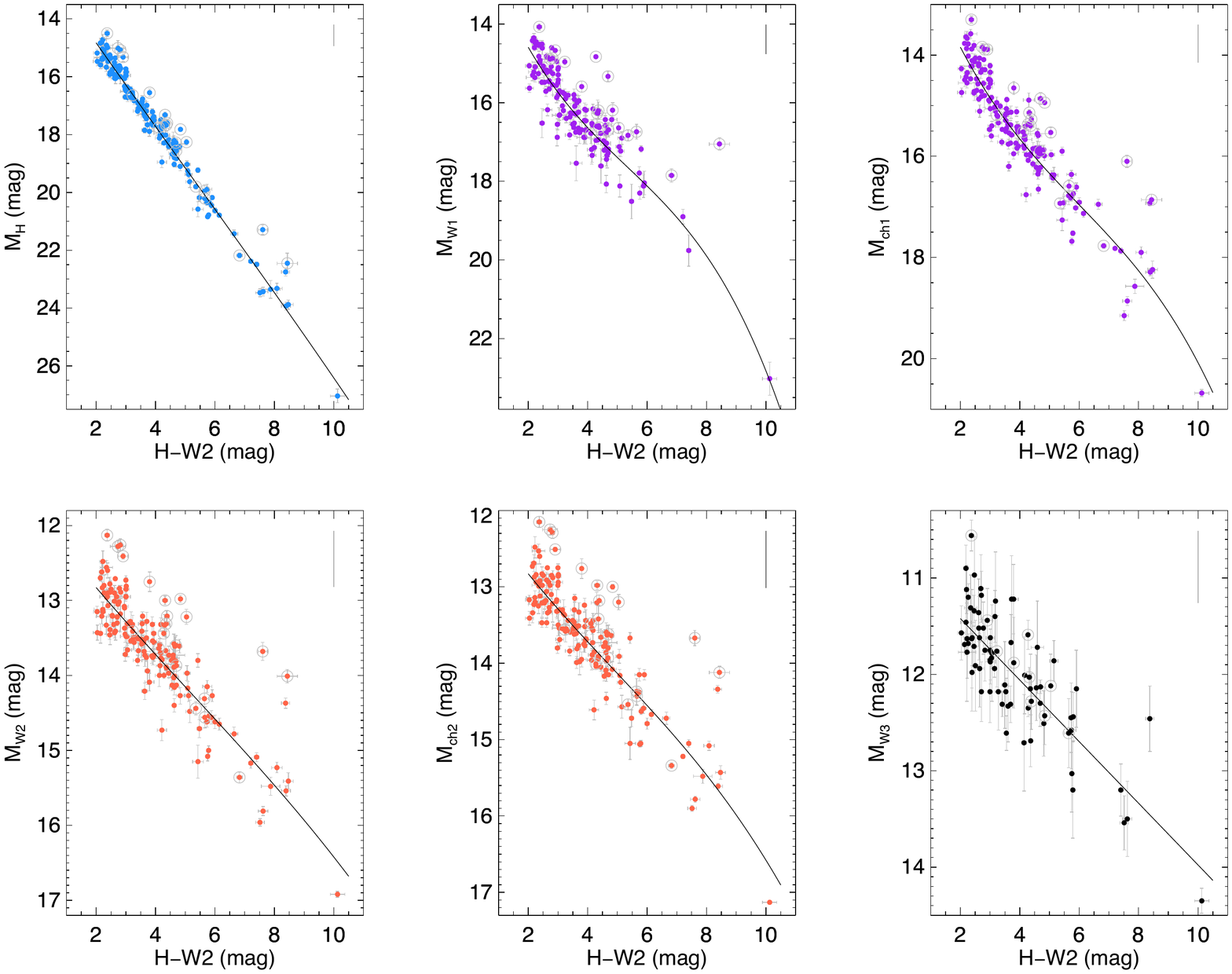}
\caption{Trends of absolute magnitude with $H-$W2 color for all objects from Table~\ref{absmag_vtan_table} that fall within 20 pc of the Sun ($\pi_{abs} \ge 50$ mas). The top row shows trends with absolute $H$-band magnitude (light blue) and the similar absolute W1 and ch1 magnitudes (purple). The bottom row shows trends with the similar absolute W2 and ch2 magnitudes (orange red) and the absolute W3 magnitude (black). (Plots of absolute magnitude with $H-$ch2 color look virtually identical to the plots shown here.) Polynomial fits to the relations are shown by the black curves, and the functional forms are presented in Table~\ref{equations}. Points ignored by the fitting (see Section~\ref{fits_to_trends}) are circled in gray. The light gray bar at the upper right of each panel shows the size of the 0.75-mag offset expected for an equal-magnitude binary.
\label{trend_abs_mags_vs_HW2}}
\end{figure*}

\begin{figure}
\figurenum{8}
\includegraphics[scale=0.40,angle=0]{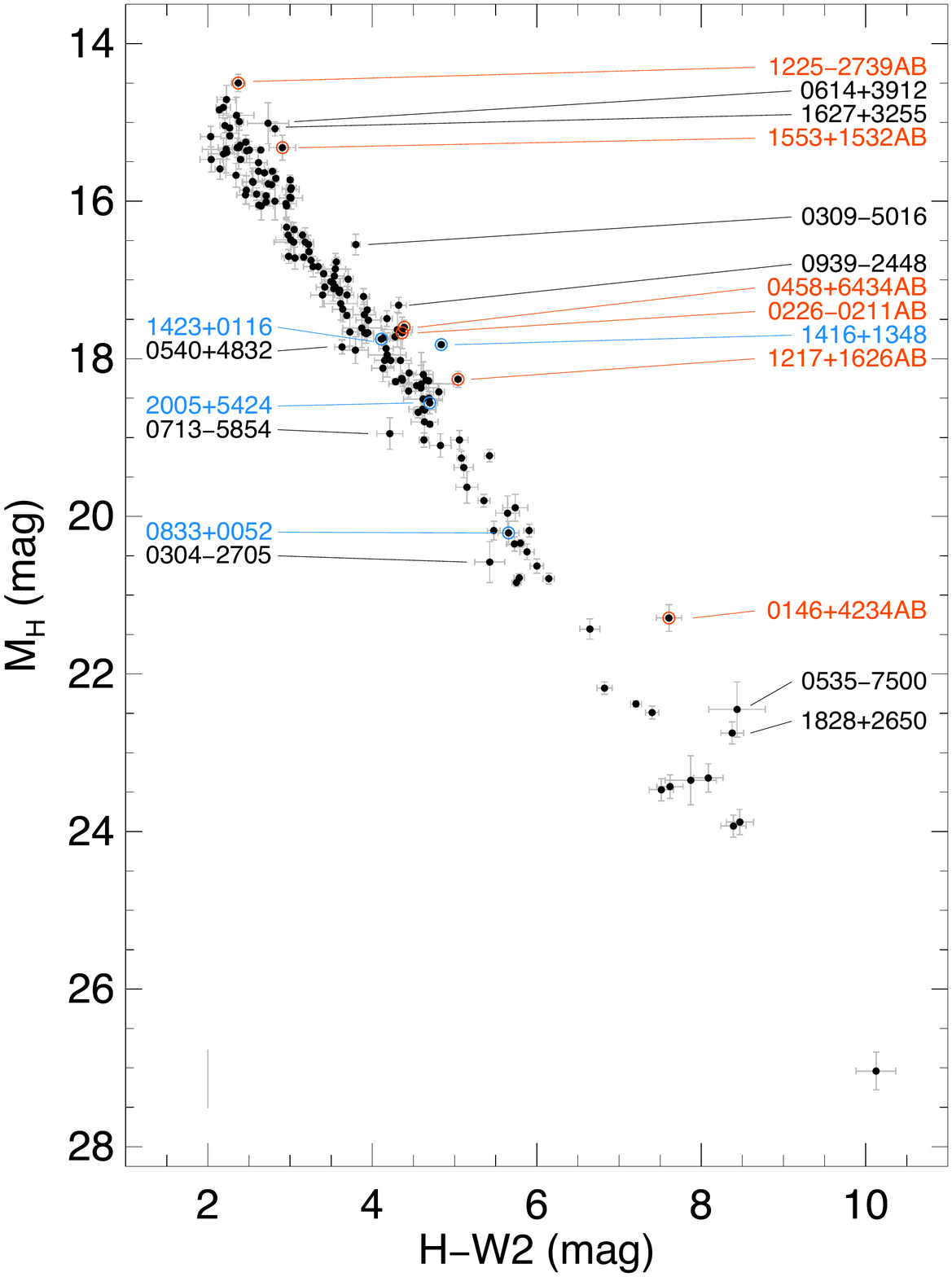}
\caption{Trends of absolute $H$-band magnitude with $H-$W2 color for all objects from Table~\ref{absmag_vtan_table} that fall within 20 pc of the Sun ($\pi_{abs} \ge 50$ mas). Known binary systems are circled and labeled in red, and known subdwarfs are circled and labeled in blue. Other objects falling significantly above or below the mean trend are labeled in black. The light gray bar at the lower left shows the size of the 0.75-mag offset expected for an equal-magnitude binary.
\label{trend_MH_vs_HW2}}
\end{figure}

Figure~\ref{trend_abs_mags_vs_HW2} shows the trend of absolute magnitudes with $H-$W2 color. These trends appear more linear than those with ch1$-$ch2 color above in that there is not a clear inflection point at the T/Y transition, corresponding to $H-$W2 $\approx$ 6.5 mag, although data redder than this value are not plentiful. The plot of $M_H$ vs.\ $H-$W2 color shows a tight correlation between quantities, so we examine this plot in detail in 
Figure~\ref{trend_MH_vs_HW2}. The tight correlation allows us to identify close binary systems that may or may not already be known. Unresolved, equal-magnitude doubles would be expected to fall 0.75 mag above the canonical sequence, and there are a number of objects that appear to lie at roughly this level above the mean trend. For reference, we label and circle in red all of the known doubles. Other objects, not known to be doubles, exhibit the same overluminosity. 
We also label the locations of the known subdwarfs on this diagram with blue circles. This shows that -- with the exception of ULAS 1416+1348, which may itself be an unresolved double -- these objects fall on the main locus. There are a few objects that also inhabit an area below the main trend.

\begin{figure*}
\figurenum{9}
\includegraphics[scale=0.65,angle=0]{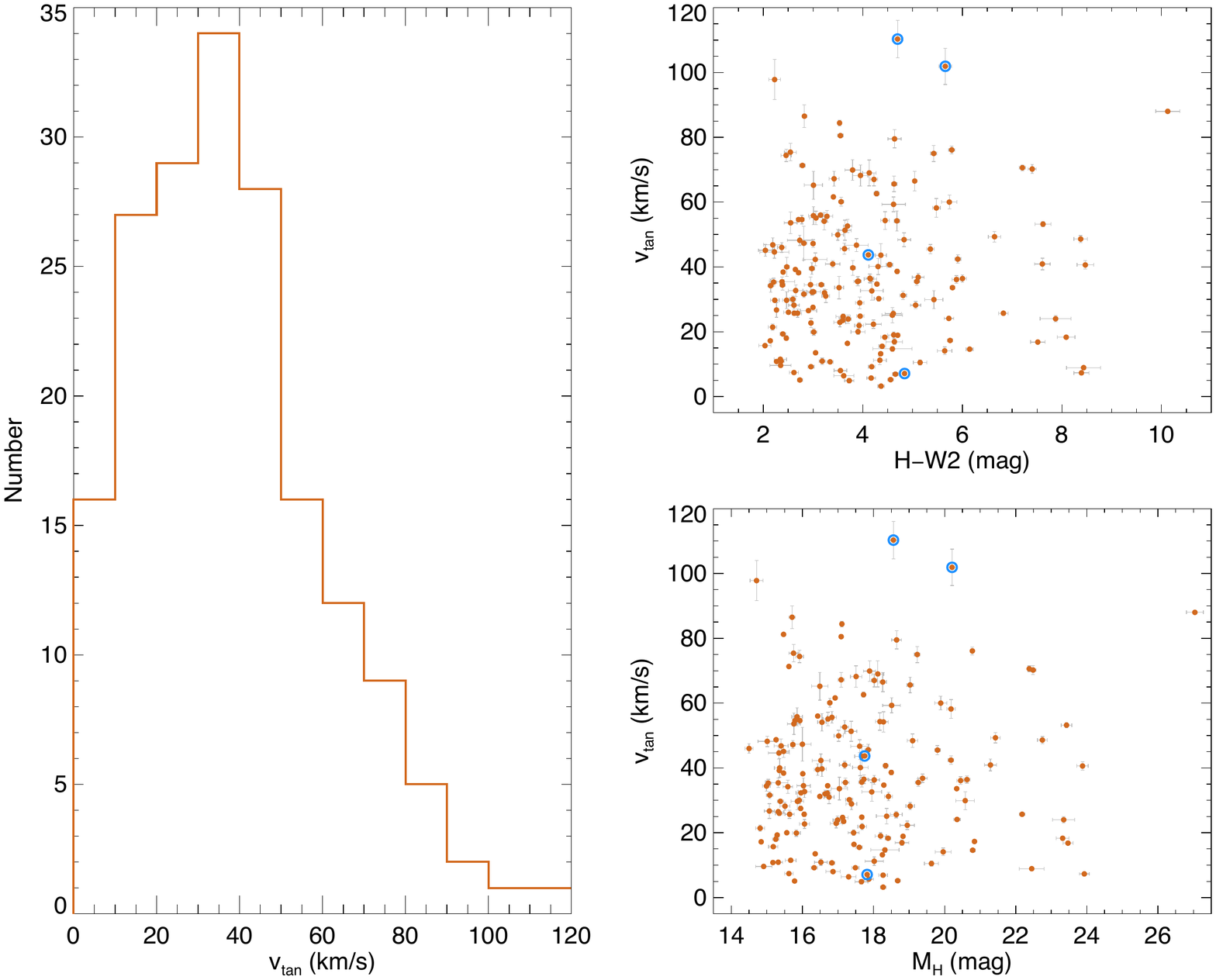}
\caption{Distributions in number, color, and absolute magnitude for our tangential velocity measurements for objects in Table~\ref{absmag_vtan_table} that fall within 20 pc of the Sun ($\pi_{abs} \ge 50$ mas). On the left is the histogram of the $v_{tan}$ values. At upper right is the distribution with respect to $H-$W2 color, and at bottom right is the distribution with respect to $M_H$. In these rightmost panels, spectroscopically identified subdwarfs (WISE 0833+0052, ULAS 1416+1348, Gl 547B, and WISE 2005+5424) are circled in blue.
\label{trend_vtan}}
\end{figure*}

Finally, we plot a histogram of the measured tangential velocities in Figure~\ref{trend_vtan} along with trends of $v_{tan}$ as a function of $H-$W2 color and $M_H$. Spectroscopically identified subdwarfs are circled in blue in the right-hand panels. The two highest $v_{tan}$ values belong to the subdwarfs WISE 2005+5424 (110 km s$^{-1}$) and WISE 0833+0052 (102 km s$^{-1}$). In comparison, for a sample of 841 late-M, L, and T dwarfs, \cite{faherty2009} find a median $v_{tan}$ value of 26 km s$^{-1}$ and only 1.7\% (14/841) with $v_{tan}$ values in excess of 100 km s$^{-1}$, similar to the percentage we find at late-T and Y types (2/180 = 1.1\%).

Objects appearing as outliers on any of these plots or having high reduced $\chi^2$ values from our {\it Spitzer}-based astrometric fitting are discussed below.

\subsection{Notes of Interest on Individual Objects within 20 pc\label{notes_on_individual_objects}}

\subsubsection{WISE 0146+4234AB\label{interest_0146+4234}}
As noted by \cite{dupuy2015}, this is a tight binary of separation 0$\farcs$0875$\pm$0$\farcs$0021 with components believed to be of types T9 and Y0. The reduced $\chi^2$ value of 1.579 for our {\it Spitzer} astrometric solution does not indicate any serious issue with the single-object fit despite the double nature of the system and the relatively short orbital period, estimated to be $\sim$6 yr (\citealt{dupuy2015}). As seen in Figure~\ref{trend_MH_vs_HW2}, its overluminosity is nonetheless evident relative to other late-T and early-Y dwarfs.

\subsubsection{WISE 0226$-$0211AB\label{interest_0226-0211}}
\begin{figure}
\figurenum{10}
\includegraphics[scale=0.425,angle=0]{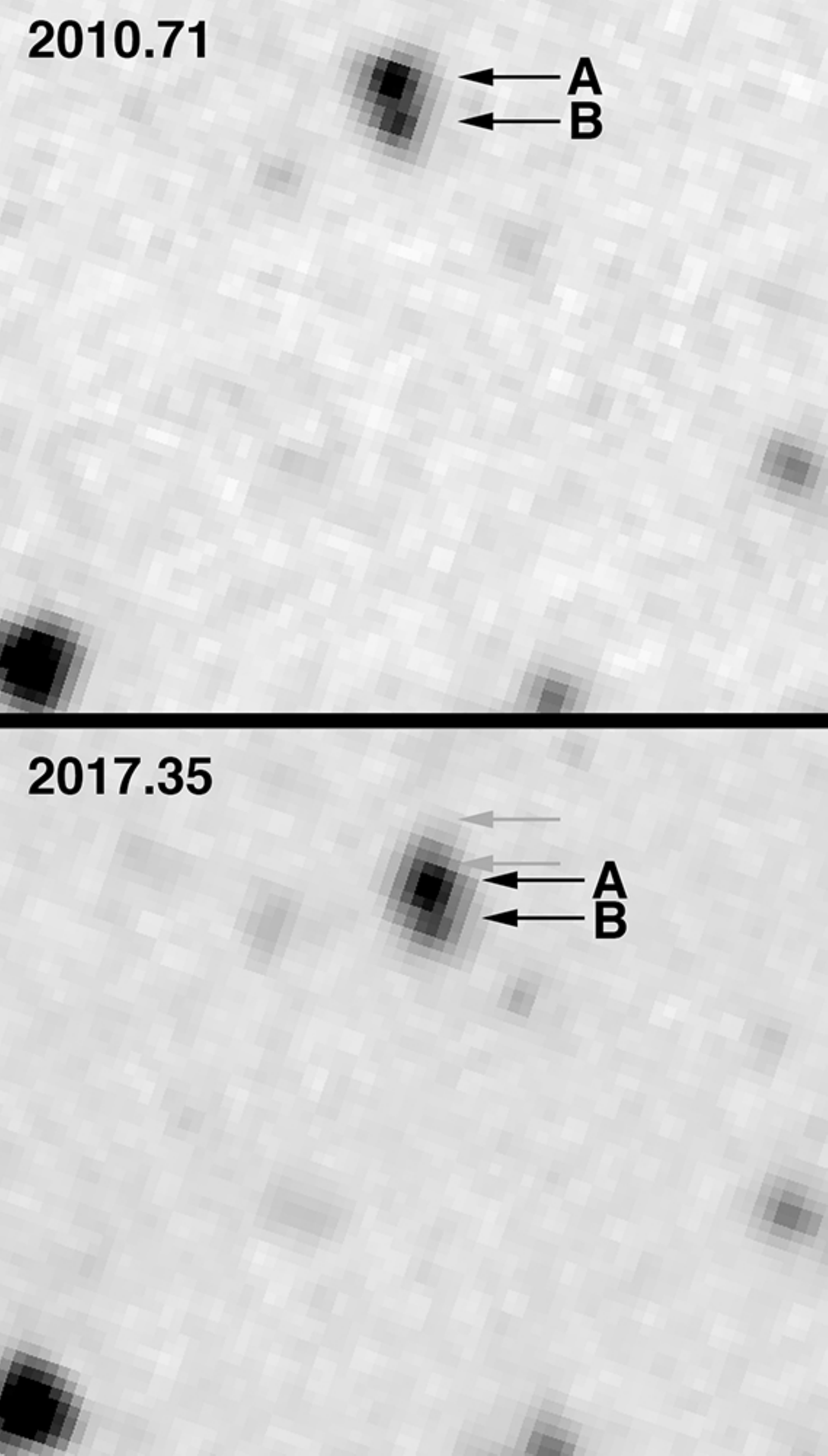}
\caption{A 33$^{\prime\prime}\times$38$^{\prime\prime}$ section of the IRAC ch2 mosaics from AOR 40824064 (epoch 2010.71, top) and 61480192 (epoch 2017.35, bottom) illustrating the common proper motion and binary nature of the WISE 0226$-$0211AB system. In the bottom panel, the two grey arrows indicate the location of the binary at the earlier epoch.
\label{WISE0226_spitzer_image}}
\end{figure}

This is a previously unreported common-proper-motion double with a separation of $\sim$2$\farcs$1 arcsec. This object has the worst reduced $\chi^2$ value of any of our {\it Spitzer} targets (23.645) because the MOPEX/APEX software is unable to successfully centroid on either the A or the B component individually. Figure~\ref{WISE0226_spitzer_image} shows the motion of the pair over a 6.6-yr baseline. This pair was observed with Keck/NIRC2 at $J$ and $H$ bands on dates 21 Sep 2013 and 22 Aug 2013 (UT), respectively, from which we measure PRF-fit magnitudes of $J = 18.57{\pm}0.05$ and $H = 18.88{\pm}0.10$ mag for the northern A component and $J = 22.33{\pm}0.07$ and $H = 22.33{\pm}0.12$ mag for the southern B component. Aperture photometry was measured on the first-epoch {\it Spitzer} AOR images (which had contemporaneous ch1 and ch2 images), and we obtained measurements of ch1 = 16.67$\pm$0.17 and ch2 = 14.86$\pm$0.05 mag for A and ch1 = 18.38$\pm$0.92 and ch2 = 15.59$\pm$0.08 mag for B. The corresponding $H-$ch2 colors are 4.02$\pm$0.11 mag for A and 6.74$\pm$0.14 mag for B, suggesting types of $\sim$T8-T8.5 for A and $\sim$Y0 for B (Figure~\ref{trend_colors_vs_type}). Using our measured parallax, which is somewhat suspect given our inability to properly centroid the pair, we find absolute magnitudes of $M_H = 17.67$ and $M_{ch2} = 13.65$ mag for A and $M_H = 21.12$ and $M_{ch2} = 14.38$ mag for B, suggesting types of $\sim$T8-T8.5 and $\sim$T9.5-Y0 for the two components (Figure~\ref{trend_abs_mags_vs_type}), consistent with the type derived above from the measured colors. Despite the agreement between the colors and the absolute magnitudes, the combined-light spectrum is that of a T7 dwarf (\citealt{kirkpatrick2011}), when expectations would suggest a type closer to T8 or T9. This spectral type determination, based on a noisy $H$-band-only spectrum, should be revisited in light of the dual nature of the source. This source is the reddest T7 in $H-$W2 color and is very faint in $M_H$ for its spectral type despite being a binary, suggesting that the type itself is probably in error.

\subsubsection{WISE 0254+0223\label{interest_0254+0223}}
This T8 dwarf has a higher-than-average $v_{tan}$ value of 84.4$\pm$0.9 km s$^{-1}$, but it shows no oddities in the color and magnitude trends. High quality spectra from 0.8 to 2.5 $\mu$m (Figure 22 of \citealt{kirkpatrick2011}; Figure 3 of \citealt{liu2011}) show it to be a normal T8 dwarf of presumably solar metallicity.

\subsubsection{WISE 0304$-$2705\label{interest_0304-2705}}
The {\it Spitzer} astrometric fit for this object has a reduced $\chi^2$ value of 2.812. Additional observations of this object are warranted, particularly since this is a peculiar Y0 dwarf. \cite{pinfield2014b} argued that the peculiar spectrum was most likely caused by an unusual metallicity and/or gravity. However, the fact that the object is too dim in $M_H$ for its $H-$W2 color (Figure~\ref{trend_MH_vs_HW2}) while other known subdwarfs do not share this effect may suggest that metallicity is not the sole cause. Continued astrometric monitoring is needed to determine if binarity may be complicating the analysis, but in our subsequent analysis we consider this to be a single Y dwarf.

\subsubsection{WISE 0309$-$5016\label{interest_0309-5016}}
This source is much brighter in $M_H$ than other objects of similar $H-$W2 color, and also much brighter in $M_{W1}$ and $M_{ch1}$ than other objects of similar ch1$-$ch2 color. This evidence points at WISE 0309$-$5016 being an unresolved double. There is no spectrum yet of this object, but methane on/off imaging suggests a spectral type of $\sim$T7 (\citealt{tinney2018}).

\subsubsection{WISE 0316+4307\label{interest_0316+4307}}
In $H-$W2 and $H-$ch2 colors, this is an abnormally red T8, and it is faint in $M_H$ for its type. The spectral type is robust, since it is based on separate, high-quality spectra at $J$ and $H$ bands (\citealt{mace2013}), so the reasons for these peculiarities are not known.

\subsubsection{WISE 0323$-$5907\label{interest_0323-5907}}
This object is too faint in $M_{W1}$ and $M_{ch1}$ for its ch1$-$ch2 color. There is no spectrum yet of this object, but the ch1$-$ch2 color of 1.244$\pm$0.034 mag suggests a type of $\sim$T6 (Figure~\ref{trend_colors_vs_type}), in agreement with the photometric type assigned by \cite{schneider2016} using 2MASS and AllWISE photometry.


\subsubsection{WISE 0335+4310\label{interest_0335+4310}}
The poor reduced $\chi^2$ value of 3.227 for this object's {\it Spitzer} astrometric fit is driven primarily by large residuals in the final year of data. For these later epochs, the T9 has moved within $\sim$4$^{\prime\prime}$ of a fainter background star, but that should not be complicating the astrometric measurement.

\subsubsection{WISE 0350$-$5658\label{interest_0350-5658}}
This Y1 dwarf is too faint for its spectral type in both $M_{W2}$ and $M_{ch2}$. It should be acknowledged, however, that there are very few Y dwarfs as late as this one, so judging observational trends at these late types is still difficult. \cite{leggett2017} argue that this object's placement on the $J-$ch2 vs.\ $M_{ch2}$ diagram relative to model predictions suggest that it may be metal rich.

\subsubsection{WISE 0359$-$5401\label{interest_0359-5401}}
This Y0 dwarf's {\it Spitzer} astrometric fit has a reduced $\chi^2$ value of 2.512, and our {\it Spitzer} images show no complications. \cite{opitz2016} used adaptive optics imaging to rule out a binary nature for this object, but only for a companion of equal magnitude more distant than 1.9 AU from the primary. Given that this is one of the few Y dwarfs known, further astrometric monitoring might prove fruitful. Based on its location on the $J-$ch2 vs.\ ch1$-$ch2 plot and the trends shown by model atmosphere calculations, \cite{leggett2017} argue that this object may be metal poor.

\subsubsection{WISE 0430+4633\label{interest_0430+4633}}
In $H-$W2 and $H-$ch2 colors, this is an abnormally red T8, and it is faint in $M_H$ for its type. The spectral type is based on single $J$-band spectrum (\citealt{mace2013}), but this type appears robust since these data have high S/N.

\subsubsection{WISE 0458+6434AB\label{interest_0458+6434}}
Despite being a known binary with 0$\farcs$5 separation (\citealt{gelino2011}), this object has a {\it Spitzer} astrometric fit that is remarkably good. We note, however, that the estimated period $\sim$70 yr suggest very little orbital motion over the timespan of our {\it Spitzer} observations. The reduced $\chi^2$ value of 0.190 -- the lowest of any of our targets -- indicates either that our measurement uncertainties are inflated or that this is a statistical fluke. We note that this object is nonetheless clearly indicated as a binary because it is abnormally bright in $M_{W1}$ and $M_{ch1}$ for its ch1$-$ch2 color. Resolved spectroscopy of the two components gives spectral types of T8.5 and T9.5 (\citealt{burgasser2012}).

\subsubsection{UGPS 0521+3640\label{interest_0521+3640}}
This T8.5 dwarf appears as an outlier on a number of diagrams. It is unusually blue for a T8.5 in both W1$-$W2 and ch1$-$ch2 color; it has a blue W1$-$ch1 color unlike that of most other late-T dwarfs and a W2$-$ch2 color significantly off zero; and it appears as an outlier on various absolute magnitude vs.\ color diagrams. As noted Table~\ref{20pc_sample_photometry}, the W1 photometry for this source is contaminated by the halo of a much brighter star. Moreover, the only extant {\it Spitzer} imaging comes from the shallow GLIMPSE survey (program 61070; PI: B. Whitney). This object is located at a Galactic latitude of only $b = -0.04$ deg, so source confusion in the Galactic Plane is likely an issue in bands other than just W1.

\subsubsection{WISE 0535$-$7500\label{interest_0535-7500}}
This $\ge$Y1: dwarf may be abnormally bright in $M_H$, $M_{W1}$, and $M_{W2}$ for its spectral type and $H-$W2 and ch1$-$ch2 colors. However, there are very few Y dwarfs known that are as late as this one, so judging observational trends is still difficult. The overluminosity may indicate an unresolved double, as was pointed out by \cite{tinney2014}. \cite{opitz2016} used high-resolution adaptive optics imaging in the near-infrared to rule out an equal-magnitude binary with a separation $>$1.9 AU. This source is located in a dense field at the outskirts of the Large Magellanic Cloud, so contamination of some of the filter measurements is a possibility. The small motion of this source (125.2$\pm$1.6 mas yr$^{-1}$) means that the object has not yet moved significantly in the {\it WISE} and {\it Spitzer} imaging since its discovery to reveal whether a background source may have complicated earlier photometry. \cite{leggett2017} favor the binary hypothesis, although \cite{tinney2014} suggests that an unusual atmosphere with thick clouds might provide an alternative explanation.

\subsubsection{WISE 0540+4832\label{interest_0540+4832}}
This T8.5 dwarf is abnormally dim in $M_H$ for its $H-$W2 color. The spectral type is based on a $J$-band spectrum from \cite{mace2013}, but it has high S/N. The reason for this peculiarity is unknown.

\subsubsection{WISE 0614+3912\label{interest_0614+3912}}
This T6 dwarf shows overluminosity in $M_H$ (Figure~\ref{trend_MH_vs_HW2}), $M_{W1}$, $M_{ch1}$, $M_{W2}$, and $M_{ch2}$ for its spectral type and for its ch1$-$ch2 and $H-$W2 colors. Keck/NIRC2 data obtained on 2013 Jan 17 UT (PI: M. Liu) and downloaded from the Keck Observatory Archive\footnote{See {\url https://koa.ipac.caltech.edu}.} (KOA) show this to be a binary with separation 195 mas and $\Delta{H} \approx 2.4$ mag. The KOA also includes Keck/NIRC2 imaging obtained on 2017 Mar 20 UT (PI: M. Liu) that confirms the pair to be a physical, co-moving system. Using the composite $H$-band magnitude from Table~\ref{20pc_sample_photometry} and the parallax listed in Table~\ref{absmag_vtan_table}, this gives magnitudes of $H \approx 16.5$ mag and $M_H \approx 15.2$ mag for the A component and $H \approx 18.9$ mag and $M_H \approx 17.6$ mag for the B component. These absolute magnitudes suggest individual spectral types of T6 and T8, based on the relation between spectral type and $M_H$ in Table~\ref{equations}. 

\subsubsection{WISE 0647$-$6232\label{interest_0647-6232}}
The {\it Spitzer} astrometric fit for this Y1 dwarf has a reduced $\chi^2$ value of 2.309. Given that this is one of the coolest Y dwarfs currently known, continued astrometric monitoring would be instructive.

\subsubsection{WISE 0713$-$5854\label{interest_0713-5854}}
This T9 dwarf is abnormally dim in $M_H$ for its $H-$W2 color. The spectral type is based on a high-quality 0.95-1.65 $\mu$m spectrum from \cite{tinney2018} and shows no peculiarites with respect to a standard T9 spectrum. The reason for this underluminosity is unknown.

\subsubsection{WISE 0723+3403\label{interest_0723+3403}}
Very little follow-up has been done on this T9: dwarf, whose plot of the {\it Spitzer} astrometric fit residuals (Figure~\ref{plots_0005p3737}) and reduced $\chi^2$ value of 2.204 may be suggestive of a companion. The possible cyclical variation in the residuals is, however, based on only nine data points. We note, however, that there is other evidence for binarity given its overluminosity in $M_H$ for its spectral type.

\subsubsection{WISE 0734$-$7157\label{interest_0734-7157}}
This Y0 dwarf has a {\it Spitzer} astrometric fit with a reduced $\chi^2$ value of 3.049. As pointed out by \cite{martin2018}, this is likely one of the warmest Y0 dwarfs known, based on colors and its absolute $J$-band magnitude. Even though {\it HST}/WFC2 images at F105W and F125W show no evidence for binarity (\citealt{schneider2015}), more detailed analysis into the possibility of its being a tighter double is warranted.

\subsubsection{WISE 0751$-$7634\label{interest_0751-7634}}
This T9 dwarf is underluminous in $M_H$, $M_{W1}$, and $M_{ch1}$ for its ch1$-$ch2 color. Another object that shows this same behavior is the sdT8 WISE 2005+5424, which might indicate that WISE 0751$-$7634 is also metal poor. On the other hand, the three other T subdwarfs in the 20-pc sample do not share these traits. \cite{leggett2017} also argue that a metal poor explanation would better explain this object's location on the $J-$ch2 vs.\ ch1$-$ch2 diagram based on atmospheric model trends.

\subsubsection{WD 0806$-$661B\label{interest_0806-661B}}
This object is a very cold companion to a white dwarf, and its spectral type is currently an educated guess. This object is sufficiently faint ($H = 25.29{\pm}0.14$ mag) that no spectrum has yet been acquired. \cite{leggett2017} argue that the object may be metal poor based on its location on the $J-$ch2 vs.\ ch1$-$ch2 diagram.

\subsubsection{WISE 0811$-$8051\label{interest_0811-8051}}
This T9.5: dwarf is one of the objects with a strangely blue W1$-$ch1 color, which is almost certainly indicative of a corrupted measurement, most likely through contamination by a background source in W1. The object, however, does not distinguish itself as an outlier on any other plots. 


\subsubsection{WISE 0825+2805\label{interest_0825+2805}}
This Y0.5 dwarf may be underluminous in $M_H$ for it spectral type, but there are very few known Y dwarfs with types this late, so accurately gauging trends is difficult.

\subsubsection{WISE 0833+0052\label{interest_0833+0052}}
This is an (sd)T9 with a high $v_{tan}$ value of 101.9$\pm$5.6 km s$^{-1}$. It does not appear as an outlier on any of our color or absolute magnitude plots.


\subsubsection{WISE 0855$-$0714\label{interest_0855-0714}}
This, the least luminous brown dwarf known, has a relatively high tangential velocity of 88.0$\pm$0.6 km s$^{-1}$. The only extant spectra of this object are low-resolution data at $L$ and $M$ bands from \cite{skemer2016} and \cite{morley2018}. Fits to model atmospheres by \cite{morley2018} indicate that the object may have slightly subsolar abundance.

\subsubsection{2MASS 0939$-$2448\label{interest_0939-2448}}
Our analysis indicates that this T8 dwarf is overluminous in $M_H$ for its $H-$W2 color (Figure~\ref{trend_MH_vs_HW2}) and overluminous in $M_{W1}$ and $M_{ch1}$ for its ch1$-$ch2 color. This has been a suspected binary for a decade: \cite{burgasser2008} noted its overluminosity early and favored a binary hypothesis and slightly subsolar metallicity, and \cite{leggett2009} reached the same conclusion in analyzing the same data set. \cite{line2017} note that the object appears to be single down to 0$\farcs$07 resolution with near-infrared adaptive optics imaging, and their own analysis reaches the same conclusion of binarity and slightly subsolar abundance. For subsequent analyses, we consider this object to be an unresolved double.

\subsubsection{2MASS 1047+2124\label{interest_1047+2124}}
This T6.5 has a higher-than-average $v_{tan}$ value of 86.5$\pm$3.5 km s$^{-1}$, but it appears to be a normal, solar metallicity T dwarf (\citealt{burgasser2002}). Its claim to fame is that it is the first T dwarf with detected bursts in the radio (\citealt{williams2015} and references therein).

\subsubsection{WISE 1050+5056\label{interest_1050+5056}}
The reduced $\chi^2$ value for this T8 dwarf's {\it Spitzer} astrometric fit is 2.940, but this is based on only seven epochs of information. Additional astrometric measurements of this object are needed not only to further study the residuals but also to determine whether or not this object falls inside of or outside of the 20-pc sample, as our current fit gives $\pi_{abs} = 49.7{\pm}5.0$ mas. For subsequent analyses, this object is assumed to fall beyond 20 pc.

\subsubsection{2MASS 1114$-$2618\label{interest_1114-2618}}
This T7.5 dwarf has a relatively high $v_{tan}$ value of 86.5$\pm$3.5 km s$^{-1}$. Both \cite{burgasser2006} and \cite{leggett2007} agree that this object's near-infrared spectrum is best explained by slightly subsolar metallicity ([M/H] $\approx -0.3$).

\subsubsection{WISE 1141$-$3326\label{interest_1141-3326}}
This Y0 dwarf is anomalously blue in ch1$-$ch2 color for its type, and has a W2$-$ch2 color that's distinctly non-zero. Moreover, it is anomalously blue in W1$-$ch2 color, is too bright in $M_{W1}$ for its $H-$W2 color, is too faint in both $M_{W2}$ and $M_{ch2}$ for its ch1$-$ch2 color, and is an outlier on the $M_H$ vs.\ ch1$-$ch2 diagram. It is quite likely that most, if not all, of these anomalies can be attributed to confusion at early epochs by a background galaxy (see Figure 2 of \citealt{tinney2018}). A new round of photometry, now that the Y dwarf has moved clear of the background source, would be informative.

\subsubsection{ULAS 1152+1134\label{interest_1152+1134}}
The fit to this object's {\it Spitzer} astrometry has a reduced $\chi^2$ value of 2.551, but this is based on only six data points. Additional astrometric measurements of this object are needed not only to further study the residuals but also to determine whether or not this object falls within the 20-pc sample. Our current fit gives $\pi_{abs} = 49.7{\pm}5.1$ mas. For subsequent analyses, this object is assumed to fall beyond 20 pc.

\subsubsection{WISE 1217+1626AB\label{interest_1217+1626}}
This T9 dwarf is overluminous in $M_{W1}$ and $M_{ch1}$ for its ch1$-$ch2 color and clearly sits above the standard sequence in the $M_H$ vs.\ $H-$W2 plot (Figure~\ref{trend_MH_vs_HW2}). This object was found to be a 0$\farcs$8 binary with component spectral types of T9 and Y0 by \cite{liu2012}. Our {\it Spitzer} astrometric fit for this object has a reduced $\chi^2$ value of 5.776, which also strongly hints at its binary nature.

\subsubsection{2MASS 1225$-$2739\label{interest_1225-2739}}
This T6 dwarf is overluminous in all measured bands for its spectral type and colors. \cite{burgasser2003c} showed that the source was a binary with separation of 0$\farcs$28 on {\it HST}/WFPC2 images at F814W and F1042M. Using the parallax of the system and photometry from the {\it HST} images, \cite{vrba2004} estimate spectral types of T6: and T8: for the pair. This was later revised with additional photometric data to T5.5 and T8 by \cite{dupuy2012}.

\subsubsection{2MASS 1231+0847\label{interest_1231+0847}} 
This T5.5 dwarf has a relatively high $V_{tan}$ value of 97.8$\pm$6.3 km s$^{-1}$. \cite{burgasser2004} noted that this object has broad \ion{K}{1} lines, indicating that it is either rapidly rotating or has a higher photospheric pressure, the latter of which could point to a higher surface gravity that is the consequence of somewhat subsolar metallicity. However, this object, which was originally given a spectral type of T6 by \cite{burgasser2004}, was revised to T5.5 by \cite{burgasser2006} and thus falls outside of the spectral type sample we consider in further analyses.


\subsubsection{WISE 1322$-$2340\label{interest_1322-2340}}
This is an abnormally blue T8 in both $H-$W2 and $H-$ch2 color, but it is not noted as peculiar on any other plots. \cite{gelino2011} did not find any evidence for binarity using adaptive optics near-infrared imaging, ruling out any companions with $\Delta{H} < 4$ mag outside of 0{\farcs}2. No peculiarities were noted in the 0.8-2.5 $\mu$m spectrum by \cite{kirkpatrick2011}.

\subsubsection{WISE 1405+5534\label{interest_1405+5534}}
This object is an outlier in $M_{W1}$ for its type, but it is known to have an unusual spectrum for an early-Y dwarf. Its $H$-band emission peak is shifted to longer wavelengths by $\sim$60\AA\ relative to other late-T and early-Y dwarfs (\citealt{cushing2011}).

\subsubsection{ULAS 1416+1348\label{interest_1416+1348}}
This is the reddest T7.5 dwarf in $H-$W2 and $H-$ch2 color. It is also overluminous in $M_{W1}$ and $M_{ch1}$ for its ch1$-$ch2 color and in $M_H$ for its $H-$W2 color. This object is a common-proper-motion companion to SDSS J141624.08+134826.7, which is often regarded as an sdL7 or d/sdL7 (\citealt{kirkpatrick2011}, \citealt{burningham2010b}). ULAS 1416+1348 has somewhat conflicting spectral types of T7.5 (blue), T7.5 pec, and sdT7.5 (\citealt{burgasser2010}, \citealt{burningham2010b}, \citealt{kirkpatrick2011}), all of which highlight its unusual spectrum. Given that its peculiarities and those of its primary are often attributed to subsolar metallicity, a designation of sdT7.5 seems appropriate. Whereas the three other late-T subdwarfs known (WISE 0833+0052, Gl 547B, and WISE 2005+5424) show no overluminosity relative to normal dwarfs, ULAS 1416+1348 very clearly does. We therefore conclude that this object is a close, unresolved double and consider it as such in subsequent analyses.

\subsubsection{WISE 1501$-$4004\label{interest_1501-4004}}
The {\it Spitzer} astrometric fit for this T6 dwarf has a reduced $\chi^2$ value of 2.493, but that is based on only seven epochs of data. Continued monitoring is indicated to see if this may indicate the presence of a lower-mass companion. The object does not, however, appear as an outlier in any of our color/magnitude/type plots.

\subsubsection{WISE 1523+3125\label{interest_1523+3125}}
This T6.5pec dwarf is the reddest T6.5 in $H-$W2 and $H-$ch2 colors. It is also underluminous in $M_H$ for its spectral type. \cite{mace2013} showed that the $J$-band flux best matched that of a T7 standard whereas the $H$-band flux best matched that of a T6 standard, which is similar to the spectral behavior of 2MASS 0937+2931, which may be slightly metal poor itself (\citealt{burgasser2002}). However, this same spectral behavior is not seen in the T subdwarfs such as Gl 547B, and 2MASS 0937+2931 does not stand out as an outlier on any of our plots. The cause for the peculiar spectra and photometry of WISE 1523+3125 is unknown.

\subsubsection{WISE 1541$-$2250\label{interest_1541-2250}}
This Y1 dwarf is another of the objects with a strangely blue W1$-$ch1 color. This is almost certainly due to an erroneous W1 measurement (see discussion in \citealt{kirkpatrick2012}) caused by its passage near a brighter background source.

\subsubsection{WISE 1542+2230\label{interest_1542+2230}}
The reduced $\chi^2$ value for this T9.5 object's {\it Spitzer} astrometric fit is 2.181. The only high-resolution imaging reported for this object is from {\it HST}/WFC2 by \cite{schneider2015} at F105W, F125W, and F140W. No evidence of a companion is seen. Continued monitoring is suggested.

\subsubsection{2MASS 1553+1532\label{interest_1553+1532}}
This T7 dwarf is overluminous in $M_{W2}$ and $M_{ch2}$ for its spectral type. This object was confirmed as a close binary by \cite{burgasser2002} using {\it HST}/NICMOS imaging. Using available photometry and combined-light spectroscopy, \cite{dupuy2012} estimate spectral types for the individual components of T6.5 and T7.5.

\subsubsection{WISE 1627+3255\label{interest_1627+3255}}
This T6 dwarf is consistently overluminous in all bands for it colors. It is a dramatic outlier, for example, on the $M_H$ vs.\ $H-$W2 plot of Figure~\ref{trend_MH_vs_HW2}. \cite{gelino2011} found no evidence for a companion down to $\Delta{H} \approx 5$ mag for separations larger than 0$\farcs$2. Because of the strong evidence favoring binarity, we consider this object to be a tight, unresolved double in subsequent analyses.


\subsubsection{WISE 1804+3117\label{interest_1804+3117}}
This T9.5: dwarf is overluminous in $M_H$, $M_{W2}$, and $M_{ch2}$ for its spectral type. Because the type is uncertain, this discrepancy would disappear if a slightly earlier spectral type were indicated. We note that the object is not overluminous on the absolute magnitude vs.\ color plots, so we will assume this object is single for subsequent analyses.

\subsubsection{WISE 1813+2835\label{interest_1813+2835}}
This T8 is unusually blue in both W1$-$W2 and W1$-$ch1 color. This indicates that the W1 measurement from WISE is contaminated (made too bright) by a background object. Indeed, recent imaging from {\it NEOWISE} shows the T dwarf has now moved away from a bluer source behind it. \cite{mace2013} do not note anything peculiar about the spectrum.

\subsubsection{WISE 1828+2650\label{interest_1828+2650}}
This $\ge$Y2 dwarf is overluminous for its spectral type in $M_{W2}$ and $M_{ch2}$ and is also overluminous in $M_H$ for its $H-$W2 color (Figure~\ref{trend_MH_vs_HW2}). On the $M_{W2}$ vs.\ $H-$W2 plot, it falls $\sim$1 mag above the trend, and in the $M_{ch1}$ vs.\ $H-$W2 plot, it falls at least 1 mag above the trend. However, WISE 1828+2650 falls in line with other Y dwarfs on the plot of $M_H$ vs.\ ch1$-$ch2. This Y dwarf has been an object of much speculation since its broad-band colors cannot be fit by any of the current suite of models (e.g., \citealt{beichman2013}). \cite{leggett2017} speculate that the object may be an equal-magnitude binary (which explains only 0.75 mag worth of overluminosity) as well as having a subsolar metallicity ($[M/H] \approx -0.5$), the latter based on model atmosphere trends seen in the $M_{ch2}$ vs.\ $J-$ch2 diagram. Inexplicably, though, their best fitting atmospheric model suggests a young system ($\sim$1.5 Gyr). Such a young age is hard to reconcile with the subsolar metallicity, so the more likely explanation is that the current suite of atmospheric models simply fails to contain the physics necessary to explain this object's near-infrared spectrum and broadband colors.

\subsubsection{WISE 1928+2356\label{interest_1928+2356}}
This T6 dwarf lies near the Galactic Plane at $l$ = 58.3 deg and $b$ = +3.1 deg, and our {\it Spitzer} astrometric fit has a reduced $\chi^2$ value of 2.347. The motion over our 5.2 yr of {\it Spitzer} observations shows that there are no background sources that might be confusing the astrometric measurements. Continued monitoring is suggested to test for the presence of an unseen companion.

\subsubsection{WISE 2005+5424\label{interest_2005+5424}}
This object has a relatively high $v_{tan}$ value of 110.3$\pm$5.8 km s$^{-1}$, and appears as an outlier on some of our other plots. This is a known sdT8 (\citealt{mace2013b, mace2018}) and is a distant, common-proper-motion companion to the Wolf 1130 (Gl 781) system.

\subsubsection{ULAS 2146$-$0010\label{interest_2146-0010}}
This object is also known as Wolf 940B, so the near-solar metallicity of the primary star can be assumed to apply to the companion as well (\citealt{burningham2009}, \citealt{leggett2010b}). Surprisingly, this T8 dwarf is an outlier on many of our plots. Because of the high proper motion of this late-T dwarf, we now know that during the time of its {\it Spitzer}/IRAC observations in 2008 (see the {\it Spitzer} Heritage Archive), which represents the only ch1 observation that exists, it was contaminated by a background object now clearly visible in later {\it NEOWISE} images at these same wavelengths (see {\it NEOWISE} data at IRSA). Thus, we conclude that the odd placement of this object on many of our diagrams is simply due to contaminated measurements.

\subsubsection{WISE 2209+2711\label{interest_2209+2711}}
This is the faintest Y0 dwarf in $M_H$, $M_{ch1}$, $M_{ch2}$, and $M_{W2}$. Its spectral type is uncertain, though. \cite{schneider2015} suggests that the noisy {\it HST}/WFC3 spectrum on which the type is based may be, despite its poor signal, hinting at sharper $Y$-, $J$-, and $H$-band peaks that point to a type closer to Y1. A later spectral type would solve the discrepancy.

\subsubsection{WISE 2212$-$6931\label{interest_2212-6931}}
The reduced $\chi^2$ value for the {\it Spitzer} astrometric fit of this T9 dwarf is 2.422. High-resolution {\it HST}/WFC3 imaging at F105W and F125W by \cite{schneider2015} shows no evidence of binarity. However, this object also shows an abnormally blue W1$-$ch1 color. This suggests possible contamination by a background object in the W1 photometry, which later NEOWISE imaging strongly suggests. This contamination at W1 should not, however, affect the ch2 astrometry, so continued monitoring is still warranted.

\subsubsection{WISE 2226+0440\label{interest_2226+0440}}
This T8 is abnormally blue in $H$-W2 and $H-$ch2 color and overluminous in $M_H$ for its type. It does not, however, appear overluminous in $M_H$ for its $H-$W2 color. We consider this to be a single object. 

\subsubsection{WISE 2344+1034\label{interest_2344+1034}}
The reduced $\chi^2$ value for this T9 dwarf's {\it Spitzer} astrometric fit is 3.157. No high-resolution imaging has been published for this source. Astrometric monitoring will continue in an attempt to determine whether the poor fit is due to an unseen companion.

\subsection{Fits to the Trends\label{fits_to_trends}}

With the examination of the outliers complete, we fit polynomial relations to trends seen in Figures~\ref{trend_colors_vs_type}, \ref{trend_abs_mags_vs_type}, \ref{trend_abs_mags_vs_ch1ch2}, and \ref{trend_abs_mags_vs_HW2} using a standard least-squares approach. Removed from consideration from the fits, and shown by the circled points in each of the four figures, are known binaries (WISE 0146+4324AB, WISE 0226$-$0211AB, WISE 0458+6434AB, WISE 1217+1626AB, 2MASS 1225$-$2739AB, and 2MASS 1553+1532AB), objects with strong indication of binarity even if this is not yet proven (WISE 0309$-$5016, WISE 0614+3912, 2MASS 0939$-$2448, ULAS 1416+1348, and WISE 1627+3255), subdwarfs (WISE 0833+0052, Gl 547B, and WISE 2005+5424), and objects proven to have poor photometry in any band (UGPS 0521+3640, WISE 0535$-$7500, WISE 0811$-$8051, WISE 1141$-$3326, WISE 1541$-$2250, WISE 1813+2835, ULAS 2146$-$0010, WISE 2212$-$6931). The coefficients for the resulting fitted polynomials are given in Table~\ref{equations}. 

\startlongtable
\begin{deluxetable*}{cccccccc}
\tabletypesize{\footnotesize}
\tablenum{8}
\tablecaption{Coefficients for the Fitted Relations in Figures~\ref{trend_colors_vs_type}, ~\ref{trend_abs_mags_vs_type}, ~\ref{trend_abs_mags_vs_ch1ch2}, ~\ref{trend_abs_mags_vs_HW2}, and ~\ref{teff_trends} \label{equations}}
\tablehead{
\colhead{$x$} & 
\colhead{$y$} &
\colhead{$c_0$} &
\colhead{$c_1$} &
\colhead{$c_2$} &
\colhead{$c_3$} &
\colhead{Valid} &
\colhead{RMS\tablenotemark{a}} \\
\colhead{} & 
\colhead{} &
\colhead{} &
\colhead{} &
\colhead{} &
\colhead{} &
\colhead{Range} &
\colhead{} \\
\colhead{(1)} &                          
\colhead{(2)} &  
\colhead{(3)} &  
\colhead{(4)} &
\colhead{(5)} &
\colhead{(6)} &
\colhead{(7)} &
\colhead{(8)}  \\
}
\startdata
SpT & H$-$W2   &    19.7444$\pm$3.43923& 
                   $-$6.76528$\pm$1.18653& 
                      0.806149$\pm$0.132423& 
                   $-$0.0264937$\pm$0.00477809&   6 $\le$ SpT $\le$ 14 & 0.51\\
SpT & W1$-$W2  & $-$1.29441$\pm$3.12275 &
                      0.812400$\pm$1.08525 &
                   $-$0.0609403$\pm$0.121097 &
                      0.00290939$\pm$ 0.00431632  & 6 $\le$ SpT $\le$ 14 & 0.39\\              
SpT & H$-$ch2  &    21.3744$\pm$3.46898 &
                   $-$7.36894$\pm$1.19729 &
                      0.876955$\pm$0.133654 &
                   $-$0.0291438$\pm$0.00482272    & 6 $\le$ SpT $\le$ 14 & 0.51\\ 
SpT & ch1$-$ch2 &   0.769877$\pm$1.27185 &
                   $-$0.229115$\pm$0.438774 &
                      0.0645336$\pm$0.0489775 &
                   $-$0.00245289$\pm$0.00176859   & 6 $\le$ SpT $\le$ 14 & 0.19\\ 
SpT & $M_H$ &        
      36.9714$\pm$4.51031 &  
     -8.66856$\pm$1.55879 &  
      1.05122$\pm$0.174229 &  
   -0.0344809$\pm$0.00629456      & 6 $\le$ SpT $\le$ 14 & 0.67\\ 
SpT & $M_{W1}$ &      
      13.8175$\pm$4.00383 &
    -0.276439$\pm$1.39146 &
    0.0844831$\pm$0.155265 &
  -0.00124337$\pm$0.00553417     & 6 $\le$ SpT $\le$ 14 & 0.50\\ 
SpT & $M_{ch1}$ &      
      17.0849$\pm$2.61678 &
     -1.78453$\pm$0.902762 &
     0.267153$\pm$0.100769 &
  -0.00878279$\pm$0.00363881     & 6 $\le$ SpT $\le$ 14 & 0.39\\
SpT & $M_{W2}$ &       
      16.3585$\pm$1.95871 &
     -1.59820$\pm$0.675466 &
     0.211474$\pm$0.0753817 &
  -0.00682090$\pm$0.00272193    & 6 $\le$ SpT $\le$ 14 & 0.29\\ 
SpT & $M_{ch2}$ &      
      16.3304$\pm$1.89109 &
     -1.56047$\pm$0.652409 &
     0.203183$\pm$0.0728241 &
  -0.00635074$\pm$0.00262970    & 6 $\le$ SpT $\le$ 14 & 0.28\\
SpT & $M_{W3}$ &       
      10.7315$\pm$0.856890 &
    0.0319455$\pm$0.201454 &
    0.0159426$\pm$0.0115214 &
    \nodata                     & 6 $\le$ SpT $\le$ 14 & 0.41\\
ch1$-$ch2 & $M_H$ &    
      18.9087$\pm$1.86865 &
     -8.38783$\pm$2.95532 &
      5.32802$\pm$1.47233 &
    -0.667961$\pm$0.231272     & 0.9 $\le$ ch1$-$ch2 $\le$ 3.7 & 0.69\\ 
ch1$-$ch2 & $M_{W1}$ & 
      11.9192$\pm$1.04102 &
      3.39859$\pm$1.66875 &
    -0.992429$\pm$0.838003 &
     0.258689$\pm$0.131855     & 0.9 $\le$ ch1$-$ch2 $\le$ 3.7 & 0.38\\
ch1$-$ch2 & $M_{ch1}$ & 
      12.9757$\pm$0.772255 &
     0.381900$\pm$1.22714 &
     0.565159$\pm$0.613775 &
   -0.0221107$\pm$0.0968472    & 0.9 $\le$ ch1$-$ch2 $\le$ 3.7 & 0.30\\
ch1$-$ch2 & $M_{W2}$ & 
      13.0115$\pm$0.769282 &
    -0.766700$\pm$1.22244 &
     0.684058$\pm$0.611271 &
   -0.0469633$\pm$0.0964210    & 0.9 $\le$ ch1$-$ch2 $\le$ 3.7 & 0.30\\
ch1$-$ch2 & $M_{ch2}$ & 
      12.9611$\pm$0.771287 &
    -0.595238$\pm$1.22560 &
     0.554636$\pm$0.613005 &
   -0.0206577$\pm$0.0967258    & 0.9 $\le$ ch1$-$ch2 $\le$ 3.7 & 0.30\\
ch1$-$ch2 & $M_{W3}$ & 
      10.9722$\pm$0.333633 &
     0.214034$\pm$0.353496 &
     0.194726$\pm$0.0877464 &
    \nodata                    & 0.9 $\le$ ch1$-$ch2 $\le$ 3.7 & 0.32\\
H$-$W2 & $M_H$ &    
      11.8526$\pm$0.421256 &
      1.51647$\pm$0.275830 &
   -0.0165129$\pm$0.0551863 &
   0.00105023$\pm$0.00336992   & 2.0 $\le$ H$-$W2 $\le$ 10.5 & 0.30\\ 
H$-$W2 & $M_{W1}$  & 
      11.1276$\pm$0.734096 &
      2.22600$\pm$0.493686 &
    -0.283468$\pm$0.100002 &
    0.0177808$\pm$0.00596974   & 2.0 $\le$ H$-$W2 $\le$ 10.5 & 0.44\\
H$-$W2 & $M_{ch1}$ & 
      11.0097$\pm$0.550438 &
      1.75712$\pm$0.360520 &
    -0.191409$\pm$0.0720912 &
    0.0106380$\pm$0.00439955   & 2.0 $\le$ H$-$W2 $\le$ 10.5 & 0.39\\ 
H$-$W2 & $M_{W2}$  & 
      11.8484$\pm$0.421319 &
     0.520033$\pm$0.275872 &
   -0.0174249$\pm$0.0551946 &
   0.00111599$\pm$0.00337042   & 2.0 $\le$ H$-$W2 $\le$ 10.5 & 0.30\\ 
H$-$W2 & $M_{ch2}$ & 
      11.7507$\pm$0.423322 &
     0.607561$\pm$0.277263 &
   -0.0398259$\pm$0.0554427 &
   0.00273370$\pm$0.00338353   & 2.0 $\le$ H$-$W2 $\le$ 10.5 & 0.30\\
H$-$W2 & $M_{W3}$  & 
      10.7836$\pm$0.276352 &
     0.318603$\pm$0.123162 &
  8.03680e-05$\pm$0.0117702 &
    \nodata                    & 2.0 $\le$ H$-$W2 $\le$ 10.5 & 0.37\\ 
$M_H$    & $T_{\rm eff}$ & 11610.9$\pm$1165.61 &
                        $-$1424.48$\pm$177.512 &
                           61.2423$\pm$8.88772 &
                        $-$0.892386$\pm$0.146389 & 14.5 $\le M_H \le$ 27.0 & 41\\
$M_{W2}$ & $T_{\rm eff}$ & 40669.7$\pm$16715.8 &
                        $-$7272.51$\pm$3476.93 &
                           440.398$\pm$240.300 &
                        $-$8.96515$\pm$5.51811 &   12.5 $\le M_{W2} \le$ 17.0 & 70\\
$M_{ch2}$& $T_{\rm eff}$ & 35476.5$\pm$15647.4 &
                        $-$6198.65$\pm$3231.62 &
                           366.839$\pm$221.694 &
                        $-$7.29548$\pm$5.05173 &   12.5 $\le M_{ch2} \le$ 17.2 & 73\\
H$-$W2   & $T_{\rm eff}$ & 1821.84$\pm$109.420 &
                        $-$490.635$\pm$68.2886 &
                           59.1937$\pm$12.9725 &
                        $-$2.55531$\pm$0.760092&   2.0 $\le$ H$-$W2 $\le$ 10.5 & 57\\
SpT      & $T_{\rm eff}$ & 2335.64$\pm$139.416 &
                        $-$286.401$\pm$31.8992 &
                           9.89240$\pm$1.78104 &   
                           \nodata             &   6 $\le$ SpT $\le$ 14  & 68\\
ch1$-$ch2& $T_{\rm eff}$ & 1603.55$\pm$94.2015 &
                        $-$658.290$\pm$94.2278 &
                           79.4503$\pm$22.5429 &   
                           \nodata             &   0.9 $\le$ ch$-$ch2 $\le$ 3.6  & 81\\
\enddata
\tablenotetext{a}{The units are magnitudes for all but the last six entries, whose units are K.}
\tablecomments{These are simple polynomial equations of the form $$y = \sum_{i=0}^{n}c_ix^i.$$ For spectral types, SpT = 6 for T6, SpT = 7 for T7,... SpT = 14 for Y4.}
\end{deluxetable*}

\section{Analysis of the Underlying Mass Function\label{underlying_mf}}

\subsection{Transforming from the Theoretical to the Observational Plane}

Determining the form of the stellar mass function enlightens us about the physical mechanisms underlying star formation and the creation efficiency for objects of all masses. For hydrogen-burning stars in a well-defined observational sample, this is a relatively straightforward measurement: the spectral type or broadband color of each star determines its location on the HR diagram's main sequence, which in turn determines its mass.

The lowest mass end of the mass function, which is comprised solely of brown dwarfs, is, however, much more difficult to measure. Unlike stars, brown dwarfs lack a simple relationship between spectral type/color and mass because they never settle onto a main sequence. Instead, they cool with time, so mass cannot be deduced without some determination of the age, which is a notoriously difficult measurement for field objects.

For brown dwarfs, an alternate method is to assume various functional forms of the mass function and to draw objects from those distributions according to an assumed rate of star formation. Using an assigned birth time, each brown dwarf is allowed to cool to the present time as predicted from an assumed suite of evolutionary models. Then the distribution of these objects as a function of color, spectral type, or other quantity is compared to the empirical distribution.

\cite{burgasser2004b} undertook this kind of analysis to produce predicted empirical distributions assuming several different power-law mass functions and a single version of the log-normal function favored by \cite{chabrier2001}. For evolutionary models, \cite{burgasser2004b} employed models by \cite{burrows1997} and the new (at the time) models of \cite{baraffe2003}. Here we attempt to expand that formalism and to produce empirical distributions over a wider range of assumed mass function models. We also include a newer suite of evolutionary models from \cite{saumon2008} that better predicts behavior across the L/T transition.

\subsubsection{Formalism}

Each assumed mass function, once properly normalized, can be thought of as a probability distribution function (PDF) describing the likelihood of finding a randomly selected object of mass $M$ (in units of $M_\odot$). Integrating the area under this PDF gives the cumulative distribution function (CDF), which gives the probability of a randomly selected object having a mass less than or equal to $M$. In order to sample from this distribution, one can use the inverse transform sampling method, which is done by switching the dependent and independent variables in the CDF and re-solving for the dependent variable, thus giving the inverse CDF. Random seeds can then be generated from a uniform distribution over the interval [0,1] and passed through the inverse CDF to produce samples of objects whose mass distribution is described by the assumed mass function. 

As an example, consider a power-law mass function with an exponent of $\alpha$: $$f(M) = PDF = CM^{-\alpha},$$ where $C$ is a constant and $M$ is the mass. If we wish to sample this distribution over the range of masses $m_1 < M < m_2$, then the cumulative distribution function would be 

\begin{align}
CDF &= C\int_{m_1}^{m_2} M^{-\alpha} dM \nonumber\\ 
    &= \begin{cases} \displaystyle\frac{M^{1-\alpha} - {m_1}^{1-\alpha}}{{m_2}^{1-\alpha} - {m_1}^{1-\alpha}}, \text{ for } \alpha \neq 1 \nonumber\\
    \\
        \displaystyle\frac{ln(M) - ln(m_1)}{ln(m_2) - ln(m_1)}, \text{ for } \alpha = 1. \nonumber
        \end{cases}
\end{align}
Note that at the low-mass limit of $M = m_1$ we have $CDF = 0$, and at the high-mass limit of $M = m_2$ we have $CDF = 1$. The inverse CDF would then be

\begin{align}
CDF^{-1} &= \begin{cases} \displaystyle [x({m_2}^{1-\alpha} - {m_1}^{1-\alpha}) + {m_1}^{1-\alpha}]^{\frac{1}{1-\alpha}}, \text{ for } \alpha \neq 1 \nonumber\\
     \\
     \displaystyle e^{x[ln(m_2) - ln(m_1)] + ln(m_1)}, \text{ for } \alpha = 1. \nonumber
            \end{cases}
\end{align}
Note that at $x=0$ we have $CDF^{-1} = m_1$ and at $x=1$ we have $CDF^{-1} = m_2$.

\subsubsection{Functional Forms\label{functional_forms}}

We considered several different forms of the mass function, as summarized in Table~\ref{mass_functions_used}: (1) a single power law, (2) a log-normal distribution, and (3) a bi-partite power law:

\begin{enumerate}

\item For the single power law, we ran simulations with six different values of the exponent ranging from $\alpha = -1.0$ to $\alpha = 1.5$ in half-integer increments. This encompasses the range of $\alpha$ values generally discussed in the literature for the substellar mass function (see Figure 2 of \citealt{bastian2010}).

\item For the log-normal function shown in Table~\ref{mass_functions_used}, we chose three different values of the mean, $\mu$, and standard deviation, $\sigma$, corresponding to published values in the literature\footnote{We quote these values for the natural (base $e$) logarithmic version of the log-normal function that the $R$ software environment uses.}: $(\mu, \sigma) = (ln(0.079), 0.69ln(10))$ from \cite{chabrier2003-review}, $(ln(0.10), 0.627ln(10))$ from \cite{chabrier2001}, and $(ln(0.22), 0.57ln(10))$ from \cite{chabrier2003}. The first two of these are applicable to the single-object mass function, and the third is applicable to the system mass function (which includes unresolved binaries). 

\item For the bi-partite power law, we took the functional form shown in Table~\ref{mass_functions_used}, taken from equation 55 of \cite{kroupa2013}. The two power laws overlap in the mass range $0.07 M_\odot < M < 0.15 M_\odot$, with the high-mass component having $\alpha_1 = 1.3\pm0.3$ and the low-mass component having $\alpha_2 = 0.3\pm0.4$. We have taken three versions of this bi-partite law, the first of which is taken at the canonical values of $\alpha_1$ and $\alpha_2$. We have also considered two other versions -- one in which the contribution from the lower-mass portion is increased (to $\alpha_2 - \sigma_{\alpha_2}$) while the higher-mass portion is decreased (to $\alpha_1 + \sigma_{\alpha_1}$), and another for which we did the opposite (using $\alpha_1 - \sigma_{\alpha_1}$ and $\alpha_2 + \sigma_{\alpha_2}$) -- to account for the full range of variation within the quoted $\alpha$ uncertainties.

\end{enumerate}

\begin{center}
\begin{deluxetable*}{cccc}
\tablenum{9}
\tablecaption{Mass Functions Considered for Masses $< 0.1M_\odot$\label{mass_functions_used}}
\tablehead{
\colhead{Name} & 
\colhead{Functional} & 
\colhead{Parameters} &
\colhead{Notation} \\
\colhead{} & 
\colhead{Form} &
\colhead{Used} &
\colhead{in Figures}\\
\colhead{(1)} &                          
\colhead{(2)} &
\colhead{(3)} &
\colhead{(4)} \\
}
\startdata
Power law & $CM^{-\alpha}$ & $\alpha = -1.0$& A\\
\nodata   & \nodata        & $\alpha = -0.5$& B\\
\nodata   & \nodata        & $\alpha = 0.0$& C\\
\nodata   & \nodata        & $\alpha = 0.5$& D\\
\nodata   & \nodata        & $\alpha = 1.0$& E\\
\nodata   & \nodata        & $\alpha = 1.5$& F\\
Log-normal& $Ce^{-(ln(M) - \mu)^2/2\sigma^2}$ & $\mu = ln(0.079)$, $\sigma = 0.69ln(10)$& G \\
\nodata   & \nodata                           & $\mu = ln(0.10)$, $\sigma = 0.627ln(10)$& H \\
\nodata   & \nodata                           & $\mu = ln(0.22)$, $\sigma = 0.57ln(10)$& I \\
Bi-partite power law&  $C\big(\frac{M}{0.07}\big)^{-\alpha_1}$ for $M > 0.07M_\odot$, 
                       $\frac{C}{3}\big(\frac{M}{0.07}\big)^{-\alpha_2}$ for $M < 0.15M_\odot$  
                                              & $\alpha_1 = 1.3$, $\alpha_2 = 0.3$& J\\
\nodata   & \nodata                           & $\alpha_1 = 1.0$, $\alpha_2 = 0.7$& K\\ 
\nodata   & \nodata                           & $\alpha_1 = 1.6$, $\alpha_2 = -0.1$& L\\ 
\enddata
\end{deluxetable*}
\end{center}

\subsubsection{Evolutionary Models}

Objects with a variety of birth times are randomly selected from the above mass functions and allowed to evolve to the present day. This latter step requires that we use evolutionary models to predict the object's current effective temperature and luminosity. For each form of the mass function, we considered two different sets of evolutionary models, both of which are for objects of solar metallicity. The first is the set of COND models from \cite{baraffe2003}, which neglect dust opacity and are most applicable in the spectral ranges mid-M and mid- to late-T, where atmospheres are believed to be largely cloud-free. The second set is from \cite{saumon2008} -- specifically, their hybrid suite of models. These hybrid models should be applicable to our full range of $T_{\rm eff}$ because they include clouds for objects warmer than the L/T transition region (a zone in which rapid cloud growth and subsequent clearing are believed to occur; \citealt{burrows2006}) and are cloudless for temperatures cooler than this.

Both sets of models have restrictions that need to be taken into account when drawing object samples. The model grids of \cite{baraffe2003} are presented at five different ages (0.1, 0.5, 1, 5, and 10 Gyr) and fully sample the range $125K \lesssim T_{\rm eff} \lesssim 2800K$, corresponding roughly to masses $0.01 M_\odot < M < 0.10M_\odot$. Objects younger than 100 Myr were not included. Values of $T_{\rm eff}$ were obtained through linear interpolation of grid points.

The models of \cite{saumon2008} are presented at twenty-six different ages covering 3 Myr $<$ age $<$ 10 Gyr and cover the temperature range $300K \lesssim T_{\rm eff} \lesssim 2400K$ for objects with masses $0.002M_\odot < M < 0.085M_\odot$. As with the Baraffe models, values of $T_{\rm eff}$ for our sample objects were obtained through linear interpolation of grid points.

\subsubsection{Birthrates}

Both \cite{allen2005} and \cite{burgasser2004} have studied the effect of differing birthrates on the shape of the resulting luminosity function. \cite{allen2005} noted that most empirically derived birthrate distributions for field stars are roughly consistent with a constant formation rate over the past 10 Gyr. They found that differences in the assumed birthrate produced small variations in the shape of the resulting luminosity function, but these were much smaller than the effects produced by the assumed functional form of the mass function itself. \cite{burgasser2004} expanded this discussion to more extreme birthrate scenarios and found that the resulting empirical luminosity function in the T dwarf regime is relatively insensitive to the assumed star formation rate, whereas the L dwarf regime is not. Nonetheless, we will simply assume constant star formation over the last 10 Gyr and leave analysis of other star formation rates to a future study.

\subsubsection{Sampling Method}

Using the above assumptions, one of us (AJC) wrote code in the $R$ software environment\footnote{See {\url https://www.r-project.org/}.} to produce simulated versions of the $T_{\rm eff}$ distributions. A random number generator was used to draw $3{\times}10^6$ values in the interval [0,1], and each was assigned an age depending upon the order of its selection. The $n$th draw was given an age of $n \times 3333.{\overline{3}}$ yr so that the full sample covered an age spread of 0-10 Gyr uniformly. Its random seed value in the [0,1] interval was passed through the inverse CDF to provide a mass. This mass and age were then compared to the evolutionary model grids to determine the present-day effective temperature of the object.

\subsection{Simulation Results}

Results are shown in the following figures. Figure~\ref{histograms_mf} shows the twelve forms of the mass function from Table~\ref{mass_functions_used}. Overplotted on each panel are three different versions, one for each of three different low-mass cutoffs assumed: 10$M_{\rm Jup}$, 5$M_{\rm Jup}$, and 1$M_{\rm Jup}$.  Subsequent figures show the resulting $T_{\rm eff}$ histograms for each of these mass functions when paired with evolutionary code of \cite{baraffe2003} in Figure~\ref{histograms_teff_baraffe2003} or \cite{saumon2008} in Figure~\ref{histograms_teff_saumon2008}. 

The simulations in Figure~\ref{histograms_teff_baraffe2003} probe sufficiently low masses that the effects of varying the low-mass cutoff can be studied. This has already been explored by \cite{burgasser2004b}, but we reiterate here that the lowest $T_{\rm eff}$ bins, particularly the one from 150-300K, is very sensitive to the mass cutoff. In Model D (power law of $\alpha$ = 0.5), for example, the space densities vary wildly between the 450-300K and 150-300K bins depending upon the cutoff mass. From the 450-300K bin to the 150-300K bin, the space density drops by $\sim$16$\times$ for the 10$M_{\rm Jup}$ cutoff, by $\sim$2$\times$ for the 5$M_{\rm Jup}$ cutoff, and is roughly identical for the 1$M_{\rm Jup}$ cutoff. This is a consequence of the fact that for the 10$M_{\rm Jup}$ cutoff, there has not been enough time in the 10 Gyr alloted for any but the lowest mass, oldest objects to cool to these temperatures. For lower-mass cutoffs, there are larger populations of low-mass objects capable of cooling to these values. Measuring the space density of these coldest objects thus provides a powerful tool in determining the cutoff mass itself.

Note also that the $T_{\rm eff}$ simulations using the \cite{baraffe2003} models are generally smooth whereas those from \cite{saumon2008} show substructure along the range of $T_{\rm eff}$. Namely, all simulations show a bump in the number counts in the 1200-1350K bin. This was noted in Figure 13 of \cite{saumon2008} and is a consequence of the hybrid models in which the cooling timescale increases from 1400K to 1200K, at the transition between L and T dwarfs. This is the $T_{\rm eff}$ zone at which objects are transitioning from cloudy to clear photospheres. It is worth noting that the 1050-1500K range over which this overdensity is seen corresponds to $\sim$L6 to $\sim$T6, a wide spectral type range that itself reflects the breadth of change in spectral morphology as the clouds clear.

\begin{figure*}
\figurenum{11}
\includegraphics[scale=0.8,angle=0]{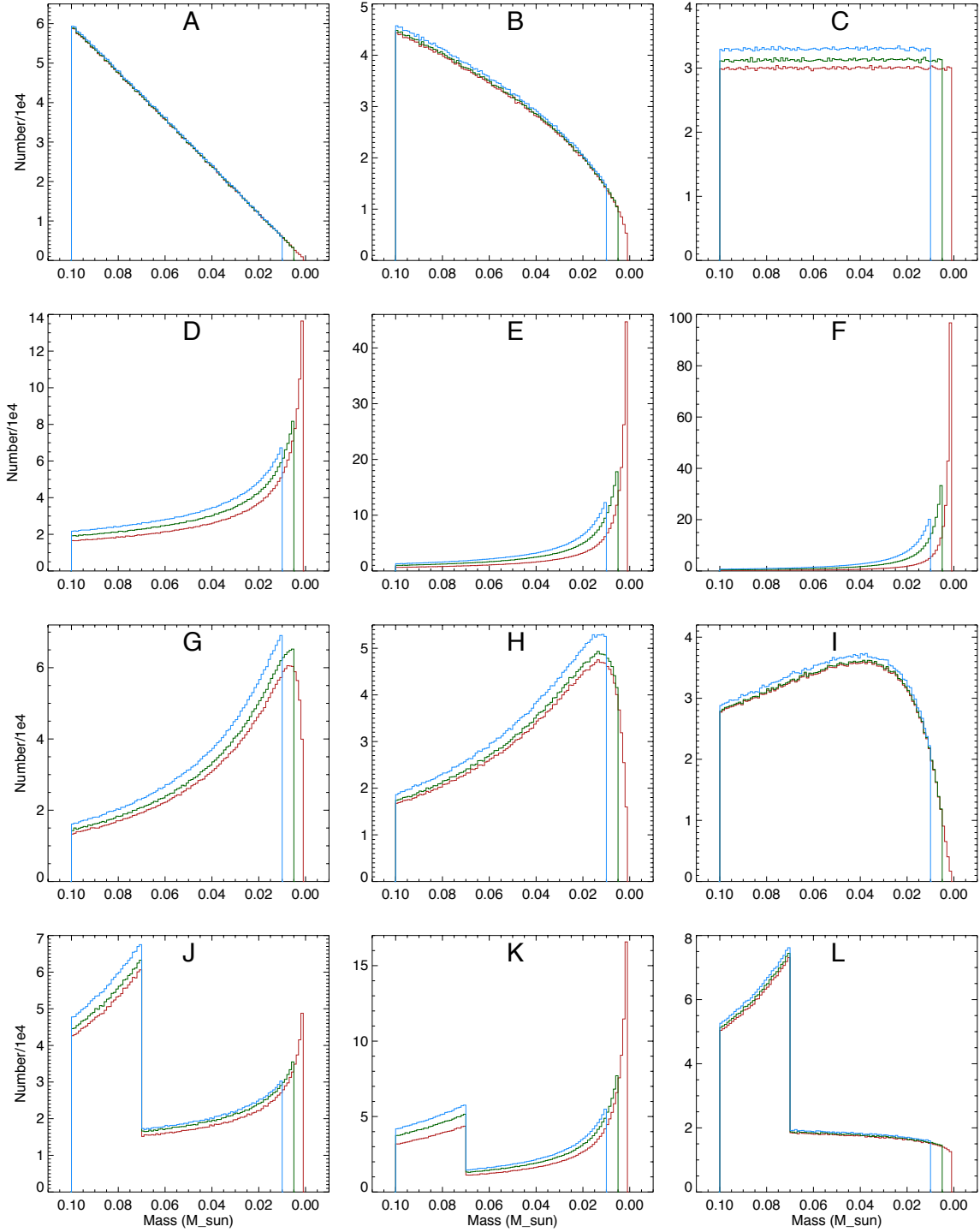}
\caption{Simulated mass functions. Each panel is labeled with a capital letter that refers back to the code in column 4 of Table~\ref{mass_functions_used} and specifies the functional form assumed. Each panel shows the results of three different assumed values of the low-mass cutoff: 10 $M_{\rm Jup}$ (blue), 5 $M_{\rm Jup}$ (green), and 1$M_{\rm Jup}$ (red). Bins are 0.001 M$_\odot$ wide.
\label{histograms_mf}}
\end{figure*}

\begin{figure*}
\figurenum{12}
\includegraphics[scale=0.8,angle=0]{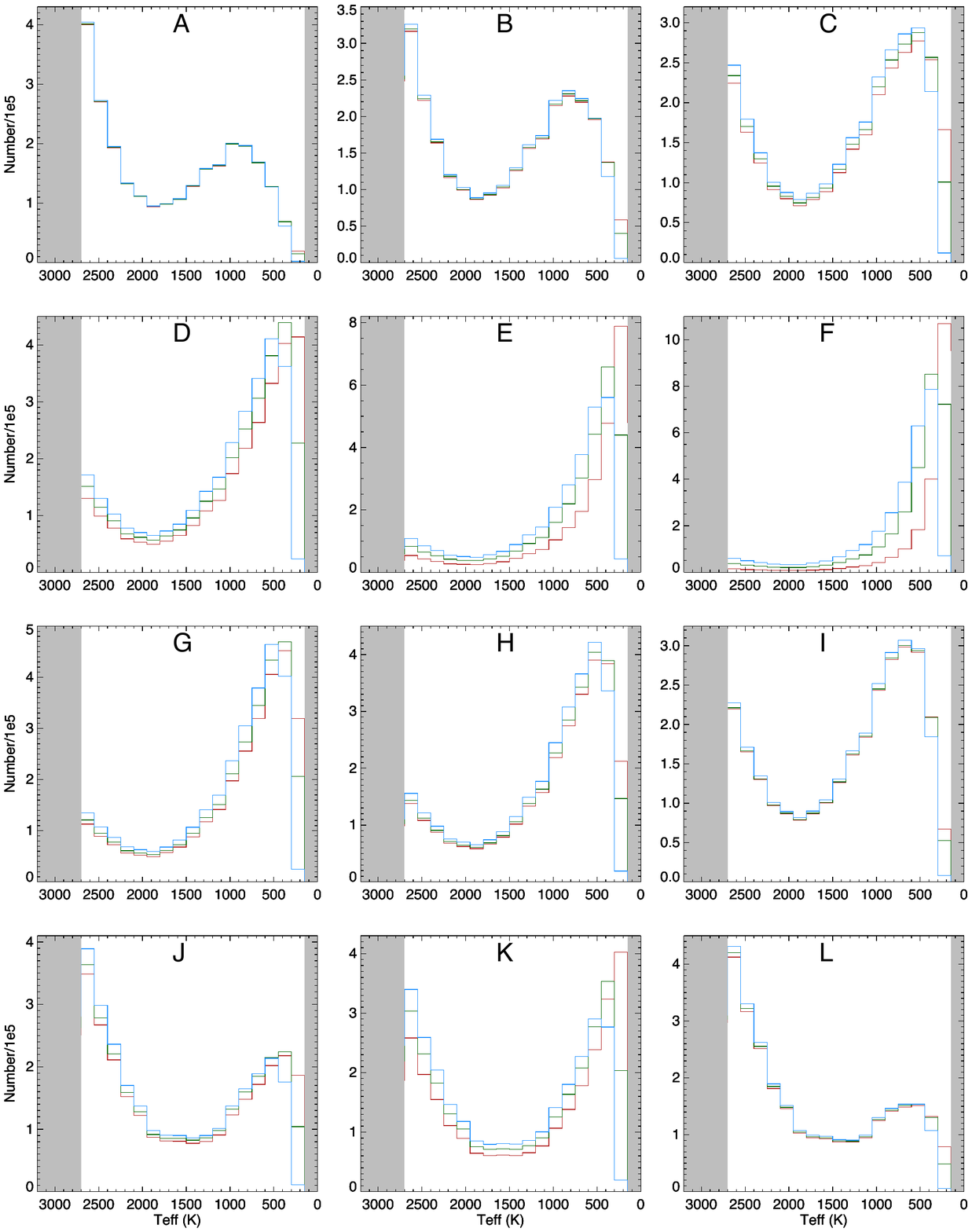}
\caption{The distribution of our simulated sample in $T_{\rm eff}$ for each of the mass function and low-mass cutoff assumptions shown in Figure~\ref{histograms_mf} and passed through the evolutionary models of \cite{baraffe2003}. Color coding is the same as in Figure~\ref{histograms_mf}. Grey zones mark areas in $T_{\rm eff}$ that are not covered by the \cite{baraffe2003} models. Bins are 150K wide.
\label{histograms_teff_baraffe2003}}
\end{figure*}

\begin{figure*}
\figurenum{13}
\includegraphics[scale=0.8,angle=0]{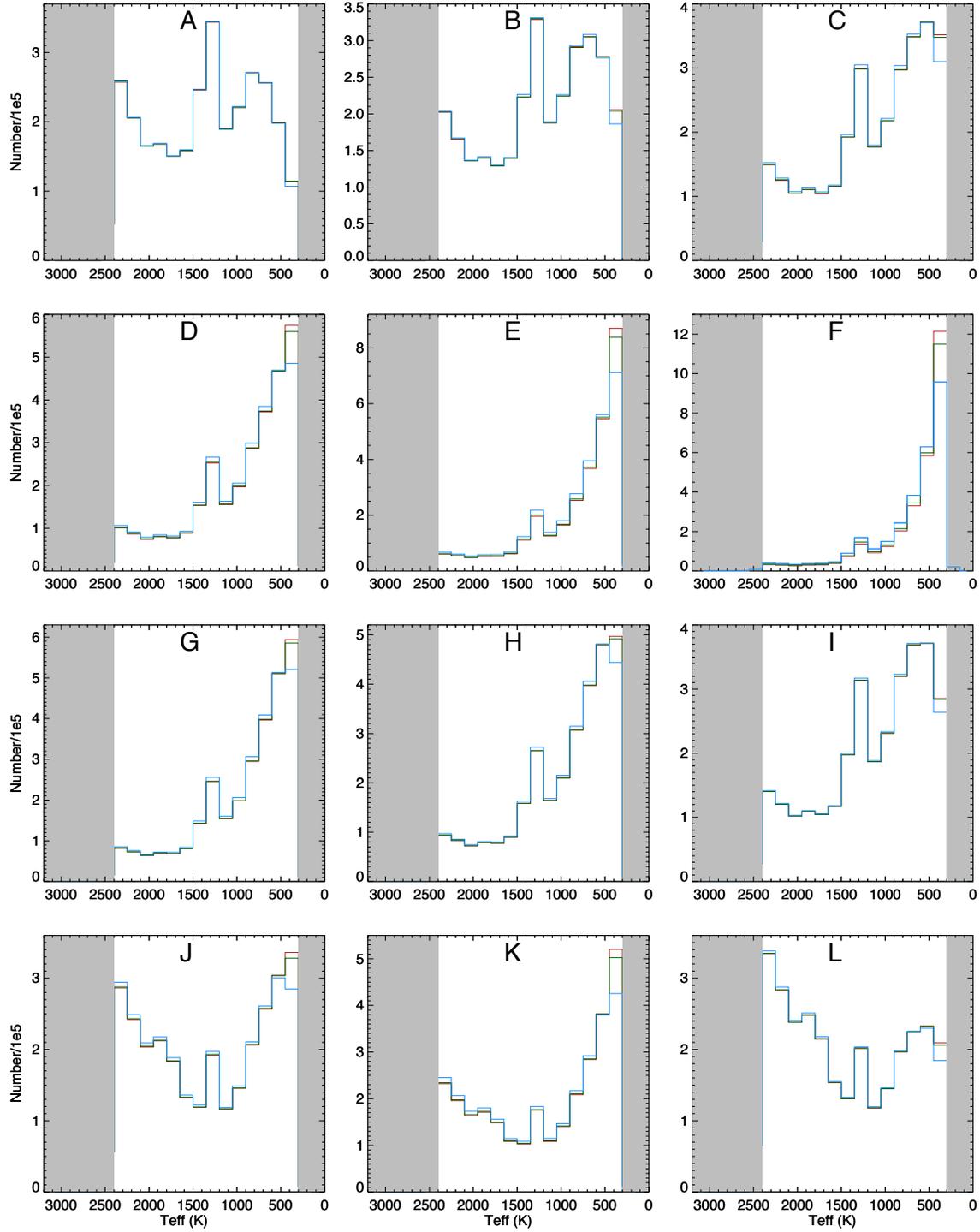}
\caption{The distribution of our simulated sample in $T_{\rm eff}$ for each of the mass function and low-mass cutoff assumptions shown in Figure~\ref{histograms_mf} and passed through the evolutionary models of \cite{saumon2008}. Color coding is the same as in Figure~\ref{histograms_mf}. Grey zones mark areas in $T_{\rm eff}$ that are not covered by the \cite{saumon2008} models. Bins are 150K wide.
\label{histograms_teff_saumon2008}}
\end{figure*}

\clearpage

\subsection{Comparison to Empirical Data\label{mass_function}}

\subsubsection{Late-T and Y Dwarfs}

To compare the output from our mass function simulations to our observational sample, we need to determine the effective temperature for each object within the 20-pc volume. Using the subset of objects from Table~\ref{absmag_vtan_table} that have effective temperature estimates in the literature (Table~\ref{teff_measures}), we can estimate $T_{\rm eff}$ values for the entire set. 

\startlongtable
\begin{deluxetable*}{llcccc}
\tabletypesize{\footnotesize}
\tablenum{10}
\tablecaption{Effective Temperature Determinations for Objects from Table~\ref{absmag_vtan_table}\label{teff_measures}}
\tablehead{
\colhead{Name} & 
\colhead{Spec.\ Type} & 
\colhead{$T_{\rm eff}$ (K) from} &
\colhead{$T_{\rm eff}$ (K) from} &
\colhead{$T_{\rm eff}$ (K) from} &
\colhead{Adopted} \\
\colhead{} & 
\colhead{} & 
\colhead{Forward Modeling} &
\colhead{Retrieval Analysis} &
\colhead{Stefan-Boltzmann} &
\colhead{$T_{\rm eff}$ (K)} \\
\colhead{} & 
\colhead{} & 
\colhead{(Ref.)} &
\colhead{(Ref.)} &
\colhead{(Ref.)} &
\colhead{} \\
\colhead{(1)} &                          
\colhead{(2)} &
\colhead{(3)} &
\colhead{(4)} &
\colhead{(5)} & 
\colhead{(6)}   
}
\startdata
  ULAS  0034-0052  &  T8.5    &  625, 540 (5,10)             & \nodata   &    \nodata & 583 \\
  2MASS 0034+0523  &  T6.5    &  840 (11)                    & \nodata   &    899 (4) & 870 \\
           Gl 27B  &  T8      &  \nodata                     & 719 (3)   &    \nodata & 719 \\
  2MASS 0050-3322  &  T7      &  980 (11)                    & 815 (3)   &    836 (4) & 836 \\
  CFBDS 0059-0114  &  T8.5    &  573, 520 (5,10)             & \nodata   &    \nodata & 547 \\
  WISE  0146+4234AB&  Y0      &  570 (7)                     & \nodata   &    \nodata & 570 \\
  WISE  0146+4234A &  T9      &  333 (16)                    & \nodata   &    \nodata & 330 \\
  WISE  0146+4234B &  Y0      &  415, 320 (2,16)             & \nodata   &    \nodata & 368 \\
  WISE  0148-7202  &  T9.5    &  600 (5)                     & \nodata   &    \nodata & 600 \\
  ULAS  0150+1359  &  T7.5    &  775 (6)                     & \nodata   &    \nodata & 775 \\
  2MASS 0243-2453  &  T6      &  1070 (11)                   & \nodata   &    972 (4) & 1021\\
  WISE  0254+0223  &  T8      &  690 (5)                     & \nodata   &    621 (4) & 656 \\
  WISE  0304-2705  &  Y0pec   &  475 (2)                     & \nodata   &    \nodata & 475 \\
  WISE  0313+7807  &  T8.5    &  662 (7)                     & \nodata   &    \nodata & 662 \\
  WISE  0325-5044  &  T8      &  575 (1)                     & \nodata   &    \nodata & 575 \\
  WISE  0335+4310  &  T9      &  525, 605 (1,7)              & \nodata   &    \nodata & 565 \\  
  2MASS 0348-6022  &  T7      &  950 (8)                     & \nodata   &    \nodata & 950 \\
  WISE  0350-5658  &  Y1      &  325, 325, 253 (1,2,5)       & \nodata   &    \nodata & 325 \\
  WISE  0359-5401  &  Y0      &  400, 435, 268 (1,2,5)       & \nodata   &    \nodata & 400 \\
  WISE  0404-6420  &  T9      &  575 (1)                     & \nodata   &    \nodata & 575 \\
  WISE  0410+1502  &  Y0      &  400, 425, 415, 491 (1,2,5,7)& \nodata   &    \nodata & 420 \\
  2MASS 0415-0935  &  T8      &  704, 750 (5,11)             & 680 (3)   &    677 (4) & 692 \\  
  WISE  0458+6434A &  T8.5    &  765 (5)                     & \nodata   &    \nodata & 765 \\     
  WISE  0458+6434B &  T9.5    &  615 (5)                     & \nodata   &    \nodata & 615 \\ 
  UGPS  0521+3640  &  T8.5    &  625 (9)                     & \nodata   &    \nodata & 625 \\
  WISE  0535-7500  &  >=Y1:   &  475, 375, 595 (1,2,5)       & \nodata   &    \nodata & 475 \\
          Gl 229B  &  T7pec   &  \nodata                     & \nodata   &    927 (4) & 927 \\   
  WISE  0647-6232  &  Y1      &  375, 335 (1,2)              & \nodata   &    \nodata & 355 \\ 
  WISE  0713-2917  &  Y0      &  450, 513 (2,7)              & \nodata   &    \nodata & 482 \\
  UGPS  0722-0540  &  T9      &  521 (5)                     & \nodata   &    569 (4) & 545 \\
  2MASS 0727+1710  &  T7      &  920 (11)                    & 807 (3)   &    845 (4) & 845 \\   
  2MASS 0729-3954  &  T8pec   &  755 (5)                     & 737 (3)   &    752 (4) & 752 \\   
  WISE  0734-7157  &  Y0      &  450, 450 (1,2)              & \nodata   &    \nodata & 450 \\ 
         WD 0806B  &  [>=Y3]  &  338, 353 (2,5)              & \nodata   &    \nodata & 346 \\ 
  DENIS 0817-6155  &  T6      &  \nodata                     & \nodata   &    1004 (4)& 1004\\
  WISE  0825+2805  &  Y0.5    &  400, 325 (1,2)              & \nodata   &    \nodata & 363 \\ 
  WISE  0836-1859  &  T8pec   &  765 (7)                     &\nodata    &    \nodata & 765 \\
  WISE  0855-0714  &  [>=Y4]  &  250 (2)                     & \nodata   &    \nodata & 250 \\ 
  ULAS  0901-0306  &  T7.5    &  670 (10)                    & \nodata   &    \nodata & 670 \\
  2MASS 0937+2931  &  T6pec   &  810 (11)                    & \nodata   &    881 (4) & 846 \\   
  2MASS 0939-2448  &  T8      &  709 (5)                     & 611 (3)   &    686 (4) & 686 \\  
  WISE  0943+3607  &  T9.5    &  475 (1)                     & \nodata   &    \nodata & 475 \\   
  2MASS 1047+2124  &  T6.5    &  \nodata                     & \nodata   &    880 (4) & 880 \\
  2MASS 1114-2618  &  T7.5    &  \nodata                     & 678 (3)   &    669 (4) & 674 \\
  WISE  1118+3125  &  T8.5    &  560 (5)                     & \nodata   &    \nodata & 560 \\
  WISE  1141-3326  &  Y0      &  425 (2)                     & \nodata   &    \nodata & 425 \\  
  WISE  1206+8401  &  Y0      &  425, 435 (1,2)              & \nodata   &    \nodata & 430 \\
  2MASS 1217-0311  &  T7.5    &  870 (11)                    & 726 (3)   &    885 (4) & 870 \\   
  WISE  1217+1626A &  T9      &  575, 583 (5,13)             & \nodata   &    \nodata & 579 \\
  WISE  1217+1626B &  Y0      &  435, 435, 401 (2,5,13)      & \nodata   &    \nodata & 435 \\
  2MASS 1225-2739B &  T8      &  850 (5)                     & \nodata   &    \nodata & 850 \\   
  2MASS 1231+0847  &  T5.5    &  1070 (11)                   & \nodata   &    \nodata & 1070\\      
  2MASS 1237+6526  &  T6.5    &  \nodata                     & \nodata   &    851 (4) & 851 \\ 
          Gl 494C  &  T8      &  671 (5)                     & \nodata   &    721 (4) & 696 \\ 
  WISE  1311+0122  &  T9:     &  672 (7)                     & \nodata   &    \nodata & 672 \\  
  ULAS  1315+0826  &  T7.5    &  580 (10)                    & \nodata   &    \nodata & 580 \\
  ULAS  1335+1130  &  T8.5    &  565, 550 (5,10)             & \nodata   &    \nodata & 558 \\
  SDSS  1346-0031  &  T6.5    &  990 (11)                    & \nodata   &    1011 (4)& 1001\\   
  WISE  1405+5534  &  Y0.5pec?&  375, 385, 415  (1,2,5)      & \nodata   &    \nodata & 385 \\
  ULAS  1416+1348  &  (sd)T7.5&  \nodata                     & 605 (3)   &    656 (4) & 631 \\
          Gl 547B  &  sdT8,T8 &  618 (5)                     & \nodata   &    \nodata & 618 \\
          Gl 570D  &  T7.5    &  800 (11)                    & 715 (3)   &    759 (4) & 759 \\   
  2MASS 1503+2525  &  T5      &  \nodata                     & \nodata   &    1016 (4)& 1016\\  
  SDSS  1504+1027  &  T7      &  \nodata                     & \nodata   &    992 (4) & 992 \\
  WISE  1541-2250  &  Y1      &  400, 375, 550, 441 (1,2,5,7)& \nodata   &    \nodata & 421 \\
  WISE  1542+2230  &  T9.5    &  475, 563  (1,7)             & \nodata   &    \nodata & 519 \\
  2MASS 1553+1532AB&  T7      &  \nodata                     & 803 (3)   &    \nodata & 803 \\ 
  2MASS 1615+1340  &  T6      &  \nodata                     & \nodata   &    906 (4) & 906 \\ 
  WISE  1617+1807  &  T8      &  600 (12)                    & \nodata   &    \nodata & 600 \\
  SDSS  1624+0029  &  T6      &  1010 (11)                   & \nodata   &    936 (4) & 973 \\  
  WISE  1639-6847  &  Y0pec   &  400, 375, 255  (1,2,5)      & \nodata   &    \nodata & 375 \\
  WISE  1711+3500A &  T8      &  761 (13)                    & \nodata   &    \nodata & 761 \\ 
  WISE  1711+3500B &  T9.5    &  465 (13)                    & \nodata   &    \nodata & 465 \\
  WISE  1738+2732  &  Y0      &  400, 425, 440, 514 (1,2,5,7)& \nodata   &    \nodata & 433 \\
  WISE  1741+2553  &  T9      &  615 (5)                     & \nodata   &    \nodata & 615 \\
  SDSS  1758+4633  &  T6.5    &  980 (11)                    & \nodata   &    \nodata & 980 \\
  WISE  1804+3117  &  T9.5:   &  620, 706 (5,7)              & \nodata   &    \nodata & 663 \\
  WISE  1812+2721  &  T8.5:   &  620 (12)                    & \nodata   &    \nodata & 620 \\
  WISE  1828+2650  &  >=Y2    &  325, 540, 400 (2,5,7)       & \nodata   &    \nodata & 400 \\
 SCR   1845-6357B  &  T6      &  1000 (14)                   & \nodata   &    \nodata & 1000\\
  WISE  2005+5424  &  sdT8    &  750 (15)                    & \nodata   &    \nodata & 750 \\
  WISE  2018-7423  &  T7      &  710 (12)                    & \nodata   &    \nodata & 710 \\
  WISE  2056+1459  &  Y0      &  425, 425, 425, 488 (1,2,5,7)& \nodata   &    \nodata & 425 \\
  ULAS  2146-0010  &  T8.5    &  554, 560 (5,12)             & \nodata   &    \nodata & 557 \\
          Gl 845C  &  T6      &  952 (17)                    & \nodata   &    \nodata & 952 \\
  WISE  2209+2711  &  Y0:     &  525, 325, 400  (1,2,7)      & \nodata   &    \nodata & 400 \\
  WISE  2212-6931  &  T9      &  550 (1)                     & \nodata   &    \nodata & 550 \\ 
  WISE  2220-3628  &  Y0      &  425, 425, 525  (1,2,7)      & \nodata   &    \nodata & 425 \\
  2MASS 2228-4310  &  T6      &  891, 1110 (4,11)            & \nodata   &    \nodata & 1001\\
  WISE  2313-8037  &  T8      &  600 (12)                    & \nodata   &    \nodata & 600 \\
  ULAS  2321+1354  &  T7.5    &  800 (6)                     & \nodata   &    \nodata & 800 \\
  WISE  2354+0240  &  Y1      &  350, 350  (1,2)             & \nodata   &    \nodata & 350 \\
\enddata
\onecolumngrid
\tablecomments{References to discovery and spectral classification papers: 
(1) \citealt{schneider2015},
(2) \citealt{leggett2017},
(3) \citealt{line2017},
(4) \citealt{filippazzo2015},
(5) \citealt{dupuy2013},
(6) \citealt{leggett2010},
(7) \citealt{beichman2014},
(8) \citealt{manjavacas2016},
(9) \citealt{burningham2011},
(10) \citealt{marocco2010},
(11) \citealt{burgasser2006b},
(12) \citealt{burgasser2010},
(13) \citealt{liu2012},
(14) \citealt{vigan2012},
(15) \citealt{mace2013b},
(16) \citealt{dupuy2015},
(17) \citealt{king2010}.}
\tablecomments{The $T_{\rm eff}$ values taken from \cite{beichman2014} are those using the models of \cite{morley2012}. \cite{dupuy2013} and \citealt{dupuy2015} used model-dependent bolometric corrections to a limited set of photometry to derive bolometric luminosities, which in turn were used to derive $T_{\rm eff}$ value using model-derived radii. Because this method is so dependent upon a model grid -- albeit an evolutionary grid as opposed to an atmospheric one -- we categorize these estimates under the forward modeling technique rather than the Stefan-Boltzmann one.}
\end{deluxetable*}

As the division in Table~\ref{teff_measures} shows, three main methods have been used in the determinations. Method 1 is forward modeling, which has been the traditional method for determining temperatures. In this method, a grid of model atmospheres is calculated using a complete theoretical description wherein a few vital physical parameters are varied to simulate those likely to be encountered in real atmospheres. Observational spectrophotometric data are then compared to the grid to find the model that best reproduces the observations. Method 2 is inverse modeling, which is also referred to as retrieval analysis. This method strives to deduce physical parameters directly from the data. As explained in \cite{line2014}, this method uses fewer assumptions than forward modeling and has the potential of discerning physical processes (e.g., non-equilibrium chemistry) missing from the forward models, or revealing poor assumptions (e.g., solar elemental abundances) made in those models. Method 3 uses the Stefan-Boltzmann Law to compute the value of $T_{\rm eff}$ directly. In this method, the absolute bolometric luminosity -- computed using a measured parallax and spectrophotometry over a wide range of wavelengths -- is used with an assumed radius ($\sim$1 R$_{Jup}$ for old brown dwarfs) to determine the temperature directly from $T_{\rm eff} = (L/(4{\pi}R^2{\sigma}))^{0.25}$.

For objects in Table~\ref{absmag_vtan_table}, we have searched for $T_{\rm eff}$ measures in the literature. If an object from Table~\ref{absmag_vtan_table} is not listed in Table~\ref{teff_measures}, that indicates that no $T_{\rm eff}$ measures were readily found for that source. However, if measurements were found, only a sample of the measurements from forward modeling are reproduced in the table, since some of the brighter objects and most of the Y dwarfs have many such references. However, we have listed all instances of the $T_{\rm eff}$ measures from retrieval analysis (all from \citealt{line2017}) and from the Stefan-Boltzmann method (all from \citealt{filippazzo2015}\footnote{In actuality, \cite{filippazzo2015} do not assume 1$R_{Jup}$ radius for all of their T dwarfs but use the radius predicted by evolutionary models matching their age estimates. They use gravity diagnostics, or knowledge of the primary if the T dwarf is a companion, to rule out youth for these T dwarfs, which generally sets the ages to a very large range of 0.5-10 Gyr. However, brown dwarfs at these ages are predicted to have contracted to their final radii, so all of the \citealt{filippazzo2015} sources in Table~\ref{teff_measures} were found to have radii of essentially 1$R_{Jup}$ anyway (0.94 to 1.06$R_{Jup}$).}). Many of the $T_{\rm eff}$ measures from forward modeling lack quoted uncertainties, partly due to the fact that the sparse sampling of the model grids makes interpretation of the actual uncertainty difficult, particularly when the perceived uncertainty is comparable to the $T_{\rm eff}$ step size in the grid itself. For uncertainties in the other values, the reader is referred to the cited paper.

For the nine objects having $T_{\rm eff}$ measures from both retrieval analysis and the Stefan-Boltzmann method, the retrieval values tend to run cooler by $\sim$40K on average. For the seven objects having $T_{\rm eff}$ measures from both forward modeling and retrieval analysis, the retrieval values tend to run cooler by $\sim$90K on average. Indeed, \citealt{line2017} has noted that the retrieval values generally run cooler than the values obtained by other methods. Finally, for the fifteen objects having $T_{\rm eff}$ measures from both forward modeling and Stefan-Boltzmann, nine are warmer in the forward modeling value and six are warmer in the Stefan-Boltzmann value. The mean offset for these is $\sim$25K on average, with the $T_{\rm eff}$ measure from forward modeling generally being the warmer one. Because of the sparseness of these comparison data, we have not attempted a correction from one method to the other since there are likely to be correlations with spectral type and/or color as well. So, for each object we merely take a median of all available measures and list that as the adopted $T_{\rm eff}$ value in Table~\ref{teff_measures}. These adopted values are the ones used in the analysis below.

\begin{figure*}
\figurenum{14}
\includegraphics[scale=0.65,angle=0]{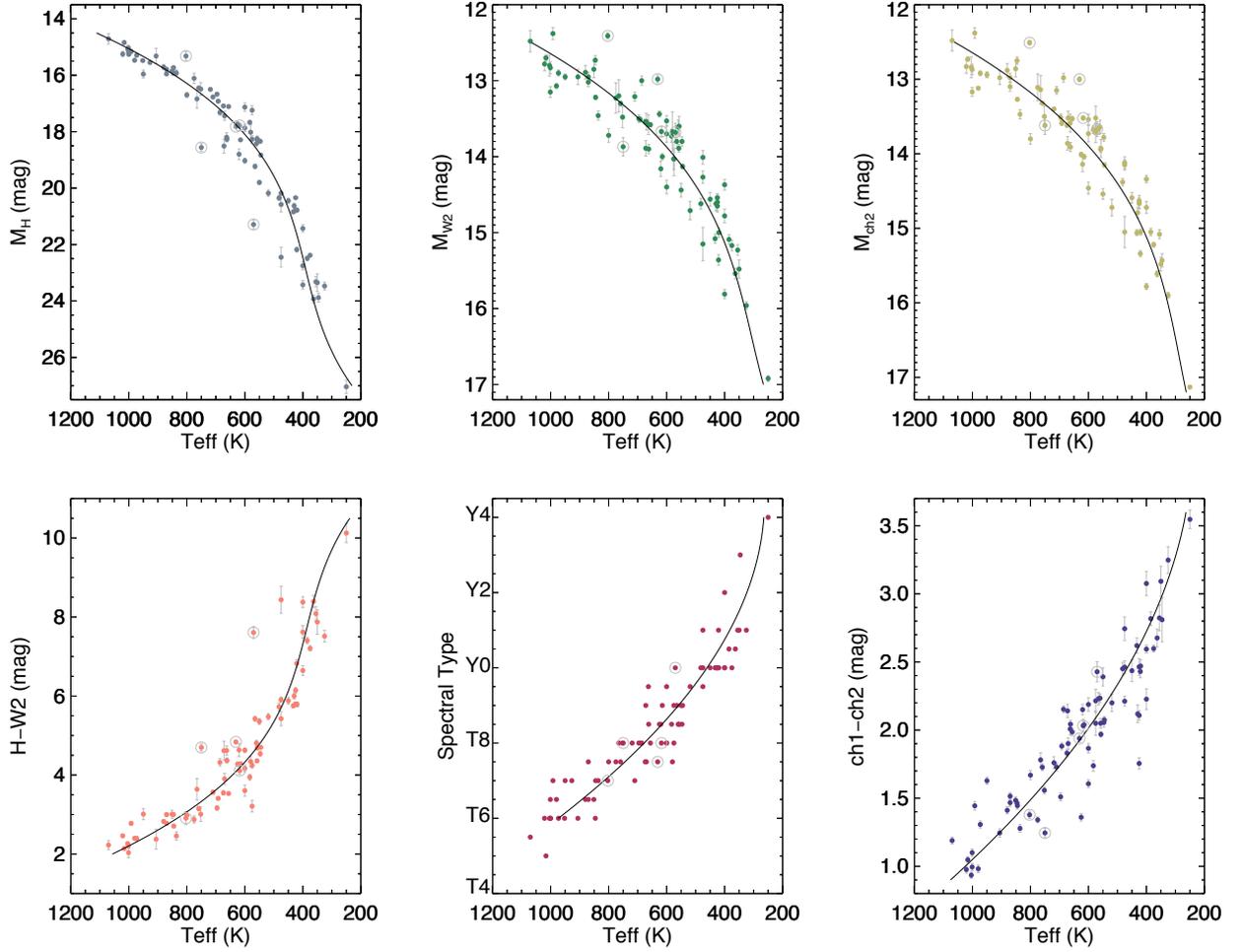}
\caption{Trends of adopted $T_{\rm eff}$ with $M_H$, $M_{W2}$, and $M_{ch2}$ magnitudes, $H-$W2 color, spectral type, and ch1$-$ch2 color, as taken from the data in Table~\ref{absmag_vtan_table} and Table~\ref{teff_measures}. Polynomial fits to the data are shown on the plots and described in Table~\ref{equations}. Open circles in these three panels indicate data points that were excluded from the fits because the objects are binary or known to be of low metallicity.
\label{teff_trends}}
\end{figure*}

Figure~\ref{teff_trends} shows the adopted $T_{\rm eff}$ values from Table~\ref{teff_measures} and the collected photometry and parallax information from Table~\ref{absmag_vtan_table} to illustrate the trend of temperature with absolute magnitudes, colors, and spectral type. Fits to the relations, excluding known binaries and subdwarfs, are shown in the figure and described in Table~\ref{equations}.  

Using these polynomial relations, we can assign $T_{\rm eff}$ values to all objects in the 20-pc sample. Table~\ref{teffs_20pc_census} gives these values and the method used in their determination. We use all relations if the supporting data have been measured, and take the median of the individual results as the adopted $T_{\rm eff}$ value. Table~\ref{teffs_20pc_census} lists which of the relations were used for each source. In the case of two of the subdwarfs, measured values from the literature were used instead.

\startlongtable
\begin{deluxetable*}{lcccccl}
\tabletypesize{\footnotesize}
\tablenum{11}
\tablecaption{Effective Temperature Determinations for the 20-pc Census of T6 and Later Dwarfs\label{teffs_20pc_census}}
\tablehead{
\colhead{Name} & 
\colhead{Adopted} & 
\colhead{$\pi_{abs}$} &
\colhead{$T_{\rm eff}$} &
\colhead{$T_{\rm eff}$} &
\colhead{($l, b$)} &
\colhead{Note}\\
\colhead{} & 
\colhead{Spec.\ Type} & 
\colhead{(mas)} &
\colhead{(K)} &
\colhead{Methods\tablenotemark{a}} &
\colhead{(deg)} &
\colhead{} \\
\colhead{(1)} &                          
\colhead{(2)} &
\colhead{(3)} &
\colhead{(4)} &
\colhead{(5)} &                          
\colhead{(6)} & 
\colhead{(7)}  
}
\startdata
\multicolumn{7}{c}{Warmer objects ($>$1050K)}\\		            
   2MASS 1231+0847   &    5.5&  75.8$\pm$4.8& 1064& 1,2,3,4,5,6&  287.8   71.1& Sec.~\ref{interest_1231+0847}\\   
   2MASS 1225-2739A  &    5.5&  76.0$\pm$2.5& 1060& 5          &  296.0   34.9& Sec.~\ref{interest_1225-2739}\\   
\hline
\multicolumn{7}{c}{900-1050K}\\
   WISE  2237+7228   &    6.0&  62.3$\pm$2.5& 1014& 1,2,3,4,5,6&  113.1   12.2& \nodata\\
   2MASS 1503+2525   &    5.0& 157.2$\pm$2.2& 1009& 1,2,3,4,5,6&   37.2   60.5& \nodata\\   
   WISE  0627-1114   &    6.0&  74.8$\pm$3.6& 1008& 1,2,3,4,5,6&  220.3  -10.4& \nodata\\
   WISE  2343-7418   &    6.0&  58.8$\pm$2.7&  993& 1,2,3,4,5,6&  309.0  -42.0& \nodata\\
\enddata
\tablenotetext{a}{These are the methods used in determining the $T_{\rm eff}$ value for the source: 1 = $M_H$ relation, 2 = $M_{W2}$ relation, 3 = $M_{ch2}$ relation, 4 = $H-$W2 relation, 5 = spectral type relation, 6 = ch1$-$ch2 relation, 7 = value taken from Table~\ref{teff_measures}. All relations are those shown in Figure~\ref{teff_trends} and described in the last six lines of Table~\ref{equations}}.
\tablecomments{Spectral types and parallaxes in brackets are estimates, not actual measures. Spectral types are coded as T6 = 6.0, T6.5 = 6.5, T7 = 7.0, ... Y0 = 10.0, etc.}
\end{deluxetable*}

The objects in Table~\ref{teffs_20pc_census} represent all of the T and Y dwarfs in Table~\ref{20pc_sample_preliminary} that have absolute parallax values $\ge$50 mas. Binaries are listed as separate objects, with individual spectral types being those discussed in Section~\ref{notes_on_individual_objects}. (For suspected binaries with no other corroborating information, equal-magnitude binaries were assumed for which the individual components have the same spectral type as the component system.) Objects are ordered by our $T_{\rm eff}$ determinations and grouped into 150K bins to coincide with the binning given in Figures~\ref{histograms_teff_baraffe2003} and ~\ref{histograms_teff_saumon2008}. There are six such bins running from 150K to 1050K. The hottest of these bins, from 900-1050K, primarily includes T6 and T6.5 dwarfs. We note that of the 28 objects in this bin, 13 fall in the 50K range from 900-950K, 12 fall in the range 950-1000K, and only 3 fall in the range 1000-1050K. By performing a spectral type cut of $\ge$T6, we have removed some earlier type objects that should otherwise be counted here. In fact, one such object that inadvertently was added to the sample -- the T5 dwarf 2MASS 1503+2525 -- has a temperature determination of 1009K. Thus, we should consider this bin to be incomplete.

A similar analysis shows that the two low-temperature bins are also incomplete. The 150-300K bin contains only WISE 0855$-$0714, at d=2.3 pc and $T_{\rm eff}$=250K. Cooler and/or more distant brown dwarfs are expected to be missing from this bin because of their extreme faintness. The 300-450K bin contains 13 objects, but none of these fall in the 50K bin from 300-350K. Objects in this temperature range have also likely eluded detection.

In addition to incompletenesses related to spectral type and intrinsically dim magnitudes, we need to determine if our goal of identifying a complete 20-pc sample has been successful. The standard analysis used for this is the $V/V_{max}$ test of \cite{schmidt1968}, which checks the distribution of objects in space. In each 150K bin, each object is assigned a value $V$ that is the volume of space interior to the object at the distance corresponding to its parallax. The value $V_{max}$ is the full volume of space contained within the distance limit being considered. For a uniform sample, the average value, $\langle{V}/{V_{max}}\rangle$, should be $\sim$0.5 because half of the sample should lie in the closer half of the volume and the rest should lie in the more distant half. If the value of $\langle{V}/{V_{max}}\rangle$ is significantly different from 0.5, then the sample is either non-homogeneous or incomplete. Given that we expect old, solivagant brown dwarfs to be isotropically distributed in space, this would indicate incompleteness to the distance limit being considered.

\begin{figure*}
\figurenum{15}
\includegraphics[scale=0.80,angle=0]{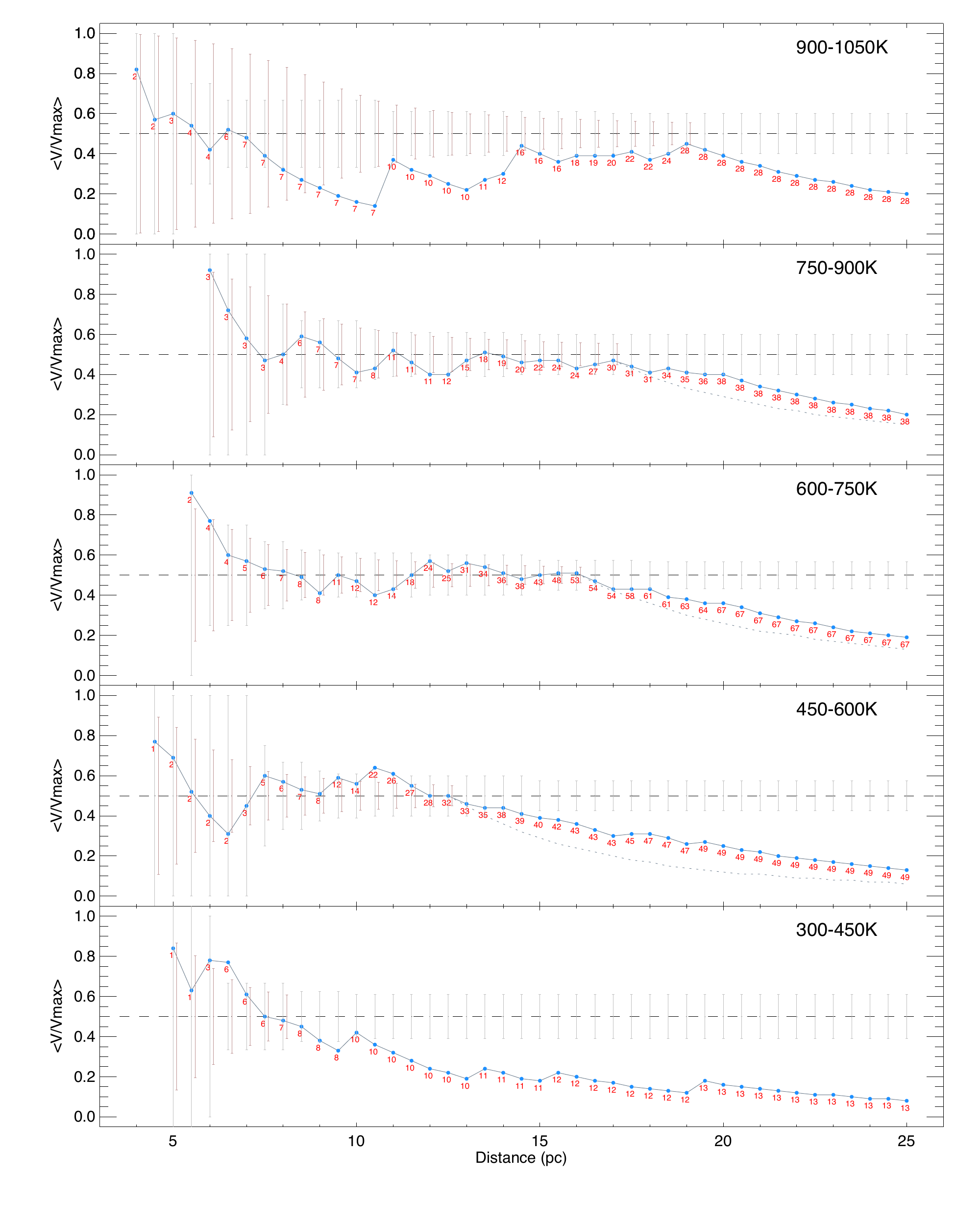}
\caption{The average $V/V_{\rm max}$ value in 0.5-pc intervals for five of the six 150-K bins encompassing our 20-pc T6-Y4 sample. (The sixth bin, from 150-300K, is not shown because it contains only one object, WISE 0855$-$0714.) Blue dots represent our empirical sample. Red labels mark the number of objects in the computation at each 0.5-pc interval. The black dashed line shows the $\langle{V}/{V_{max}}\rangle$ = 0.5 level indicating a complete sample. The grey error bars show the approximate 1$\sigma$ range around $\langle{V}/{V_{max}}\rangle$ = 0.5 that a complete sample of the size indicated by the red number would exhibit, given random statistics. The brown error bars, offset by +0.05 pc from the grey error bars for clarity, show the 1$\sigma$ variation around 0.5 obtained by 10,000 Monte Carlo simulations having the same number of objects and the same completeness limit listed in Table~\ref{space_densities}. The grey dotted lines in the middle three panels show the drop in $\langle{V}/{V_{max}}\rangle$ for a fixed number of objects, where the fixed value is set to the number of objects where $\langle{V}/{V_{max}}\rangle$ last crosses $\sim$0.5. See text for details. (Data in support of this figure can be found in Table~\ref{vmax_table}.)
\label{vmax_plot}}
\end{figure*}

Using the parallaxes listed in Table~\ref{teffs_20pc_census}, we can measure the value of $\langle{V}/{V_{max}}\rangle$ for many different assumed limiting distances. These results are shown in Figure~\ref{vmax_plot}, for five of our six 150K bins. (The coldest bin, from 150-300K, is not shown since it contains only one known object.) At distances of only a few parsec, for which there are only a handful of objects, the values of $\langle{V}/{V_{max}}\rangle$ vary wildly. For most bins, however, this value stabilizes near 0.5 as larger distances and larger number of objects are considered. At distances beyond 20 pc, the value of $\langle{V}/{V_{max}}\rangle$ drops steadily since the list in Table~\ref{teffs_20pc_census} contains no objects more distant than this limit. 

Given that there are, at most, only a few dozen objects in each 150K temperature bin, random fluctuations due to poor statistics can naturally cause the value of $\langle{V}/{V_{max}}\rangle$ to deviate from 0.5. What value of $\langle{V}/{V_{max}}\rangle$ should we take to indicate true incompleteness? To answer this question, we assume a perfectly uniform sample and examine the scale of deviations from $\langle{V}/{V_{max}}\rangle$ = 0.5 that can be imparted due to sample size. For instance, consider a sample of two marbles for which each marble is equally likely to fall in bucket A (representing the nearer half of the sky) or bucket B (representing the more distant half). There is a 50\% probability that two marbles will fall into different buckets, a 25\% probability that both will fall in bucket A, and a 25\% probability that both will fall in bucket B. By analogy with a sky distribution of brown dwarfs, $\langle{V}/{V_{max}}\rangle$ values can thus range as low as 0.0 and as high as 1.0, with only half of the values falling at or near 0.5.

We can generalize this argument to any number of marbles, although we consider only even numbers so that the average value becomes the same number of marbles in each bucket (roughly equivalent to $\langle{V}/{V_{max}}\rangle$ = 0.5). The likelihood that the same number of marbles, $n$, falls in each bucket is given by 

\begin{equation} \frac{n!}{(\frac{n}{2})!(\frac{n}{2})!2^n}.   \nonumber \end{equation}

The likelihood that we find a different number of marbles, differing by the integer $a$, is twice the following value (with the additional factor of two coming from the fact that the number in bucket A can be larger by $a$ than that in bucket B or vice versa)

\begin{equation} \frac{n!}{(\frac{n}{2}-a)!(\frac{n}{2}+a)!2^n}.  \nonumber \end{equation}

We can determine the value of $a_{max}$ at which we reach a cumulative probability, summed over all values of $a$ from $a_{min}$ = 0 (the case of equal numbers in both buckets) to $a_{max}$, of $\sim$68\%. This will allow us to determine a "most likely" range analogous to the Gaussian statistical concept of a 1$\sigma$ likelihood. Because marbles, and brown dwarfs, are quantized, so are the cumulative probabilities, so we consider the value of $a_{max}$ for which the cumulative probability first exceeds 68\%. We then take $0.5 - ((n/2)-a_{max})/n$ as the approximation for the 1$\sigma$ range on $\langle{V}/{V_{max}}\rangle$ for a uniformly distributed set of $n$ objects. We find that for $n$ = 100, the 1$\sigma$ value for $\langle{V}/{V_{max}}\rangle$ is $\sim$0.050; for $n$ = 50 it is $\sim$0.075; and for $n$ = 30 it is $\sim$0.100. For smaller values of $n$ this 1$\sigma$ value dramatically jumps: 0.11 for $n$ = 10, 0.125 for $n$ = 8, 0.167 for $n$ = 6, and 0.25 for $n$ = 4. These 1$\sigma$ values are the ones shown as the grey uncertainties around the $\langle{V}/{V_{max}}\rangle$ = 0.5 line on Figure~\ref{vmax_plot}. The 1$\sigma$ value chosen is the one corresponding to the number of objects (in red on the plot and denoted as $n$ here) being used in the computation at that distance. (When $n$ is odd, we use the 1$\sigma$ value measured for $n-1$.)

With these likelihoods in mind, we can now better interpret the curves shown in Figure~\ref{vmax_plot}. 
For the 900-1050K bin, the largest distance at which the $\langle{V}/{V_{max}}\rangle$ value is $\sim$0.5 is $d$ = 19.0 pc, and there are 28 objects out to this distance, for a space density of 0.97$\times$10$^{-3}$ pc$^{-3}$. 
For the 750-900K bin, we last reach $\langle{V}/{V_{max}}\rangle \approx 0.5$ at $d$ = 17.0 pc, and there are 30 objects out to this distance, for a space density of 1.46$\times$10$^{-3}$ pc$^{-3}$. 
For the 600-750K bin, we last reach $\langle{V}/{V_{max}}\rangle \approx 0.5$ at $d$ = 16.0 pc, and there are 53 objects out to this distance, for a space density of 3.09$\times$10$^{-3}$ pc$^{-3}$. 
For the 450-600K bin, we last reach $\langle{V}/{V_{max}}\rangle \approx 0.5$ at $d$ = 12.5 pc, and there are 32 objects out to this distance, for a space density of 3.91$\times$10$^{-3}$ pc$^{-3}$. 
For the 300-450K bin, we last reach $\langle{V}/{V_{max}}\rangle \approx 0.5$ at $d$ = 8.0 pc, and there are 7 objects out to this distance, for a space density of 3.26$\times$10$^{-3}$ pc$^{-3}$. We know that the 900-1050K and 300-450K bins are inadequately sampled over the entire temperature ranges, so the space densities in these two bins are lower limits only.
We make no claims about the completeness limit of the 150-300K bin since it has only one known object in it. 

For the three bins that are completely sampled in temperature, we can calculate uncertainties on their space densities as follows. 
For the 750-900K bin, a sample of 30 objects would still have a $\langle{V}/{V_{max}}\rangle$ value within the 1$\sigma$ envelope at $d$ = 18.0 pc, as shown by the dotted grey line in the 750-900K panel of Figure~\ref{vmax_plot}. For 30 objects, the space density for this distance would be 1.23$\times$10$^{-3}$ pc$^{-3}$. 
For the 600-750K bin, a sample of 53 objects would still have a $\langle{V}/{V_{max}}\rangle$ value within the 1$\sigma$ envelope at $d$ = 17.0 pc, corresponding to a space density of 2.58$\times$10$^{-3}$ pc$^{-3}$. 
For the 450-600K bin, a sample of 32 objects would still have a $\langle{V}/{V_{max}}\rangle$ value within the 1$\sigma$ envelope at $d$ = 13.5 pc, corresponding to a space density of 3.10$\times$10$^{-3}$ pc$^{-3}$. We adopt the difference between the space density at $\langle{V}/{V_{max}}\rangle$ $\approx$ 0.5 (from the previous paragraph) 
and that derived from the edge of the 1$\sigma$ envelope (in this paragraph) to be the 1$\sigma$ uncertainty in the space density measurement.


We can also use another method to determine when our measured value of $\langle V/V_{max} \rangle$ is significantly different from 0.5, given the size of our sample. We take the completeness limit, $d_{\rm max}$, and sample size, $N$, judged from our analysis above and determine what the 1$\sigma$ variation in $\langle V/V_{max} \rangle$ would be via Monte Carlo trials. The simulation assumes the sample is complete and uniformly but randomly distributed within $d_{\rm max}$. We generate a population of $N$ objects within $d_{\rm max}$ by simply drawing $N$ random numbers from a uniform distribution and converting them into distances $d_{\rm i} = n_{\rm i}^{1/3} d_{\rm max}$, where $i$ runs from 1 to $N$, $d_{\rm i}$ is the distance to the {\it i}th object, and $n_{\rm i}$ is the {\it i}th random number between 0 and 1. We then compute the $\langle V/V_{max} \rangle$ of this synthetic sample as a function of distance. We repeat the above process 10,000 times and at each distance take the standard deviation of the resulting distribution of $\langle V/V_{max} \rangle$ as our measure of the 1$\sigma$ value variation from the ideal value of $\langle V/V_{max} \rangle = 0.5$. These standard deviations are plotted in brown on Figure~\ref{vmax_plot}. In general, these variations are smaller than the ones determined by our method above. However, the Monte Carlo method assumes a completeness at a set distance and predicts the variations likely to be seen at smaller distances. The previous method, on the other hand, predicts the likelihood of a uniform sample at each distance interval given the number of objects internal to that distance. Because it is a more direct measurement, we will continue to use the previous method to judge our completeness limits, and we further note that it also gives us more conservative estimates on our uncertainties.

Now that we have judged completeness in distance, we must further judge whether the sky distribution itself indicates a deficiency  of objects toward the Galactic Plane. If we divide the hemisphere into eight equal-area slices cut along lines of constant $b$ = $0{\fdg}00$, ${\pm}14{\fdg}47$, ${\pm}30{\fdg}00$, and ${\pm}48{\fdg}59$ (Galactic coordinates for each object are given in column 6 of Table~\ref{teffs_20pc_census}), we find 23, 24, 23, 10, 11, 17, 23, and 21 objects starting with the northernmost piece and ending with the southernmost. Indeed, there is a clear deficit concentrated toward the Galactic Plane. To account for this deficit, we multiply our derived space density values by (4/3)(total number with $|b| > 14{\fdg}47$)/(total number at all $b$ values) = (4/3)(131)/152 = 1.15 to account for the missing objects toward the Plane. With this correction factor in hand, we derive our adopted space densities and their uncertainties as listed in Table~\ref{space_densities}.

\begin{center}
\begin{deluxetable}{cccccc}
\tablenum{12}
\tablecaption{Space Density Measurements for Early-L through Early-Y Dwarfs\label{space_densities}}
\tablehead{
\colhead{T$_{\rm eff}$} & 
\colhead{Completeness} &
\colhead{No.\ of} &
\colhead{Adjustment} &
\colhead{Adopted} \\
\colhead{Range} & 
\colhead{Limit} &
\colhead{Objects} &
\colhead{Factor} &
\colhead{Space Density} \\
\colhead{(K)} & 
\colhead{(pc)} &
\colhead{} &
\colhead{} &
\colhead{($\times$10$^{-3}$ pc$^{-3}$)} \\
\colhead{}\\
\colhead{(1)} &                          
\colhead{(2)} &
\colhead{(3)} &
\colhead{(4)} &
\colhead{(5)} \\
}
\startdata
1950-2100& 20.0& 19& 1.00&    0.57$\pm$0.20\\
1800-1950& 20.0& 16& 1.00&    0.48$\pm$0.12\\
1650-1800& 20.0& 21& 1.00&    0.63$\pm$0.09\\
1500-1650& 20.0& 26& 1.00&    0.78$\pm$0.11\\
1350-1500& \nodata& \nodata& \nodata& \nodata\\
1200-1350& \nodata& \nodata& \nodata& \nodata\\
1050-1200& \nodata& \nodata& \nodata& \nodata\\
 900-1050& 19.0& 28& 1.15&    $>$1.12\\
  750-900& 17.0& 30& 1.15&    1.68$\pm$0.23\\
  600-750& 16.0& 53& 1.15&    3.55$\pm$0.51\\
  450-600& 12.5& 32& 1.15&    4.50$\pm$0.81\\
  300-450&  8.0&  7& 1.15&    $>$3.75\\
  150-300&  ---&  1&  ---&                    ---\\
\enddata
\end{deluxetable}
\end{center}

\subsubsection{Early- to Mid-L Dwarfs}

Because our mass function simulations predict only relative space densities, our best metric for determining which simulation $T_{\rm eff}$ distributions best fit the data would be to have another set of space densities measured in a temperature range far removed from the late-T and Y dwarfs. The \cite{baraffe2003} and \cite{saumon2008} models are both valid between 300K and 2500K, so we will determine space densities for objects at the upper end of this $T_{\rm eff}$ range.

Objects at 2500K land near the boundary between late-M dwarfs and early-L dwarfs (see Figure 9.19 of \citealt{gray2009}). Fortunately, we can create a complete, volume-limited census of early- to mid-L dwarfs with reasonable ease, in no small part because these objects are much rarer than their late-T and Y brethren. We used the update to the data underlying the DwarfArchives compilation of published L, T, and Y dwarfs (see Section~\ref{target_selection}) to extract the list. All objects having either an optical or a near-infrared type from L0 to L5 were retained. Those with pre-{\it Gaia} parallaxes placing them within 20 pc of the Sun were kept. For the rest, we used the compiled $J$, $H$, and W2 magnitudes along with their spectral types to compute spectrophotometric distance estimates using the spectral type to absolute magnitude relations presented in \cite{filippazzo2015} for field dwarfs and in \cite{faherty2016} for low-gravity (young) dwarfs, whichever was applicable based on the spectral type. In order to ensure a complete 20-pc sample, we retained all objects having distance estimates of $<$23 pc. We then consulted {\it Gaia} DR2 to obtain parallaxes of these objects and dropped any whose {\it Gaia} data ruled out membership in the 20-pc sample. Of the ninety-one objects, only six lacked parallaxes from either {\it Gaia} DR2 or another published source. 

The full list is presented in Table~\ref{earlyL_20pc_census}. We list the discovery name and coordinates in column 1, the discovery reference in column 2, the adopted parallax and reference in columns 3 and 4, the optical spectral type and reference in columns 5 and 6, the near-infrared spectral type and reference in columns 7 and 8, approximate Galactic coordinates in column 9, and the final adopted type (for use in the discussion below) in column 10. Column 11 gives the $T_{\rm eff}$ bin into which each object is mapped, as discussed further below.

\startlongtable
\begin{deluxetable*}{llllllllllll}
\tabletypesize{\footnotesize}
\tablenum{13}
\tablecaption{The 20-pc Census of L0 to L5.5 Dwarfs\label{earlyL_20pc_census}}
\tablehead{
\colhead{Name and} & 
\colhead{Disc.} & 
\colhead{Parallax\tablenotemark{a}} &
\colhead{Par.} &
\colhead{Opt.} &
\colhead{Opt.} &
\colhead{NIR} &
\colhead{NIR} &
\colhead{($l$, $b$)} &
\colhead{Adopt.} &
\colhead{$T_{\rm eff}$ bin} \\
\colhead{J2000 Coords.} & 
\colhead{Ref.} & 
\colhead{(mas)} &
\colhead{Ref.} &
\colhead{Sp.\ Ty.} &
\colhead{Ref.} &
\colhead{Sp.\ Ty.} &
\colhead{Ref.} &
\colhead{(deg)} &
\colhead{Type} &
\colhead{(K)} \\
\colhead{(1)} &                          
\colhead{(2)} &
\colhead{(3)} &
\colhead{(4)} &
\colhead{(5)} &                          
\colhead{(6)} &
\colhead{(7)} &
\colhead{(8)} &
\colhead{(9)} &                          
\colhead{(10)} &
\colhead{(11)}  
}
\startdata
GJ 1001B (J0004-4044B)     & 1  &    82.0946$\pm$0.3768& 68 &   \nodata    & \nodata  & L5      & 2         &  335.3 -73.3& 5    & 1500-1650\\
GJ 1001C (J0004-4044C)     & 16 &    82.0946$\pm$0.3768& 68 &   \nodata    & \nodata  & L5      & 2         &  335.3 -73.3& 5    & 1500-1650\\
2MASS J00145575-4844171    & 3  &    50.1064$\pm$0.3898& 68 &   L2.5 pec   & 3        & L2.5$\pm$1 & 4      &  318.6 -67.2& 2.5  & 1800-1950\\
2MASSW J0015447+351603     & 5  &    58.6085$\pm$0.3664& 68 &   L2         & 5        & L1.0    & 6         &  114.7 -27.0& 1.5  & 1950-2100\\
2MASSW J0036159+182110     & 7  &   114.4167$\pm$0.2088& 68 &   L3.5       & 7        & L4$\pm$1& 8         &  117.8 -44.3& 3.5  & 1650-1800\\
2MASSW J0045214+163445     & 9  &    65.0151$\pm$0.2274& 68 &   L2$\beta$  & 10       & L2$\gamma$& 18,75   &  120.8 -46.2& 2 yng& 1800-1950\\
2MASS J01282664-5545343    & 11 &    54.0168$\pm$0.2345& 68 &   L2         & 12       & L1      & 11        &  292.3 -60.5& 1.5  & 1950-2100\\
2MASS J01443536-0716142    & 13 &    79.0319$\pm$0.6240& 68 &   L5         & 13       & L6.5    & 14        &  157.7 -66.4& 5.5  & 1500-1650\\
2MASSI J0213288+444445     & 15 &    51.6812$\pm$0.3832& 68 &   L1.5       & 15       & \nodata & \nodata   &  137.9 -15.7& 1.5  & 1950-2100\\
2MASSI J0251148-035245     & 15 &    90.62$\pm$3.02    & 69 &   L3         & 15       & L1      & 9         &  179.0 -53.1& 2    & 1800-1950\\
2MASS J03140344+1603056    & 17 &    73.4296$\pm$0.2757& 68 &   L0         & 17       & M9.4    & 6         &  165.8 -34.6& 0    & \nodata  \\
2MASS J03552337+1133437    & 17 &   109.6451$\pm$0.7368& 68 &   L5$\gamma$ & 10       & L3-L6$\gamma$ & 75  &  178.1 -30.9& 5 yng& \nodata  \\
WISE J040137.21+284951.7   & 19 &    80.2894$\pm$0.2615& 68 &   L3         & 19       & L2.5    & 19        &  165.6 -17.8& 3    & 1650-1800\\
WISE J040418.01+412735.6   & 19 &    61.7516$\pm$0.4163& 68 &   L2         & 19       & L2 pec (red)  & 19  &  157.3 -08.1& 2    & 1800-1950\\
2MASSI J0445538-304820     & 15 &    61.9685$\pm$0.1843& 68 &   L2         & 15       & \nodata & \nodata   &  231.8 -39.1& 2    & 1800-1950\\
2MASS J05002100+0330501    & 17 &    76.2093$\pm$0.3565& 68 &   L4         & 17       & L4.1    & 6         &  196.1 -22.7& 4    & 1650-1800\\
2MASSI J0512063-294954     & 15 &    {\it 54}          & 74 &   L5$\gamma$ & 75       & L5$\beta$& 75       &  232.2 -33.4& 5 yng& \nodata  \\
2MASSI J0523382-140302     & 15 &    78.3632$\pm$0.1855& 68 &   L2.5       & 15       & L5      & 9         &  216.1 -25.6& 3.5  & 1650-1800\\
SDSSp J053951.99-005902.0  & 21 &    78.5318$\pm$0.5707& 68 &   L5         & 21       & L5      & 22        &  205.4 -16.3& 5    & 1500-1650\\
LSR J0602+3910             & 23 &    85.6140$\pm$0.1663& 68 &   L1         & 23       & L2$\beta$& 75       &  172.8 +08.1& 1.5 yng& 1800-1950\\
2MASS J06244595-4521548    & 17 &    81.6233$\pm$0.4986& 68 &   L5         & 17       & L5      & 24        &  253.4 -23.3& 5    & 1500-1650\\
2MASS J06411840-4322329    & 12 &    51.2819$\pm$0.1930& 68 &   L1.5       & 12       & L2.4:   & 6         &  252.3 -19.9& 2    & 1800-1950\\
2MASSI J0652307+471034     & 15 &   109.6858$\pm$0.4380& 68 &   L4.5       & 15       & \nodata & \nodata   &  169.1 +19.7& 4.5  & 1500-1650\\
2MASS J07003664+3157266A   & 25 &    88.2790$\pm$0.3479& 68 &   L3.5       & 17       & L3$\pm$1& 2         &  184.6 +15.7& 3.5  & 1650-1800\\
WISEA J071552.38-114532.9  & 26 &    55.5855$\pm$0.3446& 68 &   \nodata    & \nodata  & L4 pec (blue)& 26   &  226.1 -00.0& 4    & 1650-1800\\
2MASS J07414279-0506464A   & 77 &    {\it 53}          & 77 &   \nodata    & \nodata  & L5      & 77,78     &  223.3 +08.7& 5    & 1500-1650\\
2MASS J07414279-0506464B   & 77 &    {\it 53}          & 77 &   \nodata    & \nodata  & L5      & 77,78     &  223.3 +08.7& 5    & 1500-1650\\
2MASSI J0746425+200032A    & 7  &    81.9$\pm$0.3      & 70 &   L0         & 28       & \nodata & \nodata   &  200.4 +20.8& 0    & \nodata  \\
2MASSI J0746425+200032B    & 27 &    81.9$\pm$0.3      & 70 &   L1.5       & 28       & \nodata & \nodata   &  200.4 +20.8& 1.5  & 1950-2100\\
DENIS-P J0751164-253043    & 29 &    56.5689$\pm$0.1555& 68 &   L1.5       & 29       & L1.1    & 6         &  242.2 +00.6& 1.5  & 1950-2100\\
SSSPM J0829-1309           & 30 &    85.5438$\pm$0.1720& 68 &   L2         & 30       & \nodata & \nodata   &  236.3 +14.6& 2    & 1800-1950\\
2MASSI J0847287-153237     & 15 &    56.9235$\pm$0.3167& 68 &   L2         & 15       & \nodata & \nodata   &  241.1 +17.0& 2    & 1800-1950\\
SIPS J0921-2104         & 17,31 &    79.3128$\pm$0.2253& 68 &   L2         & 17       & L4: (blue) & 32,33  &  250.9 +19.9& 3    & 1650-1800\\
2MASSW J1004392-333518\tablenotemark{b}  
                           & 35 &    53.2798$\pm$0.5684& 68 &   L4         & 35       & \nodata & \nodata   &  267.4 +17.5& 4    & 1650-1800\\
2MASS J10224821+5825453    & 12 &    54.3331$\pm$0.3143& 68 &   L1$\beta$  & 10       & L1      & 12        &  152.0 +49.4& 1 yng& 1950-2100\\
2MASSI J1029216+162652     & 5  &    52.3361$\pm$0.7414& 68 &   L2.5       & 5        & L2.8    & 6         &  223.4 +55.4& 2.5  & 1800-1950\\
SDSS J104523.98-014957.7   & 36 &    58.6576$\pm$0.2384& 68 &   L1         & 36       & L1      & 12        &  251.5 +48.0& 1    & 1950-2100\\
SDSS J104842.84+011158.5   & 36 &    66.4589$\pm$0.2143& 68 &   L1         & 36       & L4      & 37        &  249.1 +50.7& 2.5  & 1800-1950\\
2MASS J10511900+5613086    & 17 &    63.9956$\pm$0.1886& 68 &   L2         & 17       & L0.8    & 6         &  151.0 +53.8& 1.5  & 1950-2100\\
DENIS-P J1058.7-1548       & 38 &    54.6468$\pm$0.5213& 68 &   L3         & 39       & L3      & 22        &  267.1 +39.1& 3    & 1650-1800\\
2MASSI J1104012+195921     & 40 &    55.9160$\pm$0.4448& 68 &   L4.5       & 40       & L4      & 41        &  223.3 +64.4& 4.5  & 1500-1650\\
2MASSW J1108307+683017     & 42 &    61.3537$\pm$0.1985& 68 &   L1$\gamma$ & 75       & L1$\gamma$& 75      &  136.1 +45.8& 1 yng& 1950-2100\\
2MASS J11263991-5003550    & 43 &    59.38$\pm$1.64    & 71 &   L4.5       & 32       & L6.5$\pm$ pec & 32  &  289.2 +10.5& 5.5  & 1500-1650\\
2MASSW J1155395-372735     & 35 &    84.5693$\pm$0.1867& 68 &   L2         & 35       & L2.3    & 6         &  290.8 +24.0& 2    & 1800-1950\\
SDSSp J120358.19+001550.3  & 21 &    67.2362$\pm$0.5553& 68 &   L3         & 21       & L5.0    & 6         &  277.9 +60.8& 4    & 1650-1800\\
2MASSI J1213033-043243     & 15 &    59.4765$\pm$1.0156& 68 &   L5         & 15       & L4.2    & 6         &  285.1 +56.9& 4.5  & 1500-1650\\
2MASS J12212770+0257198    & 17 &    53.9501$\pm$0.2528& 68 &   L0         & 17       & L0.5    & 14        &  285.1 +64.7& 0    & \nodata  \\
DENIS-P J1253108-570924    & 29 &    60.0190$\pm$0.2612& 68 &   L0.5       & 29       & \nodata & \nodata   &  303.1 +05.7& 0.5  & \nodata  \\
2MASSW J1300425+191235     & 42 &    71.6755$\pm$0.2012& 68 &   L1         & 42       & L3 (blue)& 32       &  318.4 +81.8& 2    & 1800-1950\\
Kelu-1A (J1305-2541A)      & 44 &    53.8492$\pm$0.7107& 68 &   L3         & 46       & L1.5-L3 & 45        &  306.9 +37.0& 2.5  & 1800-1950\\
Kelu-1B (J1305-2541B)      & 45 &    53.8492$\pm$0.7107& 68 &   L3         & 46       & L3-L4.5 & 45        &  306.9 +37.0& 3    & 1650-1800\\
Gl499C (J1305+2046)        & 15 &    50.4348$\pm$0.8447& 68 &   L5         & 47       & L6.5    & 6         &  330.8 +82.8& 5.5  & 1500-1650\\
DENIS-P J142527.97-365023.4& 37 &    84.5181$\pm$0.3435& 68 &   L3         & 50       & L4$\gamma$& 75      &  323.1 +22.3& 3.5 yng& 1650-1800\\
2MASSW J1439284+192915     & 39 &    69.6$\pm$0.5      & 70 &   L1         & 39       & L1      & 14        &  021.5 +64.1& 1     & 1950-2100\\
G 239-25B                  & 16 &    91.7062$\pm$0.2434& 68 &   \nodata    & \nodata  & L0      & 51        &  106.7 +47.4& 0     & \nodata  \\
2MASSW J1448256+103159     & 9  &    71.2548$\pm$0.7233& 68 &   L5         & 17       & L3.5    & 24        &  007.4 +57.8& 4.5   & 1500-1650\\
Gl 564B (J1450+2354B)      & 52 &    54.9068$\pm$0.0684& 68 &   \nodata    & \nodata  & L4      & 53        &  032.7 +63.0& 4     & 1650-1800\\
Gl 564C (J1450+2354C)      & 52 &    54.9068$\pm$0.0684& 68 &   \nodata    & \nodata  & L4      & 53        &  032.7 +63.0& 4     & 1650-1800\\
DENIS-P J1454078-660447    & 29 &    93.2242$\pm$0.3013& 68 &   L3.5       & 29       & \nodata & \nodata   &  314.9 -06.1& 3.5   & 1650-1800\\
2MASSW J1506544+132106     & 42 &    85.5810$\pm$0.2883& 68 &   L3         & 42       & L4      & 24        &  016.2 +55.5& 3.5   & 1650-1800\\
2MASSW J1507476-162738     & 7  &   135.2332$\pm$0.3274& 68 &   L5         & 7        & L5.5    & 8         &  344.1 +35.2& 5     & 1500-1650\\
2MASS J15200224-4422419A   & 11 &    54.4581$\pm$0.2465& 68 &   \nodata    & \nodata  & L1.5    & 54        &  329.0 +10.8& 1.5   & 1950-2100\\
2MASS J15200224-4422419B   & 11 &    53.6580$\pm$0.6308& 68 &   \nodata    & \nodata  & L4.5    & 54        &  329.0 +10.8& 4.5   & 1500-1650\\
DENIS-P J153941.96-052042.4& 37 &    58.8245$\pm$0.4213& 68 &   L3.5       & 12       & L4.0    & 24        &  000.7 +37.9& 3.5   & 1650-1800\\
2MASSW J1555157-095605     & 35 &    73.6519$\pm$0.1870& 68 &   L1         & 35       & L1.6    & 6         &  359.5 +32.0& 1     & 1950-2100\\
2MASSW J1615441+355900     & 5  &    50.0611$\pm$0.3713& 68 &   L3         & 5        & L3.6    & 6         &  057.7 +46.0& 3     & 1650-1800\\
2MASSW J1645221-131951     & 35 &    88.8220$\pm$0.1444& 68 &   L1.5       & 35       & \nodata & \nodata   &  005.1 +20.3& 1.5   & 1950-2100\\
2MASSW J1658037+702701     & 42 &    54.1172$\pm$0.2058& 68 &   L1         & 42       & \nodata & \nodata   &  101.9 +34.8& 1     & 1950-2100\\
DENIS-P J170548.38-051645.7& 37 &    52.6734$\pm$0.3516& 68 &   L0.5       & 17       & L1.0    & 24        &  015.2 +20.6& 0.5   & \nodata  \\
2MASS J17065487-1314396    & 76 &    51.4814$\pm$0.5128& 68 &   \nodata    & \nodata  & L5 pec  & 76        &  008.3 +16.0& 5     & 1500-1650\\
2MASS J17072343-0558249B   & 56 &    85.0112$\pm$0.4386& 68 &   \nodata    & \nodata  & L3      & 56        &  014.8 +19.9& 3     & 1650-1800\\
2MASSI J1721039+334415     & 15 &    61.3203$\pm$0.2050& 68 &   L3         & 15       & L5$\pm$1 (blue) & 32&  057.3 +32.5& 4     & 1650-1800\\
VVV J172640.2-273803       & 57 &    53.9938$\pm$0.3612& 68 &   \nodata    & \nodata  & L5$\pm$1 (blue) & 57&  358.8 +04.2& 5     & 1500-1650\\
2MASS J17312974+2721233    & 17 &    83.7364$\pm$0.1182& 68 &   L0         & 17       & L0      & 12        &  050.8 +28.6& 0     & \nodata  \\
DENIS-P J1733423-165449    & 29 &    55.3156$\pm$0.3564& 68 &   L1.0       & 29       & L0.9    & 6         &  008.8 +08.6& 1     & 1950-2100\\
DENIS-P J1745346-164053    & 29 &    51.0274$\pm$0.2957& 68 &   L1.5       & 29       & L1.3    & 6         &  010.4 +06.3& 1.5   & 1950-2100\\
2MASS J17502484-0016151    & 11 &   108.2676$\pm$0.2552& 68 &   L5.0       & 59       & L5.5    & 11        &  025.5 +13.4& 5     & 1500-1650\\
2MASS J17534518-6559559    & 17 &    63.8219$\pm$0.3244& 68 &   L4         & 17       & \nodata & \nodata   &  327.5 -19.0& 4     & 1650-1800\\
WISE J180001.15-155927.2   & 60 &    80.8967$\pm$0.3389& 68 &   L4.5       & 55       & L4.3    & 6         &  012.8 +03.7& 4.5   & 1500-1650\\
2MASSI J1807159+501531     & 9  &    68.3317$\pm$0.1280& 68 &   L1.5       & 15       & L1      & 9         &  077.9 +27.3& 1.5   & 1950-2100\\
2MASS J18212815+1414010    & 61 &   106.8740$\pm$0.2518& 68 &   L4.5       & 61       & L5 pec (red) & 61   &  042.4 +12.9& 4.5   & 1500-1650\\
WISEP J190648.47+401106.8  & 62 &    59.5710$\pm$0.1363& 68 &   \nodata    & \nodata  & L1      & 62        &  071.0 +14.4& 1     & 1950-2100\\
Gl 779B (J2004+1704B)      & 63 &    56.4256$\pm$0.0690& 68 &   \nodata    & \nodata  & L4.5$\pm$1.5 & 63   &  056.5 -07.5& 4.5   & 1500-1650\\
DENIS-P J205754.1-025229   & 64 &    64.4710$\pm$0.2365& 68 &   L1.5       & 15       & L2$\beta$& 75       &  045.7 -29.2& 1.5 yng& 1800-1950\\
2MASSI J2104149-103736     & 15 &    58.1658$\pm$0.4051& 68 &   L2.5       & 3        & L2.0    & 24        &  038.5 -34.2& 2.5   & 1800-1950\\
2MASS J21373742+0808463    & 12 &    66.0620$\pm$0.8664& 68 &   L5:        & 12       & L5.0    & 24        &  062.8 -31.3& 5     & 1500-1650\\
2MASS J21580457-1550098    & 65 &    {\it 52}          & 74 &   L4:        & 3        & L5.0    & 24        &  039.5 -48.3& 4.5   & 1500-1650\\
2MASSW J2224438-015852     & 5  &    86.6169$\pm$0.7080& 68 &   L4.5       & 5        & L4.5 pec (red) & 66 &  062.2 -46.8& 4.5   & 1500-1650\\
2MASS J22551861-5713056A   & 11 &    58.8576$\pm$0.5866& 68 &   \nodata    & \nodata  & L5.5:   & 67        &  329.2 -53.5& 5.5   & 1500-1650\\
2MASS J23174712-4838501    & 12 &    50.0212$\pm$1.2656& 68 &   L4 pec     & 33       & L6.5 pec (red)& 33  &  336.7 -61.8& 5     & 1500-1650\\
2MASS J23224684-3133231    & 12 &    50.3213$\pm$0.5576& 68 &   L0$\beta$  & 12       & L2$\beta$& 18,75    &  014.6 -70.2& 0 yng & 1950-2100\\
\enddata
\onecolumngrid
\tablenotetext{a}{Values in italics are spectrophotometric distance estimates, derived as explained in the text.}
\tablenotetext{b}{This object is a possible common proper motion companion to LP 903-20 (LHS 5166).}
\tablecomments{References to discovery and spectral classification papers: (1) \citealt{goldman1999}, (2) \citealt{dupuy2012}, (3) \citealt{kirkpatrick2008}, (4) \citealt{marocco2013}, (5) \citealt{kirkpatrick2000}, (6) \citealt{bardalez2014}, (7) \citealt{reid2000}, (8) \citealt{knapp2004}, (9) \citealt{wilson2003}, (10) \citealt{cruz2009}, (11) \citealt{kendall2007}, (12) \citealt{reid2008}, (13) \citealt{liebert2003}, (14) \citealt{schneider2014}, (15) \citealt{cruz2003}, (16) \citealt{golimowski2004}, (17) \citealt{reid2006}, (18) \citealt{allers2013}, (19) \citealt{castro2013}, (20) \citealt{cruz2007}, (21) \citealt{fan2000}, (22) \citealt{geballe2002}, (23) \citealt{salim2003}, (24) \citealt{burgasser2010b}, (25) \citealt{thorstensen2003}, (26) \citealt{kirkpatrick2014}, (27) \citealt{reid2001}, (28) \citealt{bouy2004}, (29) \citealt{phan-bao2008}, (30) \citealt{scholz2002}, (31) \citealt{deacon2005}, (32) \citealt{burgasser2008b}, (33) \citealt{kirkpatrick2010}, (34) \citealt{schmidt2010}, (35) \citealt{gizis2002}, (36) \citealt{hawley2002}, (37) \citealt{kendall2004}, (38) \citealt{delfosse1997}, (39) \citealt{kirkpatrick1999}, (40) \citealt{bouy2003}, (41) \citealt{geissler2011}, (42) \citealt{gizis2000}, (43) \citealt{folkes2007}, (44) \citealt{ruiz1997}, (45) \citealt{liu2005}, (46) \citealt{koen2017}, (47) \citealt{gomes2013}, (48) \citealt{luhman2014-solar_comp}, (49) \citealt{kirkpatrick2016}, (50) \citealt{schmidt2007}, (51) \citealt{forveille2004}, (52) \citealt{potter2002}, (53) \citealt{goto2002}, (54) \citealt{burgasser2007c}, (55) \citealt{west2008}, (56) \citealt{burgasser2004}, (57) \citealt{beamin2013}, (58) \citealt{schneider2017}, (59) \citealt{metodieva2015}, (60) \citealt{folkes2012}, (61) \citealt{looper2008}, (62) \citealt{gizis2011}, (63) \citealt{liu2002}, (64) \citealt{menard2002}, (65) \citealt{gizis2003}, (66) \citealt{cushing2005}, (67) \citealt{reid2008b}, (68) Gaia Data Release 2: \citealt{gaia2016} and \citealt{gaia2018}, (69) \citealt{bartlett2017}, (70) \citealt{dahn2002}, (71) \citealt{dieterich2014}, (72) \citealt{dahn2017}, (73) Hipparcos: \citealt{vanleeuwen2007}, (74) This paper, (75) \citealt{faherty2016}, (76) \citealt{gagne2015b}, (77) \citealt{scholz2018}, (78) \citealt{cushing2018}.}
\end{deluxetable*}

\twocolumngrid

The adopted spectral types were determined as follows. For field objects, we averaged the optical and near-infrared types, but rounded any decimal type from \cite{bardalez2014} to its nearest half-type beforehand. For those averages landing at a quarter or three-quarters type, we chose the closest half type that fell in the direction of the optical value. For young objects, the optical type is used as the adopted type, unless only the near-infrared type is noted as low-gravity, in which case the averaging used above for the field objects is employed.

As was done with the late-T and Y dwarfs in the last section, we need to place these objects on a $T_{\rm eff}$ scale. \cite{faherty2016} provide a mapping from spectral type to $T_{\rm eff}$ for the L dwarfs. They find that this mapping depends upon the age of the L dwarf. In the early- to mid-L dwarf range, young (low-gravity), higher mass L dwarfs have slightly colder $T_{\rm eff}$ values compared to old field L dwarfs of the same spectral subclass (see their Figure 33). We note that the \cite{faherty2016} spectral type relation (which is the same one determined in \citealt{filippazzo2015}) for field L dwarfs gives a $\sim$150K range for each full spectral class at early L, but this  shrinks to $\sim$100K per full spectral class by mid-L. As shown in Table~\ref{earlyL_20pc_analysis}, we thus map field spectral types into $T_{\rm eff}$ as follows: L1-L1.5 is mapped to the 1950-2100K bin, L2-L2.5 to 1800-1950K, L3-L3.5-L4 to 1650-1800K, and L4.5-L5-L5.5 to 1500-1650K.

The adopted types of the young, low-gravity L dwarfs are mapped to $T_{\rm eff}$ using the "YNG" relations of \cite{faherty2016}\footnote{Note that in Table 19 of  \cite{faherty2016}, the coefficients of the $T_{\rm eff}$ YNG relation are accidentally listed in reverse order. Coefficient $c_0$ is actually $c_4$, $c_1$ is actually $c_3$, etc.} Specifically, the mapping into our 150K bins is L0-L0.5-L1 for 1950-2100K, L1.5-L2-L2.5 for 1800-1950K, L3-L3.5 for 1650-1800K, and L4 for 1500-1650K. Note that the young L dwarfs of type L4.5 and later from Table~\ref{earlyL_20pc_census} fall in a cooler bin for which later-type field L dwarfs ($\ge$L6) are expected to have poorer coverage out to 20 pc by {\it Gaia}. Likewise, old field objects in Table~\ref{earlyL_20pc_census} that are earlier in type than L0 share a temperature bin containing low-gravity late-M dwarfs, which we have not tabulated because DwarfArchives does not track objects with M dwarf types. Hence, not all objects in Table~\ref{earlyL_20pc_census} were used in the space density arguments that follow.

In Table~\ref{earlyL_20pc_analysis} we list the aforementioned mapping of types into each 150K bin along with the number of objects found in each range. Because these objects were drawn from the literature using ground-based parallaxes and spectrophotometric distance estimates, we can use them to check the completeness of the {\it Gaia} DR2 sample over our spectral type and distance range. Overall we find that DR2 is 90\% complete for early- to mid-L dwarfs within 20 pc, with the completeness dropping (slightly) at later types as expected, since {\it Gaia} has reduced sensitivity to those colder objects.

\begin{figure*}
\figurenum{16}
\includegraphics[scale=0.85,angle=0]{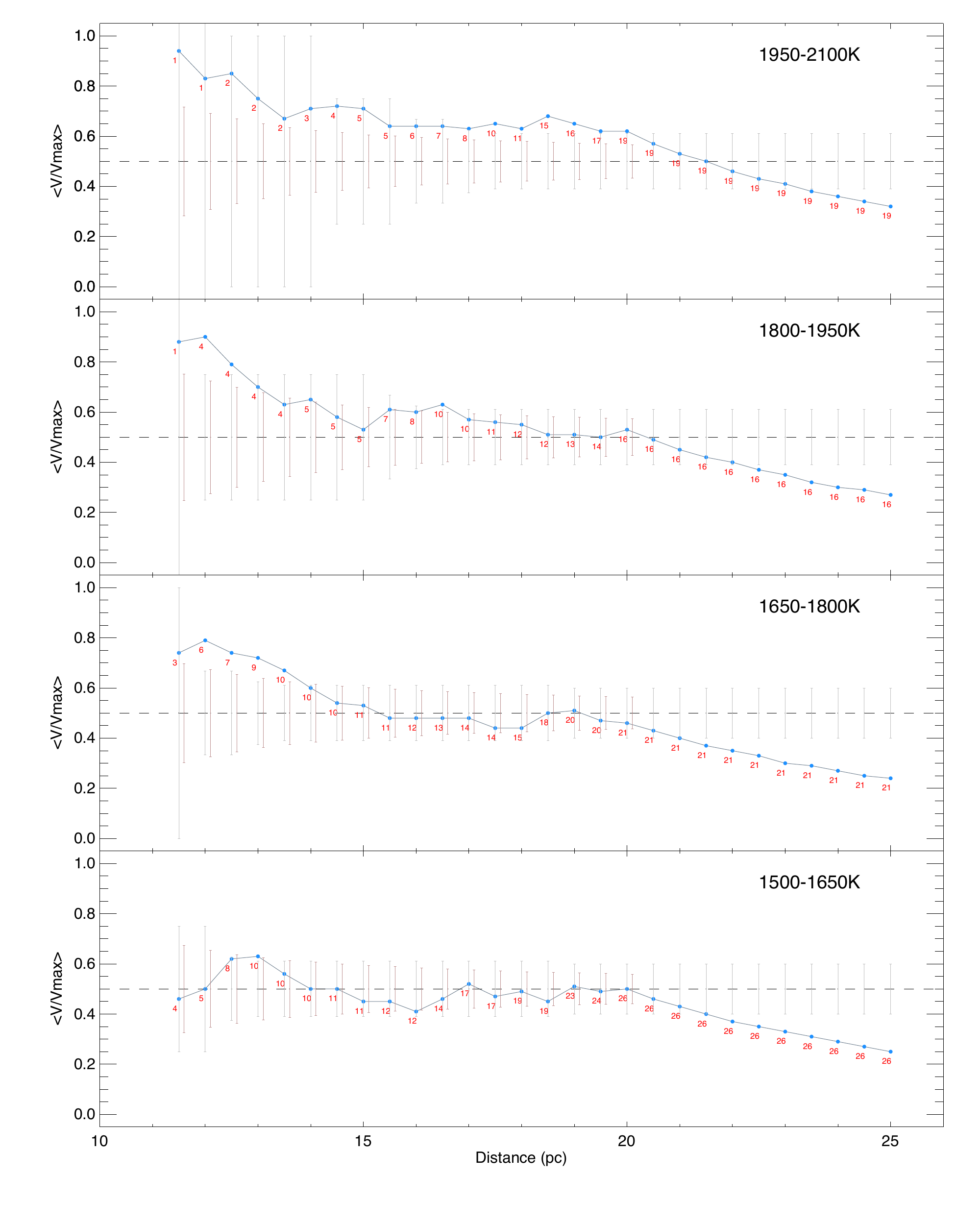}
\caption{The average $V/V_{\rm max}$ value in 0.5-pc intervals for the four 150-K bins encompassing our 20-pc early- to mid-L sample. See the caption to Figure~\ref{vmax_plot} for more details. (Data in support of this figure can be found in Table~\ref{vmax_table}.)
\label{vmax_plot2}}
\end{figure*}

Figure~\ref{vmax_plot2} shows the results of our $\langle{V}/{V_{max}}\rangle$ test on this L dwarf sample. In all four temperature bins, we are within 1$\sigma$ of the expected $\langle{V}/{V_{max}}\rangle$ value of 0.5 at $d$ = 20.0 pc, giving space densities of 0.57$\times$10$^{-3}$ pc$^{-3}$ (19 objects) for the 1950-2100K bin, 0.48$\times$10$^{-3}$ pc$^{-3}$ (16 objects) for the 1800-1950K bin, 0.63$\times$10$^{-3}$ pc$^{-3}$ (21 objects) for the 1650-1800K bin, and 0.78$\times$10$^{-3}$ pc$^{-3}$ (26 objects) for the 1500-1650K bin. The values of $\langle{V}/{V_{max}}\rangle$ are still within the 1$\sigma$ envelope at 23.0, 22.0, 21.0, and 21.0 pc for the four bins, giving densities of 0.37$\times$10$^{-3}$ pc$^{-3}$, 0.36$\times$10$^{-3}$ pc$^{-3}$, 0.54$\times$10$^{-3}$ pc$^{-3}$, and 0.67$\times$10$^{-3}$ pc$^{-3}$. We use the differences between these latter densities and the ones at our 20-pc limit as the 1$\sigma$ uncertainties in our measurements. These values are listed in Table~\ref{space_densities} along with the ones measured earlier for the late-T and Y dwarfs.

\begin{center}
\begin{deluxetable}{ccccc}
\tablenum{14}
\tablecaption{Analysis of the 20-pc Census for 1500 $< T_{\rm eff} <$ 2100K\label{earlyL_20pc_analysis}}
\tablehead{
\colhead{T$_{\rm eff}$ range} & 
\colhead{Type range} & 
\colhead{Type range} & 
\colhead{Total} & 
\colhead{Completeness} \\
\colhead{(K)} & 
\colhead{Field obj.} & 
\colhead{Young obj.} & 
\colhead{Number} &
\colhead{of Gaia DR2\tablenotemark{a}} \\
\colhead{(1)} &                          
\colhead{(2)} &
\colhead{(3)} &
\colhead{(4)} &
\colhead{(5)} \\
}
\startdata
1950-2100& L1, L1.5&       L0, L0.5, L1   &   19& 89\% \\
1800-1950& L2, L2.5&       L1.5, L2, L2.5 &   16& 94\% \\
1650-1800& L3, L3.5, L4&   L3, L3.5       &   21& 100\% \\
1500-1650& L4.5, L5, L5.5& L4             &   26& 81\%\tablenotemark{b} \\
\enddata
\tablenotetext{a}{This is the percentage of objects with parallaxes listed in the Second Data Release from Gaia.}
\tablenotetext{b}{Note that only Gl 779A, not the L dwarf companion Gl 779B, is listed in Gaia DR2.}
\end{deluxetable}
\end{center}

Unlike for the late-T and Y dwarfs, we find that this sample has no obvious incompleteness along the Galactic Plane. Because of the poorer statistics, we divide the celestial sphere into only four equal-area pieces sliced along lines of constant $b$ values of $+30^\circ$, $0^\circ$, and $-30^\circ$. (Galactic coordinates for each object are given in column 9 of Table~\ref{earlyL_20pc_census}.) We find 25, 25, 14, and 16 objects starting with the northernmost piece and ending with the southernmost. This shows that our 20-pc sample of early-L dwarfs lacks an obvious zone of avoidance in the Galactic Plane because 39/80 = 49\% of the sources are concentrated in the 50\% of the sky closest to the Plane itself. We can thus assume that the 20-pc sample of 1500-2100K dwarfs is reasonably complete both sky-wide and to the full 20-pc distance limit. Therefore, no adjustment factor (see Table~\ref{space_densities}) to the measured space densities is needed.

\subsubsection{A Uniform Distribution of the Nearest Brown Dwarfs}

\cite{bihain2016} have posited that the distribution of nearby brown dwarfs on the sky is astonishingly non-uniform. Those authors claim that most of the brown dwarfs with the 6.5-pc volume (21 out of 26 objects) are leading the Sun in its Galactic orbit, with only a handful (the remaining 5) trailing the Sun. They further state that the probability of such a distribution occurring by chance is 0.098\%, although the 2$\sigma$ error bar on this quantity pushes it as high as 11\%. They conjecture that the literature sample of brown dwarfs within 6.5 pc is incomplete due to biases in the methodology being employed to search the all-sky 2MASS and {\it WISE} data sets, although the ways in which the search criteria might have created such an odd bias are not clearly specified. Alternatively, they suggest that if the sample is complete, or nearly so, it is either caused by some "random inhomogeneity" or an unknown dynamical mechanism causing brown dwarfs to segregate from stars in this manner, as the stellar sample (136 objects within 6.5 pc) shows no such asymmetric distribution.

\begin{figure}
\figurenum{17}
\includegraphics[scale=0.55,angle=0]{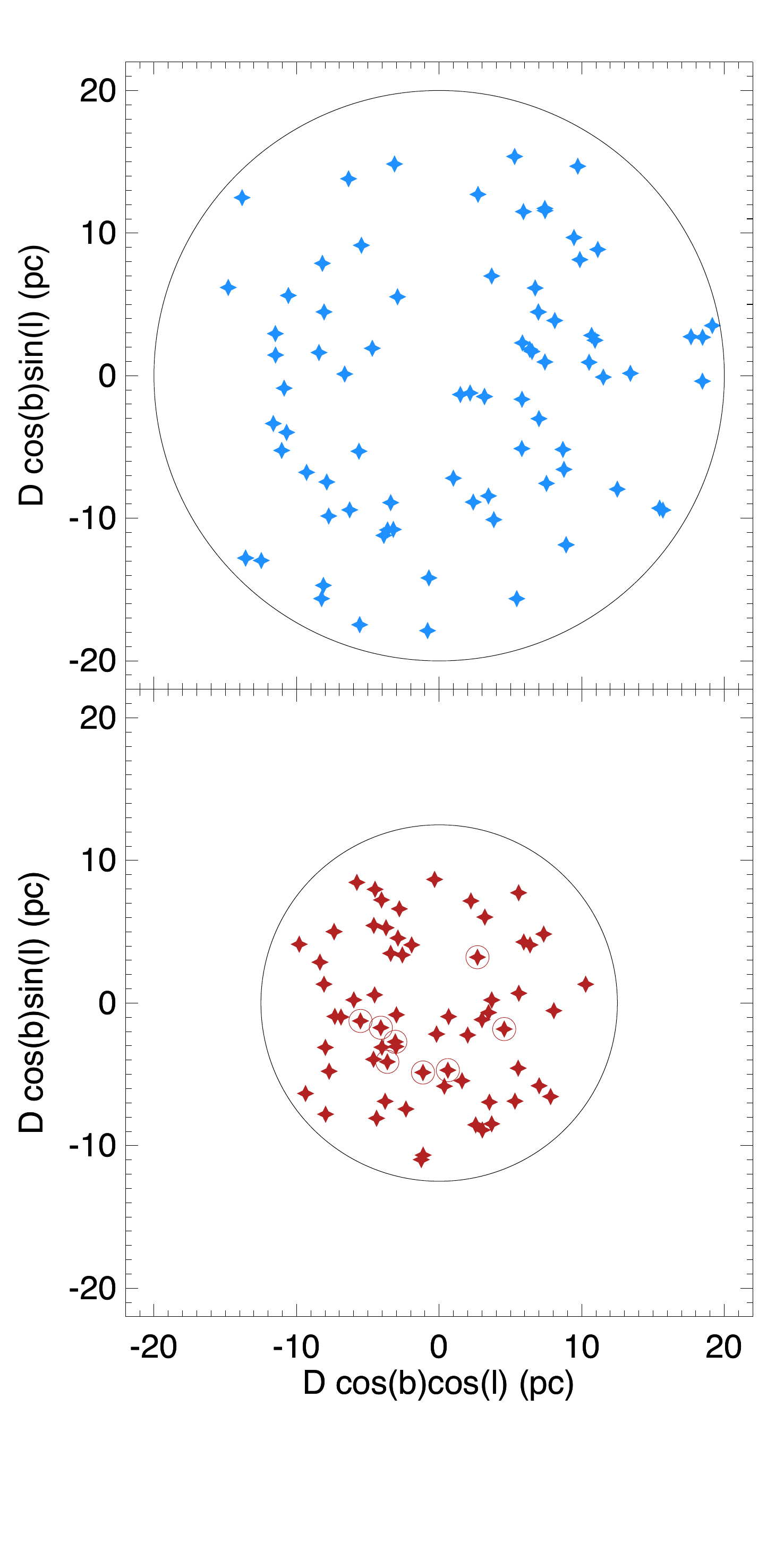}
\caption{A Galactic projection plot identical to that in Figure 1 of \cite{bihain2016} showing the distributions of our two samples. The upper panel shows the 1500-2100K sample (82 objects) out to our completeness limit of 20 pc, and the lower panel shows the 450-900K sample (69 objects) out to our completeness limit of 12.5 pc. Objects within 6.5 pc of the Sun are circled individually; no such objects appear in the upper panel.
\label{sky_distribution}}
\end{figure}

We show here, using a sample $\sim$6 times larger, that nearby brown dwarfs are distributed uniformly and that the \cite{bihain2016} observation is solely the result of the statistics of small numbers. To demonstrate this, we employ our sample of 1500-2100K and 450-900K objects used in the space density measurements of Table~\ref{space_densities} . The warmer group of objects is comprised of early- to mid-L dwarfs, which includes some low-mass stars as well as brown dwarfs, and the cooler group is comprised of late-T to Y dwarfs, all of which are brown dwarfs. We note also that we have not added any new objects to these samples that weren't already in the 6.5-pc accounting of \cite{bihain2016}, with the exception of considering 2MASS 0939$-$2448 to be an unresolved double (see Section~\ref{notes_on_individual_objects}). Figure~\ref{sky_distribution} shows the sky distribution of these objects on a Galactic projection identical to that used in Figure 1 of \cite{bihain2016}. For the 1500-2100K objects, we use the entire sample out to 20 pc since we believe that to be relatively complete, and for the 450-900K objects we use only the subsample out to 12.5 pc since the coolest sub-bin (450-600K) is complete only out to that distance. 

Both sets of objects are distributed so that roughly equal numbers are seen in the upper and lower halves of the diagram. We circle objects falling within the same 6.5-pc distance used by \cite{bihain2016} to show that these objects appear, as those authors noted, primarily in the lower half of the diagram. (Note that {\it none} of the objects in the 1500-2100K range falls within 6.5 pc.) However, this is a result of random statistics on a small sample, as the larger sample shows no such effect. It is very difficult to imagine a selection effect or a dynamical process at work within the 6.5-pc distance that would not also be operating in the distance range from 6.5 to 12.5 pc.

\subsubsection{Best Fits to the Space Densities\label{space_density_fits}}

We can now test which of the 72 models in Figures~\ref{histograms_teff_baraffe2003} and \ref{histograms_teff_saumon2008} best match our measured space densities. We use the IDL routine {\tt mpfit} (\citealt{markwardt2009}) to perform a weighted least-squares fit between the data and the model where the only adjustable parameter is the scaling between the arbitrary number counts in the models and our measured space densities. For the calculation, we use only the seven values in Table~\ref{space_densities} with measured uncertainties (four in the 1500-2100K range and three in the 450-900K range) and ignore the ranges in which we have only lower limits (900-1050K and 300-450K) and the range in which we have only one data point and no measurable space density (150-300K).

\begin{figure*}
\figurenum{18}
\includegraphics[scale=0.70,angle=0]{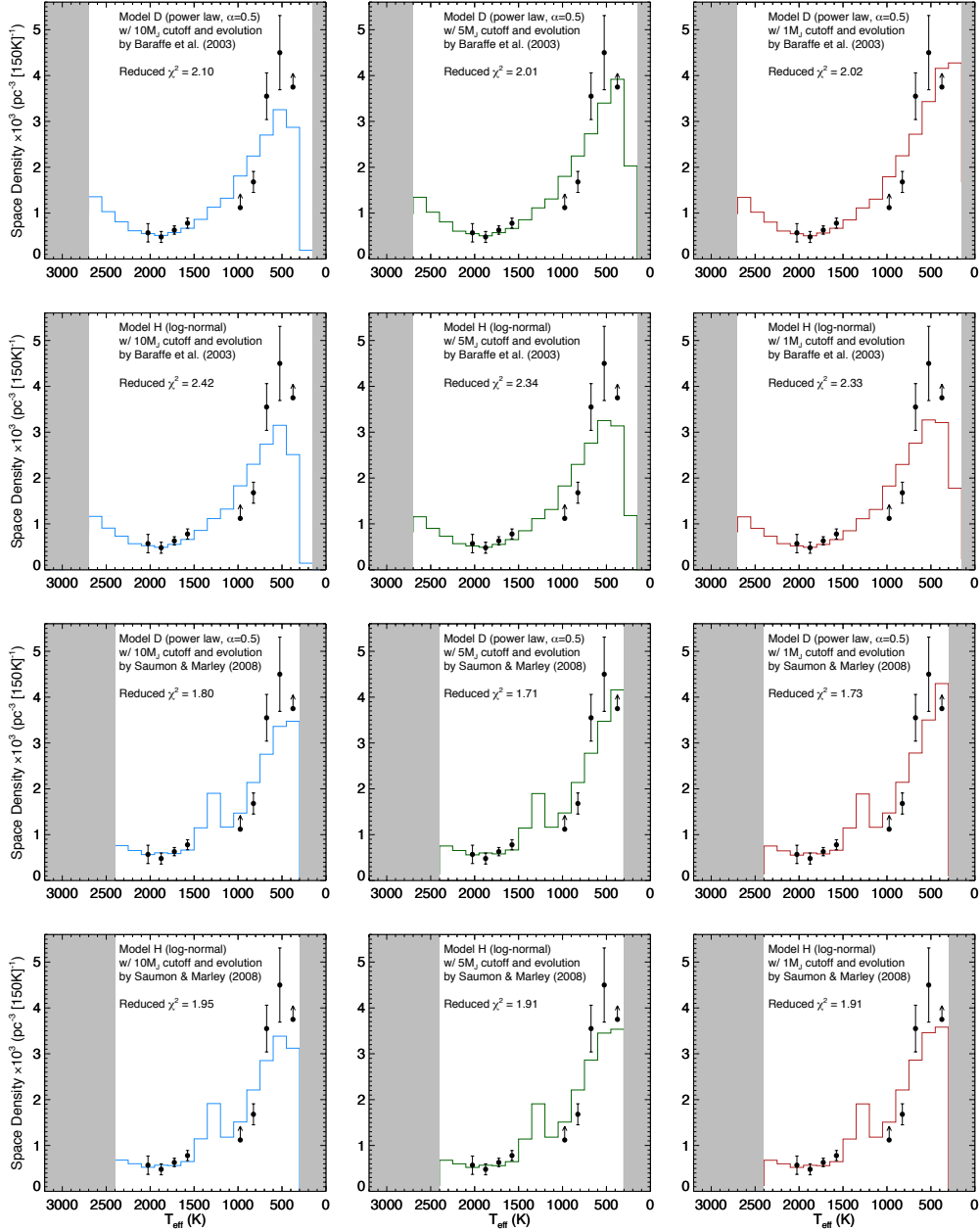}
\caption{The best fits between the simulations of Figures~\ref{histograms_teff_baraffe2003} and \ref{histograms_teff_saumon2008} and our measured space densities in Table~\ref{space_densities}. The top two rows show the two mass functions using the \cite{baraffe2003} evolutionary code that provide the smalled reduced $\chi^2$ values: the power law with $\alpha = 0.5$ (top row; Model D) and the single-object log-normal from \cite{chabrier2003} (second row; Model G). Similarly, the bottom two rows show the two mass functions using the \cite{saumon2008} evolutionary code that provide the smallest reduced $\chi^2$ values: the power law with $\alpha = 0.5$ (third row; Model D) and the single-object log-normal from \cite{chabrier2003} (bottom row; Model H). In each row the mass functions are shown with assumed low-mass cutoffs of 10$M_{Jup}$ (blue, left panels), 5$M_{Jup}$ (dark green, middle panels), and 1$M_{Jup}$ (red, right panels). Our measured spaced densities and their uncertainties are show in black on each panel.
\label{space_density_best_fits}}
\end{figure*}

The best fit to each model produces a reduced $\chi^2$ value, and we illustrate in Figure~\ref{space_density_best_fits} the fits for which this value is minimized. For mass functions paired with the evolutionary code of \cite{baraffe2003}, the best fits are given by the power law mass function with $\alpha = 0.5$ (model D). The single-object log-normal mass function of \cite{chabrier2001} (model H) is the next best model, although it consistently underpredicts the space densities seen in the 300-750K range. If we consider the lower limit to the space density in the 300-450K bin, we find that for the power-law model there is a preference for the lower-mass cutoffs, with the current data favoring a cutoff no higher than $\sim$5$M_{Jup}$. The other power law mass functions produced poorer reduced $\chi^2$ values of $\sim$7.8, $\sim$5.2, $\sim$3.0, $\sim$3.0, and $\sim$6.2 for $\alpha$ values of $-$1.0, $-$0.5, 0.0, +1.0, and +1.5 (models A, B, C, E, and F), respectively. The other log-normal mass functions give reduced $\chi^2$ values of $\sim$3.8 and $\sim$3.2 for models G and I, respectively. The bi-partite power law mass functions of \cite{kroupa2013} give much larger reduced $\chi^2$ values of $\sim$7.2, $\sim$4.1, and $\sim$10.2 for models J, K, and L, respectively.

For mass functions paired with the evolutionary code of \cite{saumon2008}, we find even better fits. As with the \cite{baraffe2003} models, the best fits are given by the power law mass function with $\alpha = 0.5$ (model D), with the single-object log-normal mass function of \cite{chabrier2001} (model H) being only slightly worse. The log-normal function again underpredicts the space densities in the 300-750K range. The $\alpha = 0.5$ power law suggests that the cutoff mass is likely no higher than $\sim$5$M_{Jup}$ given our measured lower limit to the space density in the 300-450K bin. The other power law mass functions produced poorer reduced $\chi^2$ values of $\sim$8.7, $\sim$5.7, $\sim$3.1, $\sim$2.4, and $\sim$5.6 for $\alpha$ values of $-$1.0, $-$0.5, 0.0, +1.0, and +1.5 (models A, B, C, E, and F), respectively. The other log-normal mass functions give reduced $\chi^2$ values of $\sim$2.0 and $\sim$3.1 for models G and I, respectively, showing that the log-normal function of \cite{chabrier2003} (model G) produces almost as good a match to the observed data as that from \cite{chabrier2001} (model H). The bi-partite power law mass functions of \cite{kroupa2013} give much larger reduced $\chi^2$ values of $\sim$12.0, $\sim$8.8, and $\sim$14.6 for models J, K, and L, respectively.

Three other points should be noted regarding these preliminary results:

\begin{enumerate}

\item As mentioned earlier, there is a clear discriminator present in the 1050-1500K range of the simulations that will determine whether the \cite{baraffe2003} or \cite{saumon2008} evolutionary models are preferred, from the point of view of the mass function. In the six lower panels of Figure~\ref{space_density_best_fits} based on the \cite{saumon2008} models, there is a clear overdensity relative to a simple interpolation between the space densities seen in the 750-900K and 1500-1650K bins that is not present in the six upper panels based on the \cite{baraffe2003} models. Using a literature update by CRG to the known L and T dwarfs archived at DwarfArchives along with spectrophotometric distance estimates to those still lacking parallax determinations, we find strong evidence that this overdensity is, indeed, present. This can be tested with a large, volume-limited sample of objects in this temperature range, which both the ground-based parallax program of \cite{best2018} and the Cycle 14 {\it Spitzer} program 14000 (PI: Kirkpatrick) are currently aiming to do.

\item There is a dramatic difference, slightly over a factor of two, in our measured space densities between the 750-900K bin and the 600-750K bin. The predictions, even in our best fits, are unable to model this effect. Is this effect real or an artifact of our transformation to $T_{\rm eff}$? This rapid rise in space density with falling temperature is also seen in the distribution by spectral type. Our late-T dwarf sample in Table~\ref{teffs_20pc_census} shows 34 T7-T7.5 dwarfs but a surprisingly larger number (64) of T8-T8.5 dwarfs. Performing a $V/V_{max}$ test on these objects shows that the earlier spectral types are complete to 13.5 pc with 18 objects; the later spectral types are complete to 13.5 with 34 objects. The difference in these space densities is almost a factor of two (1.89$\times$), paralleling the effect seen when binning by $T_{\rm eff}$.

We can also examine this effect in other quantities. Figure 14 suggests that between 600 and 900K, $M_H$ runs from $\sim$16 to 18 mag. If we divide the absolute magnitude range into two portions, $16 \le M_H < 17$ mag and $17 \le M_H < 18$ mag, we find 34 objects in the brighter bin and 43 objects in the fainter one. Performing a simple $V/V_{max}$ test finds that the brighter bin is complete to $\sim$15.5 pc with 28 objects; the fainter bin is complete to $\sim$13.0 pc with 20 objects. In this case, the space density of the fainter bin is only 1.21$\times$ higher than that of the brighter bin. We can perform a similar exercise on ch1$-$ch2 color, for which Figure 14 shows that a color span of 1.2 to 2.0 mag corresponds to $T_{\rm eff}$ values of 900K down to 600K. We find 49 T dwarfs with $1.2 \le$ ch1$-$ch2 $< 1.6$ mag and 61 with $1.6 \le$ ch1$-$ch2 $< 2.0$ mag. Performing a simple $V/V_{max}$ test finds that the bluer bin is complete to $\sim$14.5 pc with 25 objects; the redder bin is complete to $\sim$13.5 pc with 35 objects. In this case, the space density of the redder bin is 1.73$\times$ higher than that of the bluer bin.

We conclude that the jump in space density from 750-900K to 600-750K is likely real since it appears in other physical quantities like spectral type and color, although less so with absolute magnitude. The fact that this jump appears not to be well modeled by the predicted mass functions plus evolutionary code may suggest that the Solar Neighborhood has an unusual distribution of T dwarfs relative to the Milky Way as a whole, but this seems at odds with the fact that this peculiarity in the space densities occurs at the point where we have our largest number of objects and hence our best statistical measurement. It is possible, then, that the discrepancy with respect to predictions is real and points to missing physics in the evolutionary models.

\item As the three panels in the top row of Figure~\ref{space_density_best_fits} indicate, improving the space density measurement in the 300-450K bin as well as providing a robust lower limit to the density in the 150-300K bin would better help determine the low-mass cutoff of the mass function as the simulations show vast differences in the predicted numbers in these ranges. Current work to find more of these Y dwarfs, which at these temperatures are typically those classified as Y1 and later, is currently underway using proper motion searches afforded by the long time baseline (7+ years) now available between the classic {\it WISE} and {\it NEOWISE} Reactivation missions. The prospect for discovering more of these objects looks promising based on early results from the Backyard Worlds Citizen Science project (\citealt{kuchner2017}) and CatWISE (\citealt{meisner2018}), but characterizing the spectra of such discoveries and measuring accurate distances will be challenging given the extreme faintness of these objects. (See Section~\ref{conclusions} for more discussion on this point.)

\end{enumerate}

\begin{figure*}
\figurenum{19}
\includegraphics[scale=0.80,angle=0]{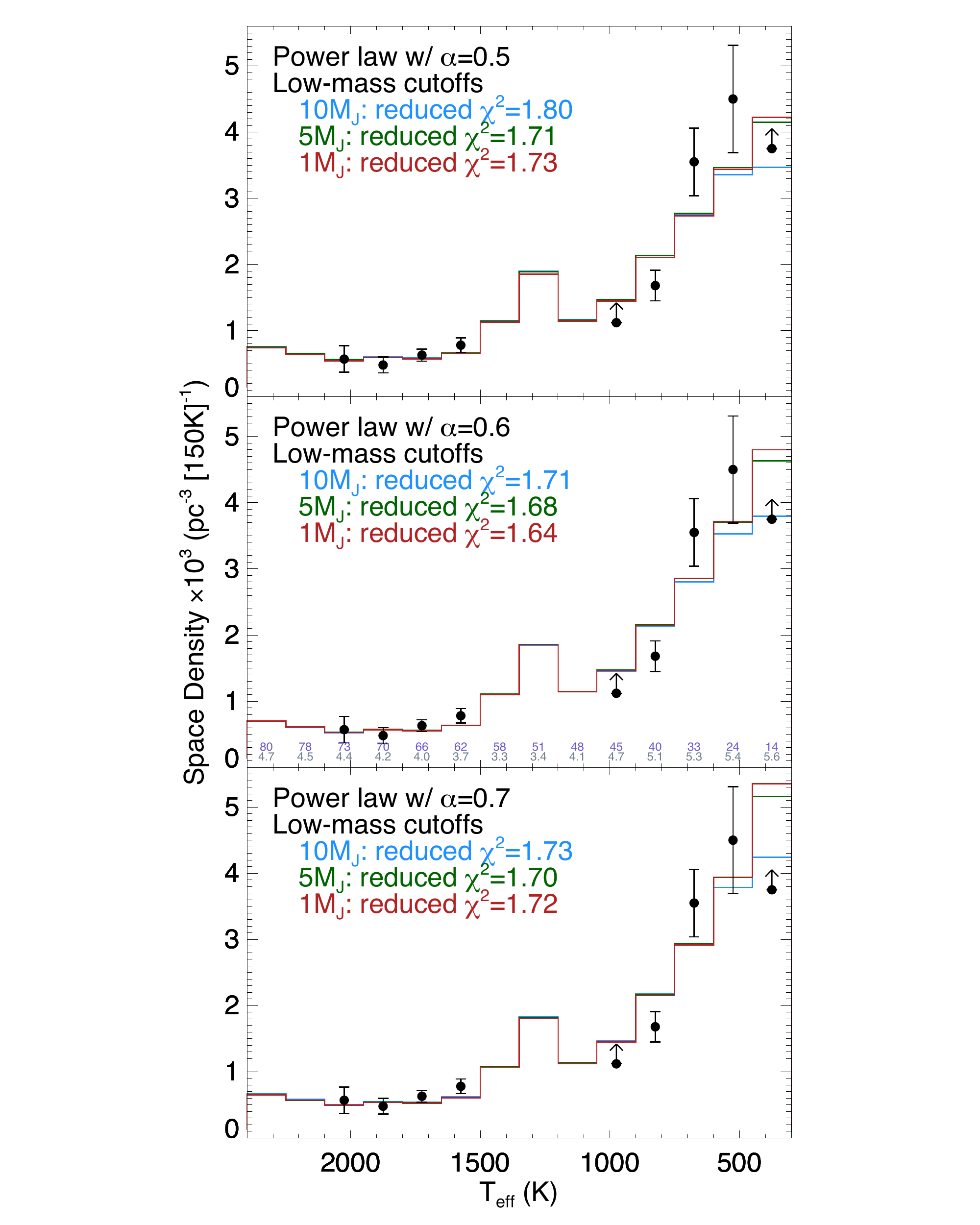}
\caption{Fits of power laws with $\alpha$ values of 0.5 (top panel), 0.6 (middle panel), and 0.7 (bottom panel) to our observational data (black points). These predicted $T_{\rm eff}$ distributions have been passed through the evolutionary models of \cite{saumon2008}. Each panel shows simulations for each of three low-mass cutoffs: 10 $M_{Jup}$ (blue), 5 $M_{Jup}$ (green), and 1 $M_{Jup}$ (red). The minimum reduced $\chi^2$ value is found for the $\alpha=0.6$ model. For the $\alpha=0.6$ model with 5 $M_{Jup}$ cutoff, we have also calculated the mean mass and mean age in each $T_{\rm eff}$ bin. These values are shown in the middle panel; the mean mass is shown in units of $M_{Jup}$ (upper value in purple), and the mean age is shown in units of Gyr (lower value in grey).
\label{alpha_0.6}}
\end{figure*}

Given the success of fitting our observational data to the power law model with $\alpha = 0.5$, we searched for even better fits using a smaller grid spacing (increments of 0.1) for $\alpha$. We find that the $\alpha=0.6$ version using the evolutionary models of \cite{saumon2008} results in the lowest overall reduced $\chi^2$, as illustrated in Figure~\ref{alpha_0.6}. The fits to the 1 and 5 $M_{Jup}$ cutoffs are slightly better than the one for the 10 $M_{Jup}$ cutoff. If we assume, based on the $T_{\rm eff}$ determinations in Table~\ref{teffs_20pc_census}, that our coldest bin on Figure~\ref{alpha_0.6} is reasonably complete to $T_{\rm eff} \approx$ 350K, then the simulations predict a mean mass of 16 $M_{Jup}$ with standard deviation of 5 $M_{Jup}$. (See the figure for the mean mass and age in each $T_{\rm eff}$ bin.) It should also be stated again that the next coldest bin -- the 150-300K bin just off the right side of the plot -- has only one known object, WISE 0855$-$0714, whose mass has been estimated at 3-10 $M_{Jup}$ (\citealt{luhman2014-0855}). Thus, our observations are already beginning to measure the field mass function into the regime below the deuterium-burning limit of $\sim$13$ M_{Jup}$. 

\section{Discussion\label{discussion}}

The fitting of our measured space densities to the suite of simulated $T_{\rm eff}$ distributions in Section~\ref{space_density_fits} was focused solely on the substellar regime. Here we place these results in context with efforts that have attempted to decribe the mass function across the entirety of star formation's mass spectrum, from $\sim$100$M_\odot$ to $\sim$1$M_{Jup}$.

Specifically, early work by \cite{salpeter1955} suggested that the stellar mass function at very high masses could be adequately described by a power law with exponent $\alpha = 2.35$. \cite{miller1979} found this same power law for $M > 10 M_\odot$, but at lower masses favored a two-piece power law with exponents of $\alpha = 1.5$ for $1 M_\odot < M < 10 M_\odot$ and $\alpha = 0.4$ for $0.1 M_\odot < M < 1 M_\odot$. \cite{kroupa2001} prefers a three-piece power law in the stellar regime having segments with $\alpha \approx 2.3$ for $M > 1.0 M_\odot$, $\alpha \approx 2.7$ for $0.5 M_\odot < M < 1.0 M_\odot$, and $\alpha \approx 1.8$ for $0.08 M_\odot < M < 0.5 M_\odot$, once the hidden members of unresolved binaries are properly accounted for. \cite{chabrier2003-review} has suggested that over the entire stellar mass range ($\sim$100$M_\odot$ to $\sim$0.1$M_\odot$), the mass function can be adequately described by either a log-normal or a two-segment power law.

With the discovery of brown dwarfs in both the field and young clusters, researchers have attempted to describe the mass function into the substellar regime. \cite{kroupa2013} suggest that brown dwarfs form in a fundamentally different way from stars and that the mass function of brown dwarfs is a power law with $\alpha \approx 0.3$. However, they state that there is significant overlap in mass between the normal ``stellar'' portion (a power law with $\alpha \approx 1.3$) of the mass function and the ``substellar'' part but that the two pieces are discontinuous. \cite{chabrier2001}, on the other hand, prefers a log-normal form, with values\footnote{See the footnote in Section~\ref{functional_forms} regarding our translation of these $\mu$ and $\sigma$ values into a natural logarithm form.} of $\mu=ln(0.10)$ and $\sigma=0.627ln(10)$. These values were later revised to $\mu=ln(0.079)$ and $\sigma=0.69ln(10)$ by \cite{chabrier2003} and later, as stated by \cite{chabrier2014}, to $\mu=ln(0.25)$ and $\sigma=0.55ln(10)$ by \cite{chabrier2005}.

We find, however, that neither the Kroupa bi-partite power law nor the Chabrier log-normal forms adequately fit the data even when we allow for the vertical axis scaling (the space density) to be adjusted as a free parameter. Specifically, the Chabrier log-normal forms underpredict the number of objects below 650K. Indeed, as \cite{chabrier2014} note, their preferred values of $\mu$ and $\sigma$ from \cite{chabrier2005} behave like a power law with a negative $\alpha$ value below $T_{\rm eff} \approx$ 1300K. Using the two $\alpha$ values and their break points in mass as suggested by \cite{kroupa2013}, the bi-partite power laws fare even worse; however, these still hold promise, as the individual mass break points and $\alpha$ values can be fine tuned to provide better fits. This should be possible since our best-fit value of $\alpha \approx 0.6$ seems to describe the 350-2100K portion of the $T_{\rm eff}$ distribution very well. As predicted by \cite{hoffmann2018}, the difference between a log-normal and power-law representation of the mass function at the smallest masses provides a powerful discriminant in distinguishing between formation theories.

Based on a theoretical perspective, \cite{hennebelle2008} argue that the full stellar + substellar mass function should be described as a power law at highest masses and a single log-normal form at lower masses. The same conclusion is reached via a somewhat different formalism by \cite{hopkins2012}, although more recent theoretical considerations by \cite{guszejnov2018} have raised questions about how both of these methods treat formation at the smallest masses. For these lowest mass objects, the log-normal form implies a single ``characteristic'' mass governing their formation. \cite{chabrier2003} argues that using multi-part power laws instead of a log-normal form to describe the mass function necessarily implies that multiple characteristic masses determine the processes of formation for low-mass objects. Multiple (or at least two) characteristic masses may, in fact, be necessary to describe the observational results for low-mass stars and brown dwarfs, as \cite{kroupa2013} argue. A theoretical formalism for this alternate scenario has been described by \cite{thies2010}.

Finally, we caution that the {\it field} mass function is the result of formation processes that have occurred in different environments over the last $\sim$10 Gyr. If the physics governing star formation in one environment is found to differ substantially from that in another, disentangling the physics from the resulting mixture may be extremely difficult. For example, the formation of brown dwarfs could be fundamentally different in a low-mass moving group such as the TW Hydrae Association than it is from that operating in a high-mass region such as the Orion Nebula Cluster, where oblation from O star winds and the higher potential for gravitational disruption of forming pre-stellar cores may play critical roles. (If so, such effects are likely to be seen only below $\sim$30$M_{Jup}$ because \cite{andersen2008} do not see strong evidence for environment-specific effects at higher masses.) The field mass function may, therefore, have utility only in predicting the outcome of galaxy-wide star formation, although it may eventually be used in determining the relative roles that each environmentally dependent process plays in the overall picture.

\startlongtable
\begin{deluxetable*}{rrrrrrrrrr}
\tabletypesize{\footnotesize}
\tablenum{15}
\tablecaption{Data in Support of Figures~\ref{vmax_plot} and \ref{vmax_plot2}\label{vmax_table}}
\tablehead{
\colhead{Dist.} & 
\multicolumn{9}{c}{Number($\langle{V}/{V_{max}}\rangle$)\tablenotemark{a}} \\
\colhead{(pc)} & 
\colhead{1950-} & 
\colhead{1800-} &
\colhead{1650-} &
\colhead{1500-} &
\colhead{900-} &
\colhead{750-} &
\colhead{600-} &
\colhead{450-} &
\colhead{300-} \\
\colhead{} & 
\colhead{2100K}&
\colhead{1950K} & 
\colhead{1800K} &
\colhead{1650K} &
\colhead{1500K} &
\colhead{900K} &
\colhead{750K} &
\colhead{600K} &
\colhead{450K} \\
\colhead{(1)} &                          
\colhead{(2)} &
\colhead{(3)} &
\colhead{(4)} &
\colhead{(5)} &                          
\colhead{(6)} &
\colhead{(7)} &
\colhead{(8)} &
\colhead{(9)} &                          
\colhead{(10)}  
}
\startdata
  4.0&    \nodata&   \nodata&   \nodata&      \nodata&     2(0.82)&    \nodata&    \nodata&    \nodata&    \nodata\\
  4.5&    \nodata&   \nodata&   \nodata&      \nodata&     2(0.57)&    \nodata&    \nodata&    1(0.77)&    \nodata\\
  5.0&    \nodata&   \nodata&   \nodata&      \nodata&     3(0.60)&    \nodata&    \nodata& 2(0.69)&    1(0.84)\\
  5.5&    \nodata&   \nodata&   \nodata&      \nodata&     4(0.54)&    \nodata&    2(0.91)&    2(0.52)&    1(0.63)\\
  6.0&    \nodata&   \nodata&   \nodata&      \nodata&     4(0.42)&    3(0.92)&    4(0.77)&    2(0.40)&    3(0.78)\\
  6.5&    \nodata&   \nodata&   \nodata&      \nodata&     6(0.52)&    3(0.72)&    4(0.60)&    2(0.31)&    6(0.77)\\
  7.0&    \nodata&   \nodata&   \nodata&      \nodata&     7(0.48)&    3(0.58)&    5(0.57)&    3(0.45)&    6(0.61)\\
  7.5&    \nodata&   \nodata&   \nodata&      \nodata&     7(0.39)&    3(0.47)&    6(0.53)&    5(0.60)&    6(0.50)\\
  8.0&    \nodata&   \nodata&   \nodata&      \nodata&     7(0.32)&    4(0.50)&    7(0.52)&    6(0.57)&    7(0.48)\\
  8.5&    \nodata&   \nodata&   \nodata&      \nodata&     7(0.27)&    6(0.59)&    8(0.49)&    7(0.53)&    8(0.45)\\
  9.0&    \nodata&   \nodata&   \nodata&      \nodata&     7(0.23)&    7(0.56)&    8(0.41)&    8(0.51)&    8(0.38)\\
  9.5&    \nodata&   \nodata&   \nodata&      \nodata&     7(0.19)&    7(0.48)&   11(0.50)&   12(0.59)&    8(0.33)\\
 10.0&    \nodata&   \nodata&   \nodata&      \nodata&     7(0.16)&    7(0.41)&   12(0.47)&   14(0.56)&   10(0.42)\\
 10.5&    \nodata&   \nodata&   \nodata&      \nodata&     7(0.14)&    8(0.43)&   12(0.40)&   22(0.64)&   10(0.36)\\
 11.0&    \nodata&   \nodata&   \nodata&      \nodata&    10(0.37)&   11(0.52)&   14(0.43)&   26(0.61)&   10(0.32)\\
 11.5&    1(0.94)&   1(0.88)&   3(0.74)&      4(0.46)&    10(0.32)&   11(0.46)&   18(0.50)&   27(0.55)&   10(0.28)\\
 12.0&    1(0.83)&   4(0.90)&   6(0.79)&      5(0.50)&    10(0.29)&   11(0.40)&   24(0.57)&   28(0.50)&   10(0.24)\\
 12.5&    2(0.85)&   4(0.79)&   7(0.74)&      8(0.62)&    10(0.25)&   12(0.40)&   25(0.52)&   32(0.50)&   10(0.22)\\
 13.0&    2(0.75)&   4(0.70)&   9(0.72)&     10(0.63)&    10(0.22)&   15(0.47)&   31(0.56)&   33(0.46)&   10(0.19)\\
 13.5&    2(0.67)&   4(0.63)&  10(0.67)&     10(0.56)&    11(0.27)&   18(0.51)&   34(0.54)&   35(0.44)&   11(0.24)\\
 14.0&    3(0.71)&   5(0.65)&  10(0.60)&     10(0.50)&    12(0.30)&   19(0.49)&   36(0.51)&   38(0.44)&   11(0.22)\\
 14.5&    4(0.72)&   5(0.58)&  10(0.54)&     11(0.50)&    16(0.44)&   20(0.46)&   38(0.48)&   39(0.41)&   11(0.19)\\
 15.0&    5(0.71)&   5(0.53)&  11(0.53)&     11(0.45)&    16(0.40)&   22(0.47)&   43(0.50)&   40(0.39)&   11(0.18)\\
 15.5&    5(0.64)&   7(0.61)&  11(0.48)&     12(0.45)&    16(0.36)&   24(0.47)&   48(0.51)&   42(0.38)&   12(0.22)\\
 16.0&    6(0.64)&   8(0.60)&  12(0.48)&     12(0.41)&    18(0.39)&   24(0.43)&   53(0.51)&   43(0.36)&   12(0.20)\\
 16.5&    7(0.64)&  10(0.63)&  13(0.48)&     14(0.46)&    19(0.39)&   27(0.45)&   54(0.47)&   43(0.33)&   12(0.18)\\
 17.0&    8(0.63)&  10(0.57)&  14(0.48)&     17(0.52)&    20(0.39)&   30(0.47)&   54(0.43)&   43(0.30)&   12(0.17)\\
 17.5&   10(0.65)&  11(0.56)&  14(0.44)&     17(0.47)&    22(0.41)&   31(0.44)&   58(0.43)&   45(0.31)&   12(0.15)\\
 18.0&   11(0.63)&  12(0.55)&  15(0.44)&     19(0.49)&    22(0.37)&   31(0.41)&   61(0.43)&   47(0.31)&   12(0.14)\\
 18.5&   15(0.68)&  12(0.51)&  18(0.50)&     19(0.45)&    24(0.40)&   34(0.43)&   61(0.39)&   47(0.29)&   12(0.13)\\
 19.0&   16(0.65)&  13(0.51)&  20(0.51)&     23(0.51)&    28(0.45)&   35(0.41)&   63(0.38)&   47(0.26)&   12(0.12)\\
 19.5&   17(0.62)&  14(0.50)&  20(0.47)&     24(0.49)&    28(0.42)&   36(0.40)&   64(0.36)&   49(0.27)&   13(0.18)\\
 20.0&   19(0.62)&  16(0.53)&  21(0.46)&     26(0.50)&    28(0.39)&   38(0.40)&   67(0.36)&   49(0.25)&   13(0.16)\\
 20.5&   19(0.57)&  16(0.49)&  21(0.43)&     26(0.46)&    28(0.36)&   38(0.37)&   67(0.34)&   49(0.23)&   13(0.15)\\
 21.0&   19(0.53)&  16(0.45)&  21(0.40)&     26(0.43)&    28(0.34)&   38(0.34)&   67(0.31)&   49(0.22)&   13(0.14)\\
 21.5&   19(0.50)&  16(0.42)&  21(0.37)&     26(0.40)&    28(0.31)&   38(0.32)&   67(0.29)&   49(0.20)&   13(0.13)\\
 22.0&   19(0.46)&  16(0.40)&  21(0.35)&     26(0.37)&    28(0.29)&   38(0.30)&   67(0.27)&   49(0.19)&   13(0.12)\\
 22.5&   19(0.43)&  16(0.37)&  21(0.33)&     26(0.35)&    28(0.27)&   38(0.28)&   67(0.26)&   49(0.18)&   13(0.11)\\
 23.0&   19(0.41)&  16(0.35)&  21(0.30)&     26(0.33)&    28(0.26)&   38(0.26)&   67(0.24)&   49(0.17)&   13(0.11)\\
 23.5&   19(0.38)&  16(0.32)&  21(0.29)&     26(0.31)&    28(0.24)&   38(0.25)&   67(0.22)&   49(0.16)&   13(0.10)\\
 24.0&   19(0.36)&  16(0.30)&  21(0.27)&     26(0.29)&    28(0.22)&   38(0.23)&   67(0.21)&   49(0.15)&   13(0.09)\\
 24.5&   19(0.34)&  16(0.29)&  21(0.25)&     26(0.27)&    28(0.21)&   38(0.22)&   67(0.20)&   49(0.14)&   13(0.09)\\
 25.0&   19(0.32)&  16(0.27)&  21(0.24)&     26(0.25)&    28(0.20)&   38(0.20)&   67(0.19)&   49(0.13)&   13(0.08)\\
\enddata
\onecolumngrid
\tablenotetext{a}{For each temperature bin listed, the first value is the number of objects out to the distance given in column (1), and the second value (in parentheses) is the value of $\langle{V}/{V_{max}}\rangle$ for objects out to that distance.}

\end{deluxetable*}

\section{Conclusions and Future Plans\label{conclusions}}

We have presented preliminary trigonometric parallaxes for 184 dwarfs with spectral types from T6 through early-Y. The vast majority of these, 142, come from a dedicated {\it Spitzer}/IRAC ch2 program, with the rest coming from programs at the USNO, NTT, and UKIRT. We use these parallaxes to produce a 20-pc sample with which we fit trends to various relationships between colors, spectral types, absolute magnitudes, and effective temperatures. We use these parallaxes to determine the distance limits at which our sample is complete for each of five 150K-wide bins ranging over $300K < T_{\rm eff} < 1050K$. We also take a sample of early- to mid-L dwarfs from the literature, supplemented with recent {\it Gaia} data, to produce a complete 20-pc sample across four hotter 150K-wide bins in the range $1500K < T_{\rm eff} < 2100K$. We compute the observed space densities in these bins and compare to simulations using various forms of the mass function passed through different evolutionary code -- either \cite{baraffe2003} or \cite{saumon2008} -- to produce predicted distributions as a function of $T_{\rm eff}$. Fits of our sample to these simulations show that a power law with $\alpha = 0.6$ provides the best match. Functions involving log-normal forms do not fit the observed space densities well at the lowest temperatures and lowest masses. We find that simulations with low-mass cutoffs of 10 $M_{Jup}$ underpredict the number of objects in these same bins, with which we conclude that the low mass cutoff for star formation, if there is one, must be lower than $\sim$5 $M_{Jup}$, a result corroborated by analysis of the low-mass constituents in nearby young clusters (e.g., \citealt{luhman2016b}). Obtaining this result for the {\it field} substellar distribution, however, confirms that the formation of objects this low in mass is not a recent phenomenon but has been occurring over the lifetime of the Milky Way. The predicted mean age (see Figure~\ref{alpha_0.6}) of objects in the 300-450K, for example, is 5.6 Gyr for a mean mass of 14 $M_{Jup}$. 

These new results represent a vast improvement upon our previous attempt (\citealt{kirkpatrick2012}) to derive the shape of the low-mass end of the mass function for several reasons: (1) There have been six additional years of follow-up by the entire community of brown dwarf researchers to uncover late-T and Y dwarfs in the Solar Neighborhood, so we have benefited in this paper from a larger sample of objects. (2) Rather than using measured parallaxes for a handful of objects (42 late-T dwarfs and 7 Y dwarfs) and extending those via spectophotometric distance estimates to the others, we now have actual trigonometric parallaxes for (almost) the entire sample -- 126 late-T dwarfs and 26 Y dwarfs. (3) With actual parallaxes in hand, we can eliminate unresolved interlopers that would fall within our volume based on their spectrophotometric distances, and for objects still within the volume, we can better access which objects are likely to be unresolved binaries themselves. (4) We have created a new suite of mass function simulations, some tied to newer and more realistic evolutionary models, that better incorporate the complexities of cooling across the brown dwarf sequence. In \cite{kirkpatrick2012}, we concluded that the brown dwarf segment of the mass function most closely approximated a power-law with $\alpha \approx 0.0$, but if we used a normalization based on the space density of low-mass {\it stellar} objects, this power law overpredicted our counts by a factor of 2-3. In this paper, we fit the brown dwarf segment independently and find that a much steeper power-law is indicated, with $\alpha \approx 0.6$. This number is much closer to the $\alpha$ values of the brown dwarf mass function found in the Pleiades ($\sim$0.6; \citealt{bouvier1998, casewell2007}) as well as other young clusters (typically $\sim$0.5; see review by \citealt{luhman2012-review}) and star forming regions ($\lesssim 0.5$; see review by \citealt{bastian2010}).

We will be able to improve upon the results in this paper in several ways. The tabulated parallaxes and their uncertainties should continue to improve for our {\it Spitzer} sample of 142 targets as there is a final year of Cycle 13 observations currently underway. The measured parallaxes for objects in the USNO, NTT, and UKIRT parallax programs will also continue to improve, as those results presented here are also regarded as preliminary. For our {\it Spitzer} targets with large ($>2$) reduced $\chi^2$ values in their astrometric fits, we will obtain an additional year of data in Cycle 14 to monitor whether the residuals in the fits show signs of periodicity related to unseen companions. In Cycle 14, we will also be acquiring astrometric data in IRAC ch2 of the 20-pc L dwarf sample in the missing 1050-1500K $T_{\rm eff}$ interval of Table~\ref{space_densities} as well as obtaining astrometry for those T dwarfs in Table~\ref{teffs_20pc_census} lacking parallax measurements. Other improvements for our continued analysis includes the use of {\it Gaia} DR2 data to tie directly to the absolute astrometric reference frame for the individual {\it Spitzer} exposures, as parallactic solutions for most of our reference stars are available; this will obviate the need for a correction from relative to absolute astrometry at a later step. Moreover, as the community of nearby star researchers continues to scour the {\it Gaia} DR2 data, volume-limited samples of the omnipresent M dwarfs will finally become published out to $\sim$20 pc or more, enabling us to check how the brown dwarf power law segment found in this paper dovetails with the low-mass end of the field stellar mass function.

Most pressing, however, is the need to uncover even colder objects that will enable us to determine more accurately the low-mass cutoff of star formation and to discern if the power-law form continues to describe the observed space density at the coldest temperatures. In this regard, we need to complete the census of the coldest members in the 300-450K bin, and find more objects that occupy the 150-300K bin with WISE 0855$-$0714 (see Table~\ref{space_densities}). For this we need an all-sky data set probing wavelengths where these objects are brightest, and $\sim$5$\mu$m images from {\it WISE} will likely be the only ones capable of providing that info for many years to come. Presently, two efforts are underway to discover more Y dwarfs using these data. The first is the Citizen Scientist project Backyard Worlds (\citealt{faherty2018}) and the second is a NASA-funded ADAP proposal led by Peter Eisenhardt called CatWISE. Backyard Worlds is taking the unWISE coadds (\citealt{lang2014, meisner2017a, meisner2017b}) spanning several years of the {\it WISE} and {\it NEOWISE} missions and creating blinking coadds on which members of the public can identify moving objects. CatWISE is taking epochal versions of the unWISE coadds (\citealt{meisner2018b}) and running an AllWISE-style processing (\citealt{cutri2013, kirkpatrick2014}) on them to measure proper motions for all detected sources. For both, the main advantage over previous {\it WISE} data sets is that motions over a long, $\sim$7-year time baseline can now be used to uncover nearby objects, freeing selections from relying on color to identify cold objects that may be detected only weakly at W2 and undetected in W1. (Previous motion searches with {\it WISE} data by \citealt{luhman2014-solar_comp, luhman2014, kirkpatrick2014, kirkpatrick2016} and \citealt{schneider2016} were limited by a small 0.5-yr time baseline, shallow individual frame depths, or both.) Obtaining imaging and spectroscopic characterization of cold discoveries from both programs will be possible using current ground-based and space-based assets, although for the coldest discoveries, given their extreme faintness, obtaining spectra will likely have to wait for the launch of the {\it James Webb Space Telescope}. Sadly, once {\it Spitzer} ceases operations there will be no obvious instrumentation with which to obtain the much needed parallaxes for these discoveries. The astronomical community, and brown dwarf researchers in particular, will be left with a need that no planned future mission fulfills. 

\acknowledgments
JDK thanks the staff of the {\it Spitzer} Science Center for their assistance in making this project possible; Frank Masci, Lee Rottler, and Steve Schurr for programming advice; Yossi Shvartzvald and Sergio Fajardo-Acosta for lucrative discussions; and the anonymous referee and the U.S.\ Naval Observatory Editorial Board for comments that improved the paper.
We also thank Trudy M.\ Tilleman and Justice E.\ Bruursema for their contributions to the USNO observing effort.
We further thank David Ciardi for acquiring Keck/NIRC2 $H$-band imaging for WISE 0226$-$0211AB. 
Research by RLS was supported by the 2015  Henri Chr\'etien International Research Grant administered by the American Astronomical Society. 
ECM is supported by an NSF Astronomy and Astrophysics Postdoctoral Fellowship under award AST-1801978. 
AJC gratefully acknowledges financial support through the Fellowships and Internships in Extremely Large Data Sets (FIELDS) Program, a National Aeronautics and Space Administration (NASA) science/technology/engineering/math (STEM) grant administered by the University of California, Riverside.
FM is supported by an appointment to the NASA Postdoctoral Program at the Jet Propulsion Laboratory (JPL), administered by Universities Space Research Association under contract with NASA. 
Work in this paper is based on observations made with the {\it Spitzer Space Telescope}, which is operated by JPL, California Institute of Technology (Caltech), under a contract with NASA. Support for this work was provided by NASA through Cycle 9 and Cycle 13 awards issued by JPL/Caltech. 
This publication makes use of data products from {\it WISE}, which is a joint project of the University of California, Los Angeles, and JPL/Caltech, funded by NASA.
This work has made use of data from the European Space Agency (ESA) mission {\it Gaia} (\url{https://www.cosmos.esa.int/gaia}), processed by the {\it Gaia} Data Processing and Analysis Consortium (DPAC, \url{https://www.cosmos.esa.int/web/gaia/dpac/consortium}). Funding for the DPAC has been provided by national institutions, in particular the institutions participating in the {\it Gaia} Multilateral Agreement.
This research has made use of IRSA, which is operated by JPL/Caltech, under contract with NASA, and KOA, which is operated by the W. M. Keck Observatory and the NASA Exoplanet Science Institute (NExScI), under contract with NASA. 
This research has also made use of the SIMBAD database, operated at CDS, Strasbourg, France

\facilities{Spitzer(IRAC), USNO:61in(ASTROCAM), NTT (SOFI), UKIRT(WFCAM), WISE, Gaia, IRSA}

\software{IDL (https://www.harrisgeospatial.com/Software-Technology/IDL),
          MOPEX/APEX (http://irsa.ipac.caltech.edu),
          R (https://www.r-project.org), 
          STILTS \cite{taylor2006}, 
          mpfit \cite{markwardt2009}}


\begin{thebibliography}{}
\bibitem[Albert et al.(2011)]{albert2011} Albert, L., Artigau, {\'E}., Delorme, P., et al.\ 2011, \aj, 141, 203 
\bibitem[Allen et al.(2005)]{allen2005} Allen, P.~R., Koerner, D.~W., Reid, I.~N., \& Trilling, D.~E.\ 2005, \apj, 625, 385 
\bibitem[Allers \& Liu(2013)]{allers2013} Allers, K.~N. \& Liu, M.~C.\ 2013, \apj, 772, 79.
\bibitem[Andersen et al.(2008)]{andersen2008} Andersen, M., Meyer, M.~R., Greissl, J., \& Aversa, A.\ 2008, \apjl, 683, L183
\bibitem[Artigau et al.(2010)]{artigau2010} Artigau, {\'E}., Radigan, J., Folkes, S., et al.\ 2010, \apjl, 718, L38 
\bibitem[Baraffe et al.(2003)]{baraffe2003} Baraffe, I., Chabrier, G., Barman, T.~S., Allard, F., \& Hauschildt, P.~H.\ 2003, \aap, 402, 701 
\bibitem[Bardalez Gagliuffi, et al.(2014)]{bardalez2014} Bardalez Gagliuffi, D.~C., Burgasser, A.~J., Gelino, C.~R., et al.\ 2014, \apj, 794, 143.
\bibitem[Bartlett, et al.(2017)]{bartlett2017} Bartlett, J.~L., Lurie, J.~C., Riedel, A., et al.\ 2017, \aj, 154, 151.
\bibitem[Bastian et al.(2010)]{bastian2010} Bastian, N., Covey, K.~R., \& Meyer, M.~R.\ 2010, \araa, 48, 339
\bibitem[Bate(2005)]{bate2005} Bate, M.~R.\ 2005, \mnras, 363, 363 
\bibitem[Beam{\'\i}n, et al.(2013)]{beamin2013} Beam{\'\i}n, J.~C., Minniti, D., Gromadzki, M., et al.\ 2013, \aap, 557, L8.
\bibitem[Bedin \& Fontanive(2018)]{bedin2018} Bedin, L.~R., \& Fontanive, C.\ 2018, \mnras, in press. 
\bibitem[Beichman et al.(2014)]{beichman2014} Beichman, C., Gelino, C.~R., Kirkpatrick, J.~D., et al.\ 2014, \apj, 783, 68 
\bibitem[Beichman et al.(2013)]{beichman2013} Beichman, C., Gelino, C.~R., Kirkpatrick, J.~D., et al.\ 2013, \apj, 764, 101 
\bibitem[Best et al.(2018)]{best2018} Best, W.~M.~J., Liu, M.~C., Magnier, E., \& Dupuy, T.\ 2018, American Astronomical Society Meeting Abstracts \#231, 231, 349.19 
\bibitem[Best et al.(2017)]{best2017} Best, W.~M.~J., Liu, M.~C., Dupuy, T.~J., \& Magnier, E.~A.\ 2017, \apjl, 843, L4 
\bibitem[Bihain \& Scholz(2016)]{bihain2016} Bihain, G., \& Scholz, R.-D.\ 2016, \aap, 589, A26 
\bibitem[Bihain et al.(2013)]{bihain2013} Bihain, G., Scholz, R.-D., Storm, J., \& Schnurr, O.\ 2013, \aap, 557, A43 
\bibitem[Biller et al.(2006)]{biller2006} Biller, B.~A., Kasper, M., Close, L.~M., Brandner, W., \& Kellner, S.\ 2006, \apjl, 641, L141 
\bibitem[Bouvier et al.(1998)]{bouvier1998} Bouvier, J., Stauffer, J.~R., Martin, E.~L., et al.\ 1998, \aap, 336, 490 
\bibitem[Bouy, et al.(2004)]{bouy2004} Bouy, H., Duch{\^e}ne, G., K{\"o}hler, R., et al.\ 2004, \aap, 423, 341.
\bibitem[Bouy, et al.(2003)]{bouy2003} Bouy, H., Brandner, W., Mart{\'\i}n, E.~L., et al.\ 2003, \aj, 126, 1526.
\bibitem[Burgasser et al.(2012)]{burgasser2012} Burgasser, A.~J., Gelino, C.~R., Cushing, M.~C., et al.\ 2012, \apj, 745, 26.
\bibitem[Burgasser et al.(2011)]{burgasser2011} Burgasser, A.~J., Cushing, M.~C., Kirkpatrick, J.~D., et al.\ 2011, \apj, 735, 116 
\bibitem[Burgasser et al.(2010)]{burgasser2010} Burgasser, A.~J., Looper, D., \& Rayner, J.~T.\ 2010, \aj, 139, 2448 
\bibitem[Burgasser, et al.(2010)]{burgasser2010b} Burgasser, A.~J., Cruz, K.~L., Cushing, M., et al.\ 2010b, \apj, 710, 1142.
\bibitem[Burgasser et al.(2008)]{burgasser2008} Burgasser, A.~J., Tinney, C.~G., Cushing, M.~C., et al.\ 2008, \apjl, 689, L53 
\bibitem[Burgasser, et al.(2008)]{burgasser2008b} Burgasser, A.~J., Looper, D.~L., Kirkpatrick, J.~D., et al.\ 2008b, \apj, 674, 451.
\bibitem[Burgasser et al.(2007)]{burgasser2007b} Burgasser, A.~J., Reid, I.~N., Siegler, N., et al.\ 2007b, Protostars and Planets V, 427
\bibitem[Burgasser(2007)]{burgasser2007} Burgasser, A.~J.\ 2007a, \apj, 659, 655
\bibitem[Burgasser, et al.(2007)]{burgasser2007c} Burgasser, A.~J., Looper, D.~L., Kirkpatrick, J.~D., et al.\ 2007, \apj, 658, 557.
\bibitem[Burgasser et al.(2006)]{burgasser2006} Burgasser, A.~J., Geballe, T.~R., Leggett, S.~K., Kirkpatrick, J.~D., \& Golimowski, D.~A.\ 2006a, \apj, 637, 1067 
\bibitem[Burgasser et al.(2006)]{burgasser2006b} Burgasser, A.~J., Burrows, A., \& Kirkpatrick, J.~D.\ 2006b, \apj, 639, 1095 
\bibitem[Burgasser et al.(2004)]{burgasser2004} Burgasser, A.~J., McElwain, M.~W., Kirkpatrick, J.~D., et al.\ 2004, \aj, 127, 2856 
\bibitem[Burgasser(2004)]{burgasser2004b} Burgasser, A.~J.\ 2004, \apjs, 155, 191 
\bibitem[Burgasser et al.(2003)]{burgasser2003a} Burgasser, A.~J., Kirkpatrick, J.~D., McElwain, M.~W., et al.\ 2003a, \aj, 125, 850
\bibitem[Burgasser et al.(2003)]{burgasser2003b} Burgasser, A.~J., McElwain, M.~W., \& Kirkpatrick, J.~D.\ 2003b, \aj, 126, 2487 
\bibitem[Burgasser et al.(2003)]{burgasser2003c} Burgasser, A.~J., Kirkpatrick, J.~D., Reid, I.~N., et al.\ 2003c, \apj, 586, 512.
\bibitem[Burgasser et al.(2002)]{burgasser2002} Burgasser, A.~J., Kirkpatrick, J.~D., Brown, M.~E., et al.\ 2002, \apj, 564, 421 
\bibitem[Burgasser et al.(2000)]{burgasser2000} Burgasser, A.~J., Kirkpatrick, J.~D., Cutri, R.~M., et al.\ 2000, \apjl, 531, L57
\bibitem[Burgasser et al.(1999)]{burgasser1999} Burgasser, A.~J., Kirkpatrick, J.~D., Brown, M.~E., et al.\ 1999, \apjl, 522, L65 
\bibitem[Burningham et al.(2013)]{burningham2013} Burningham, B., Cardoso, C.~V., Smith, L., et al.\ 2013, \mnras, 433, 457 
\bibitem[Burningham et al.(2011)]{burningham2011} Burningham, B., Lucas, P.~W., Leggett, S.~K., et al.\ 2011, \mnras, 414, L90
\bibitem[Burningham et al.(2010)]{burningham2010} Burningham, B., Pinfield, D.~J., Lucas, P.~W., et al.\ 2010, \mnras, 406, 1885 
\bibitem[Burningham et al.(2010)]{burningham2010b} Burningham, B., Leggett, S.~K., Lucas, P.~W., et al.\ 2010b, \mnras, 404, 1952.
\bibitem[Burningham et al.(2009)]{burningham2009} Burningham, B., Pinfield, D.~J., Leggett, S.~K., et al.\ 2009, \mnras, 395, 1237 
\bibitem[Burningham et al.(2008)]{burningham2008} Burningham, B., Pinfield, D.~J., Leggett, S.~K., et al.\ 2008, \mnras, 391, 320 
\bibitem[Burrows et al.(2006)]{burrows2006} Burrows, A., Sudarsky, D., \& Hubeny, I.\ 2006, \apj, 640, 1063 
\bibitem[Burrows et al.(1997)]{burrows1997} Burrows, A., Marley, M., Hubbard, W.~B., et al.\ 1997, \apj, 491, 856 
\bibitem[Cardoso et al.(2015)]{cardoso2015} Cardoso, C.~V., Burningham, B., Smart, R.~L., et al.\ 2015, \mnras, 450, 2486 
\bibitem[Carey et al.(2010)]{carey2010} Carey, S.~J., Surace, J.~A., Glaccum, W.~J., et al.\ 2010, \procspie, 7731, 77310N 
\bibitem[Casali et al.(2007)]{casali2007} Casali, M., Adamson, A., Alves de Oliveira, C., et al.\ 2007, \aap, 467, 777 
\bibitem[Casewell et al.(2007)]{casewell2007} Casewell, S.~L., Dobbie, P.~D., Hodgkin, S.~T., et al.\ 2007, \mnras, 378, 1131 
\bibitem[Castro et al.(2013)]{castro2013} Castro, P.~J., Gizis, J.~E., Harris, H.~C., et al.\ 2013, \apj, 776, 126 
\bibitem[Chabrier et al.(2014)]{chabrier2014} Chabrier, G., Johansen, A., Janson, M., \& Rafikov, R.\ 2014, Protostars and Planets VI, 619 
\bibitem[Chabrier(2005)]{chabrier2005} Chabrier, G.\ 2005, The Initial Mass Function 50 Years Later, 327, 41 
\bibitem[Chabrier(2003)]{chabrier2003-review} Chabrier, G.\ 2003, \pasp, 115, 763 
\bibitem[Chabrier(2003)]{chabrier2003} Chabrier, G.\ 2003, \apjl, 586, L133 
\bibitem[Chabrier(2001)]{chabrier2001} Chabrier, G.\ 2001, \apj, 554, 1274 
\bibitem[Chiu et al.(2006)]{chiu2006} Chiu, K., Fan, X., Leggett, S.~K., et al.\ 2006, \aj, 131, 2722 
\bibitem[Cruz et al.(2009)]{cruz2009} Cruz, K.~L., Kirkpatrick, J.~D., \& Burgasser, A.~J.\ 2009, \aj, 137, 3345 
\bibitem[Cruz, et al.(2007)]{cruz2007} Cruz, K.~L., Reid, I.~N., Kirkpatrick, J.~D., et al.\ 2007, \aj, 133, 439.
\bibitem[Cruz, et al.(2003)]{cruz2003} Cruz, K.~L., Reid, I.~N., Liebert, J., et al.\ 2003, \aj, 126, 2421.
\bibitem[Cushing et al. (2018)]{cushing2018} Cushing, M.~C., Moskovitz, N., \& Gustafsson, A., 2018, Research Notes of the American Astronomical Society, in press.
\bibitem[Cushing et al.(2016)]{cushing2016} Cushing, M.~C., Hardegree-Ullman, K.~K., Trucks, J.~L., et al.\ 2016, \apj, 823, 152 
\bibitem[Cushing et al.(2014)]{cushing2014} Cushing, M.~C., Kirkpatrick, J.~D., Gelino, C.~R., et al.\ 2014, \aj, 147, 113 
\bibitem[Cushing et al.(2011)]{cushing2011} Cushing, M.~C., Kirkpatrick, J.~D., Gelino, C.~R., et al.\ 2011, \apj, 743, 50
\bibitem[Cushing, et al.(2005)]{cushing2005} Cushing, M.~C., Rayner, J.~T. \& Vacca, W.~D.\ 2005, \apj, 623, 1115.
\bibitem[Cutri et al.(2013)]{cutri2013} Cutri, R.~M., Wright, E.~L., Conrow, T., et al.\ 2013, Explanatory Supplement to the AllWISE Data Release Products, by R.~M.~Cutri et al.~, 
\bibitem[Dahn et al.(2017)]{dahn2017} Dahn, C.~C., Harris, H.~C., Subasavage, J.~P., et al.\ 2017, \aj, 154, 147 
\bibitem[Dahn, et al.(2002)]{dahn2002} Dahn, C.~C., Harris, H.~C., Vrba, F.~J., et al.\ 2002, \aj, 124, 1170.
\bibitem[Deacon et al.(2012)]{deacon2012} Deacon, N.~R., Liu, M.~C., Magnier, E.~A., et al.\ 2012, \apj, 757, 100 
\bibitem[Deacon, et al.(2005)]{deacon2005} Deacon, N.~R., Hambly, N.~C. \& Cooke, J.~A.\ 2005, \aap, 435, 363.
\bibitem[Delfosse et al.(1997)]{delfosse1997} Delfosse, X., Tinney, C.~G., Forveille, T., et al.\ 1997, \aap, 327, L25 
\bibitem[Delorme et al.(2008)]{delorme2008} Delorme, P., Delfosse, X., Albert, L., et al.\ 2008, \aap, 482, 961 
\bibitem[Dieterich, et al.(2014)]{dieterich2014} Dieterich, S.~B., Henry, T.~J., Jao, W.-C., et al.\ 2014, \aj, 147, 94.
\bibitem[Dupuy et al.(2015)]{dupuy2015} Dupuy, T.~J., Liu, M.~C., \& Leggett, S.~K.\ 2015, \apj, 803, 102 
\bibitem[Dupuy \& Kraus(2013)]{dupuy2013} Dupuy, T.~J., \& Kraus, A.~L.\ 2013, Science, 341, 1492 
\bibitem[Dupuy \& Liu(2012)]{dupuy2012} Dupuy, T.~J., \& Liu, M.~C.\ 2012, \apjs, 201, 19 
\bibitem[EROS Collaboration, et al.(1999)]{goldman1999} EROS Collaboration, Goldman, B., Delfosse, X., et al.\ 1999, \aap, 351, L5.
\bibitem[Faherty et al.(2018)]{faherty2018} Faherty, J.~K., Kuchner, M., Schneider, A., et al.\ 2018, American Astronomical Society Meeting Abstracts \#231, 231, 158.14 
\bibitem[Faherty et al.(2016)]{faherty2016} Faherty, J.~K., Riedel, A.~R., Cruz, K.~L., et al.\ 2016, \apjs, 225, 10 
\bibitem[Faherty et al.(2012)]{faherty2012} Faherty, J.~K., Burgasser, A.~J., Walter, F.~M., et al.\ 2012, \apj, 752, 56 
\bibitem[Faherty et al.(2009)]{faherty2009} Faherty, J.~K., Burgasser, A.~J., Cruz, K.~L., et al.\ 2009, \aj, 137, 1 
\bibitem[Fan, et al.(2000)]{fan2000} Fan, X., Knapp, G.~R., Strauss, M.~A., et al.\ 2000, \aj, 119, 928.
\bibitem[Fazio et al.(2004)]{fazio2004} Fazio, G.~G., Hora, J.~L., Allen, L.~E., et al.\ 2004, \apjs, 154, 10 
\bibitem[Filippazzo, et al.(2015)]{filippazzo2015} Filippazzo, J.~C., Rice, E.~L., Faherty, J., et al.\ 2015, \apj, 810, 158.
\bibitem[Fischer et al.(2003)]{fischer2003} Fischer, J., Vrba, F.~J., Toomey, D.~W., et al.\ 2003, \procspie, 4841, 564 
\bibitem[Folkes, et al.(2012)]{folkes2012} Folkes, S.~L., Pinfield, D.~J., Jones, H.~R.~A., et al.\ 2012, \mnras, 427, 3280.
\bibitem[Folkes, et al.(2007)]{folkes2007} Folkes, S.~L., Pinfield, D.~J., Kendall, T.~R., et al.\ 2007, \mnras, 378, 901.
\bibitem[Fortney et al.(2008)]{fortney2008} Fortney, J.~J., Lodders, K., Marley, M.~S., \& Freedman, R.~S.\ 2008, \apj, 678, 1419-1435 
\bibitem[Forveille, et al.(2004)]{forveille2004} Forveille, T., S{\'e}gransan, D., Delorme, P., et al.\ 2004, \aap, 427, L1.
\bibitem[Gagn{\'e}, et al.(2015)]{gagne2015b} Gagn{\'e}, J., Faherty, J.~K., Cruz, K.~L., et al.\ 2015, \apjs, 219, 33.
\bibitem[Gagn{\'e} et al.(2015)]{gagne2015} Gagn{\'e}, J., Burgasser, A.~J., Faherty, J.~K., et al.\ 2015, \apjl, 808, L20 
\bibitem[Gaia Collaboration et al.(2018)]{gaia2018} Gaia Collaboration, Brown, A.~G.~A., Vallenari, A., et al.\ 2018, arXiv:1804.09365 
\bibitem[Gaia Collaboration et al.(2016)]{gaia2016} Gaia Collaboration, Prusti, T., de Bruijne, J.~H.~J., et al.\ 2016, \aap, 595, A1 
\bibitem[Geballe, et al.(2002)]{geballe2002} Geballe, T.~R., Knapp, G.~R., Leggett, S.~K., et al.\ 2002, \apj, 564, 466.
\bibitem[Geballe et al.(2001)]{geballe2001} Geballe, T.~R., Saumon, D., Leggett, S.~K., et al.\ 2001, \apj, 556, 373 
\bibitem[Gei{\ss}ler, et al.(2011)]{geissler2011} Gei{\ss}ler, K., Metchev, S., Kirkpatrick, J.~D., et al.\ 2011, \apj, 732, 56.
\bibitem[Gelino et al.(2011)]{gelino2011} Gelino, C.~R., Kirkpatrick, J.~D., Cushing, M.~C., et al.\ 2011, \aj, 142, 57 
\bibitem[Gillon et al.(2017)]{gillon2017} Gillon, M., Triaud, A.~H.~M.~J., Demory, B.-O., et al.\ 2017, \nat, 542, 456 
\bibitem[Gizis, et al.(2011)]{gizis2011} Gizis, J.~E., Troup, N.~W. \& Burgasser, A.~J.\ 2011, \apj, 736, L34.
\bibitem[Gizis, et al.(2003)]{gizis2003} Gizis, J.~E., Reid, I.~N., Knapp, G.~R., et al.\ 2003, \aj, 125, 3302.
\bibitem[Gizis(2002)]{gizis2002} Gizis, J.~E.\ 2002, \apj, 575, 484.
\bibitem[Gizis, et al.(2000)]{gizis2000} Gizis, J.~E., Monet, D.~G., Reid, I.~N., et al.\ 2000, \aj, 120, 1085.
\bibitem[Goldman et al.(2010)]{goldman2010} Goldman, B., Marsat, S., Henning, T., Clemens, C., \& Greiner, J.\ 2010, \mnras, 405, 1140 
\bibitem[Golimowski et al.(2004)]{golimowski2004} Golimowski, D.~A., Henry, T.~J., Krist, J.~E., et al.\ 2004, \aj, 128, 1733 
\bibitem[Gomes, et al.(2013)]{gomes2013} Gomes, J.~I., Pinfield, D.~J., Marocco, F., et al.\ 2013, \mnras, 431, 2745.
\bibitem[Goto, et al.(2002)]{goto2002} Goto, M., Kobayashi, N., Terada, H., et al.\ 2002, \apj, 567, L59.
\bibitem[Gray \& Corbally(2009)]{gray2009} Gray, R.~O., \& Corbally, C., J.\ 2009, Stellar Spectral Classification by Richard O.~Gray and Christopher J.~Corbally.~Princeton University Press, 2009.~ISBN: 
\bibitem[Green (1985)]{green1985} Green, R.M. 1985, {\it Spherical Astronomy} , Cambridge: Cambridge University Press, p.186.
\bibitem[Guszejnov et al.(2018)]{guszejnov2018} Guszejnov, D., Hopkins, P.~F., Grudi{\'c}, M.~Y., Krumholz, M.~R., \& Federrath, C.\ 2018, \mnras, 480, 182 
\bibitem[Harrington \& Dahn(1980)]{harrington1980} Harrington, R.~S., \& Dahn, C.~C.\ 1980, \aj, 85, 454 
\bibitem[Hawley, et al.(2002)]{hawley2002} Hawley, S.~L., Covey, K.~R., Knapp, G.~R., et al.\ 2002, \aj, 123, 3409.
\bibitem[Hennebelle \& Chabrier(2008)]{hennebelle2008} Hennebelle, P., \& Chabrier, G.\ 2008, \apj, 684, 395 
\bibitem[Henry et al.(2006)]{henry2006} Henry, T.~J., Jao, W.-C., Subasavage, J.~P., et al.\ 2006, \aj, 132, 2360 
\bibitem[Hoffmann et al.(2018)]{hoffmann2018} Hoffmann, K.~H., Essex, C., Basu, S., \& Prehl, J.\ 2018, \mnras, 478, 2113 
\bibitem[Hopkins(2012)]{hopkins2012} Hopkins, P.~F.\ 2012, \mnras, 423, 2037 
\bibitem[Hora et al.(2012)]{hora2012} Hora, J.~L., Marengo, M., Park, R., et al.\ 2012, \procspie, 8442, 844239 
\bibitem[Ingalls et al.(2014)]{ingalls2014} Ingalls, J.~G., Carey, S.~J., Lowrance, P.~J., Grillmair, C.~J., \& Stauffer, J.~R.\ 2014, \procspie, 9143, 91431M 
\bibitem[Ingalls et al.(2012)]{ingalls2012} Ingalls, J.~G., Krick, J.~E., Carey, S.~J., et al.\ 2012, \procspie, 8442, 84421Y 
\bibitem[Kasper et al.(2007)]{kasper2007} Kasper, M., Biller, B.~A., Burrows, A., et al.\ 2007, \aap, 471, 655 
\bibitem[Kendall et al.(2007)]{kendall2007} Kendall, T.~R., Jones, H.~R.~A., Pinfield, D.~J., et al.\ 2007, \mnras, 374, 445 
\bibitem[Kendall, et al.(2004)]{kendall2004} Kendall, T.~R., Delfosse, X., Mart{\'\i}n, E.~L., et al.\ 2004, \aap, 416, L17.
\bibitem[King et al.(2010)]{king2010} King, R.~R., McCaughrean, M.~J., Homeier, D., et al.\ 2010, \aap, 510, A99 
\bibitem[Kirkpatrick, et al.(2016)]{kirkpatrick2016} Kirkpatrick, J.~D., Kellogg, K., Schneider, A.~C., et al.\ 2016, \apjs, 224, 36.
\bibitem[Kirkpatrick et al.(2014)]{kirkpatrick2014} Kirkpatrick, J.~D., Schneider, A., Fajardo-Acosta, S., et al.\ 2014, \apj, 783, 122
\bibitem[Kirkpatrick et al.(2013)]{kirkpatrick2013-0647} Kirkpatrick, J.~D., Cushing, M.~C., Gelino, C.~R., et al.\ 2013, \apj, 776, 128
\bibitem[Kirkpatrick et al.(2012)]{kirkpatrick2012} Kirkpatrick, J.~D., Gelino, C.~R., Cushing, M.~C., et al.\ 2012, \apj, 753, 156
\bibitem[Kirkpatrick et al.(2011)]{kirkpatrick2011} Kirkpatrick, J.~D., Cushing, M.~C., Gelino, C.~R., et al.\ 2011, \apjs, 197, 19
\bibitem[Kirkpatrick, et al.(2010)]{kirkpatrick2010} Kirkpatrick, J.~D., Looper, D.~L., Burgasser, A.~J., et al.\ 2010, \apjs, 190, 100.
\bibitem[Kirkpatrick, et al.(2008)]{kirkpatrick2008} Kirkpatrick, J.~D., Cruz, K.~L., Barman, T.~S., et al.\ 2008, \apj, 689, 1295.
\bibitem[Kirkpatrick(2005)]{kirkpatrick2005} Kirkpatrick, J.~D.\ 2005, \araa, 43, 195 
\bibitem[Kirkpatrick, et al.(2000)]{kirkpatrick2000} Kirkpatrick, J.~D., Reid, I.~N., Liebert, J., et al.\ 2000, \aj, 120, 447.
\bibitem[Kirkpatrick, et al.(1999)]{kirkpatrick1999} Kirkpatrick, J.~D., Reid, I.~N., Liebert, J., et al.\ 1999, \apj, 519, 802.
\bibitem[Knapp et al.(2004)]{knapp2004} Knapp, G.~R., Leggett, S.~K., Fan, X., et al.\ 2004, \aj, 127, 3553 
\bibitem[Koen, et al.(2017)]{koen2017} Koen, C., Miszalski, B., V{\"a}is{\"a}nen, P., et al.\ 2017, \mnras, 465, 4723.
\bibitem[Kroupa et al.(2013)]{kroupa2013} Kroupa, P., Weidner, C., Pflamm-Altenburg, J., et al.\ 2013, Planets, Stars and Stellar Systems.~Volume 5: Galactic Structure and Stellar Populations, 5, 115 
\bibitem[Kroupa(2001)]{kroupa2001} Kroupa, P.\ 2001, \mnras, 322, 231 
\bibitem[Kuchner et al.(2017)]{kuchner2017} Kuchner, M.~J., Faherty, J.~K., Schneider, A.~C., et al.\ 2017, \apjl, 841, L19 
\bibitem[Lang(2014)]{lang2014} Lang, D.\ 2014, \aj, 147, 108 
\bibitem[Lawrence et al.(2007)]{lawrence2007} Lawrence, A., Warren, S.~J., Almaini, O., et al.\ 2007, \mnras, 379, 1599 
\bibitem[Leggett et al.(2017)]{leggett2017} Leggett, S.~K., Tremblin, P., Esplin, T.~L., Luhman, K.~L., \& Morley, C.~V.\ 2017, \apj, 842, 118
\bibitem[Leggett et al.(2015)]{leggett2015} Leggett, S.~K., Morley, C.~V., Marley, M.~S., \& Saumon, D.\ 2015, \apj, 799, 37
\bibitem[Leggett et al.(2013)]{leggett2013} Leggett, S.~K., Morley, C.~V., Marley, M.~S., et al.\ 2013, \apj, 763, 130 
\bibitem[Leggett et al.(2012)]{leggett2012} Leggett, S.~K., Saumon, D., Marley, M.~S., et al.\ 2012, \apj, 748, 74 
\bibitem[Leggett et al.(2010)]{leggett2010} Leggett, S.~K., Burningham, B., Saumon, D., et al.\ 2010a, \apj, 710, 1627
\bibitem[Leggett et al.(2010)]{leggett2010b} Leggett, S.~K., Saumon, D., Burningham, B., et al.\ 2010b, \apj, 720, 252.
\bibitem[Leggett et al.(2009)]{leggett2009} Leggett, S.~K., Cushing, M.~C., Saumon, D., et al.\ 2009, \apj, 695, 1517 
\bibitem[Leggett et al.(2007)]{leggett2007} Leggett, S.~K., Marley, M.~S., Freedman, R., et al.\ 2007, \apj, 667, 537 
\bibitem[Leggett et al.(2002)]{leggett2002} Leggett, S.~K., Hauschildt, P.~H., Allard, F., Geballe, T.~R., \& Baron, E.\ 2002, \mnras, 332, 78 
\bibitem[Liebert, et al.(2003)]{liebert2003} Liebert, J., Kirkpatrick, J.~D., Cruz, K.~L., et al.\ 2003, \aj, 125, 343.
\bibitem[Line et al.(2017)]{line2017} Line, M.~R., Marley, M.~S., Liu, M.~C., et al.\ 2017, \apj, 848, 83.
\bibitem[Line et al.(2014)]{line2014} Line, M.~R., Fortney, J.~J., Marley, M.~S., \& Sorahana, S.\ 2014, \apj, 793, 33 
\bibitem[Liu et al.(2013)]{liu2013} Liu, M.~C., Magnier, E.~A., Deacon, N.~R., et al.\ 2013, \apjl, 777, L20 
\bibitem[Liu et al.(2012)]{liu2012} Liu, M.~C., Dupuy, T.~J., Bowler, B.~P., Leggett, S.~K., \& Best, W.~M.~J.\ 2012, \apj, 758, 57 
\bibitem[Liu et al.(2011)]{liu2011} Liu, M.~C., Deacon, N.~R., Magnier, E.~A., et al.\ 2011, \apjl, 740, L32 
\bibitem[Liu et al.(2007)]{liu2007} Liu, M.~C., Leggett, S.~K., \& Chiu, K.\ 2007, \apj, 660, 1507 
\bibitem[Liu \& Leggett(2005)]{liu2005} Liu, M.~C. \& Leggett, S.~K.\ 2005, \apj, 634, 616.
\bibitem[Liu, et al.(2002)]{liu2002} Liu, M.~C., Fischer, D.~A., Graham, J.~R., et al.\ 2002, \apj, 571, 519.
\bibitem[Lodieu et al.(2012)]{lodieu2012} Lodieu, N., Burningham, B., Day-Jones, A., et al.\ 2012, \aap, 548, A53 
\bibitem[Lodieu et al.(2007)]{lodieu2007} Lodieu, N., Pinfield, D.~J., Leggett, S.~K., et al.\ 2007, \mnras, 379, 1423 
\bibitem[Looper, et al.(2008)]{looper2008} Looper, D.~L., Kirkpatrick, J.~D., Cutri, R.~M., et al.\ 2008, \apj, 686, 528.
\bibitem[Looper et al.(2007)]{looper2007} Looper, D.~L., Kirkpatrick, J.~D., \& Burgasser, A.~J.\ 2007, \aj, 134, 1162 
\bibitem[Low \& Lynden-Bell(1976)]{low1976} Low, C., \& Lynden-Bell, D.\ 1976, \mnras, 176, 367 
\bibitem[Lowrance et al.(2016)]{lowrance2016} Lowrance, P.~J., Carey, S.~J., Surace, J.~A., et al.\ 2016, \procspie, 9904, 99045Z 
\bibitem[Lowrance et al.(2014)]{lowrance2014} Lowrance, P.~J., Carey, S.~J., Ingalls, J.~G., et al.\ 2014, \procspie, 9143, 914358 
\bibitem[Lucas et al.(2010)]{lucas2010} Lucas, P.~W., Tinney, C.~G., Burningham, B., et al.\ 2010, \mnras, 408, L56 
\bibitem[Lucas et al.(2008)]{lucas2008} Lucas, P.~W., Hoare, M.~G., Longmore, A., et al.\ 2008, \mnras, 391, 136 
\bibitem[Luhman et al.(2016)]{luhman2016b} Luhman, K.~L., Esplin, T.~L., \& Loutrel, N.~P.\ 2016, \apj, 827, 52 
\bibitem[Luhman \& Esplin(2016)]{luhman2016} Luhman, K.~L., \& Esplin, T.~L.\ 2016, \aj, 152, 78 
\bibitem[Luhman \& Sheppard(2014)]{luhman2014} Luhman, K.~L., \& Sheppard, S.~S.\ 2014, \apj, 787, 126 
\bibitem[Luhman(2014b)]{luhman2014-0855} Luhman, K.~L.\ 2014a, \apjl, 786, L18 
\bibitem[Luhman(2014a)]{luhman2014-solar_comp} Luhman, K.~L.\ 2014b, \apj, 781, 4
\bibitem[Luhman(2012)]{luhman2012-review} Luhman, K.~L.\ 2012, \araa, 50, 65 
\bibitem[Luhman et al.(2012)]{luhman2012} Luhman, K.~L., Loutrel, N.~P., McCurdy, N.~S., et al.\ 2012, \apj, 760, 152 
\bibitem[Luhman et al.(2011)]{luhman2011} Luhman, K.~L., Burgasser, A.~J., \& Bochanski, J.~J.\ 2011, \apjl, 730, L9 
\bibitem[Luhman et al.(2007)]{luhman2007} Luhman, K.~L., Patten, B.~M., Marengo, M., et al.\ 2007, \apj, 654, 570 
\bibitem[Lutz \& Kelker(1973)]{lutz1973} Lutz, T.~E., \& Kelker, D.~H.\ 1973, \pasp, 85, 573 
\bibitem[Mace et al.(2018)]{mace2018} Mace, G.~N., Mann, A.~W., Skiff, B.~A., et al.\ 2018, \apj, 854, 145 
\bibitem[Mace et al.(2013)]{mace2013} Mace, G.~N., Kirkpatrick, J.~D., Cushing, M.~C., et al.\ 2013a, \apjs, 205, 6 
\bibitem[Mace et al.(2013)]{mace2013b} Mace, G.~N., Kirkpatrick, J.~D., Cushing, M.~C., et al.\ 2013b, \apj, 777, 36 
\bibitem[Mainzer et al.(2011)]{mainzer2011} Mainzer, A., Cushing, M.~C., Skrutskie, M., et al.\ 2011, \apj, 726, 30 
\bibitem[Manjavacas et al.(2016)]{manjavacas2016} Manjavacas, E., Goldman, B., Alcal{\'a}, J.~M., et al.\ 2016, \mnras, 455, 1341.
\bibitem[Manjavacas et al.(2013)]{manjavacas2013} Manjavacas, E., Goldman, B., Reffert, S., \& Henning, T.\ 2013, \aap, 560, A52 
\bibitem[Markwardt(2009)]{markwardt2009} Markwardt, C.~B.\ 2009, Astronomical Data Analysis Software and Systems XVIII, 411, 251 
\bibitem[Marocco, et al.(2013)]{marocco2013} Marocco, F., Andrei, A.~H., Smart, R.~L., et al.\ 2013, \aj, 146, 161.
\bibitem[Marocco et al.(2010)]{marocco2010} Marocco, F., Smart, R.~L., Jones, H.~R.~A., et al.\ 2010, \aap, 524, A38 
\bibitem[Moorwood et al.(1998)]{moorwood1998} Moorwood, A., Cuby, J.-G., \& Lidman, C.\ 1998, The Messenger, 91, 9 
\bibitem[Martin et al.(2018)]{martin2018} Martin, E.~C., Kirkpatrick, J.~D., Smart, R.~L., et al.\ 2018, \apj, submitted
\bibitem[McCaughrean et al.(2004)]{mccaughrean2004} McCaughrean, M.~J., Close, L.~M., Scholz, R.-D., et al.\ 2004, \aap, 413, 1029 
\bibitem[McElwain \& Burgasser(2006)]{mcelwain2006} McElwain, M.~W., \& Burgasser, A.~J.\ 2006, \aj, 132, 2074 
\bibitem[McMahon et al.(2013)]{mcmahon2013} McMahon, R.~G., Banerji, M., Gonzalez, E., et al.\ 2013, The Messenger, 154, 35 
\bibitem[Meisner et al.(2018b)]{meisner2018b} Meisner, A.~M., Lang, D., \& Schlegel, D.~J.\ 2018, \aj, 156, 69 
\bibitem[Meisner et al.(2018a)]{meisner2018} Meisner, A., Cushing, M.~C., Cutri, R., et al.\ 2018, Research Notes of the American Astronomical Society, 2, 140 
\bibitem[Meisner et al.(2017b)]{meisner2017b} Meisner, A.~M., Lang, D., \& Schlegel, D.~J.\ 2017, \aj, 154, 161 
\bibitem[Meisner et al.(2017a)]{meisner2017a} Meisner, A.~M., Lang, D., \& Schlegel, D.~J.\ 2017, \aj, 153, 38 
\bibitem[M{\'e}nard, et al.(2002)]{menard2002} M{\'e}nard, F., Delfosse, X. \& Monin, J.-L.\ 2002, \aap, 396, L35.
\bibitem[Mendez \& van Altena(1996)]{mendez1996} Mendez, R.~A., \& van Altena, W.~F.\ 1996, \aj, 112, 655 
\bibitem[Metodieva, et al.(2015)]{metodieva2015} Metodieva, Y., Antonova, A., Golev, V., et al.\ 2015, \mnras, 446, 3878.
\bibitem[Mr{\'o}z et al.(2017)]{mroz2017} Mr{\'o}z, P., Udalski, A., Skowron, J., et al.\ 2017, \nat, 548, 183 
\bibitem[Miller \& Scalo(1979)]{miller1979} Miller, G.~E., \& Scalo, J.~M.\ 1979, \apjs, 41, 513 
\bibitem[Monet et al.(2010)]{monet2010} Monet, D.~G., Jenkins, J.~M., Dunham, E.~W., et al.\ 2010, arXiv:1001.0305 
\bibitem[Morley et al.(2018)]{morley2018} Morley, C.~V., Skemer, A.~J., Allers, K.~N., et al.\ 2018, \apj, 858, 97.
\bibitem[Morley et al.(2012)]{morley2012} Morley, C.~V., Fortney, J.~J., Marley, M.~S., et al.\ 2012, \apj, 756, 172 
\bibitem[Mugrauer et al.(2006)]{mugrauer2006} Mugrauer, M., Seifahrt, A., Neuh{\"a}user, R., \& Mazeh, T.\ 2006, \mnras, 373, L31 
\bibitem[Murray et al.(2011)]{murray2011} Murray, D.~N., Burningham, B., Jones, H.~R.~A., et al.\ 2011, \mnras, 414, 575 
\bibitem[Mu{\v z}i{\'c} et al.(2012)]{muzic2012} Mu{\v z}i{\'c}, K., Radigan, J., Jayawardhana, R., et al.\ 2012, \aj, 144, 180 
\bibitem[Nakajima et al.(1995)]{nakajima1995} Nakajima, T., Oppenheimer, B.~R., Kulkarni, S.~R., et al.\ 1995, \nat, 378, 463 
\bibitem[Opitz et al.(2016)]{opitz2016} Opitz, D., Tinney, C.~G., Faherty, J.~K., et al.\ 2016, \apj, 819, 17.
\bibitem[Phan-Bao, et al.(2008)]{phan-bao2008} Phan-Bao, N., Bessell, M.~S., Mart{\'\i}n, E.~L., et al.\ 2008, \mnras, 383, 831.
\bibitem[Pinfield et al.(2014)]{pinfield2014a} Pinfield, D.~J., Gomes, J., Day-Jones, A.~C., et al.\ 2014a, \mnras, 437, 1009 
\bibitem[Pinfield et al.(2014)]{pinfield2014b} Pinfield, D.~J., Gromadzki, M., Leggett, S.~K., et al.\ 2014b, \mnras, 444, 1931 
\bibitem[Pinfield et al.(2012)]{pinfield2012} Pinfield, D.~J., Burningham, B., Lodieu, N., et al.\ 2012, \mnras, 422, 1922 
\bibitem[Pinfield et al.(2008)]{pinfield2008} Pinfield, D.~J., Burningham, B., Tamura, M., et al.\ 2008, \mnras, 390, 304 
\bibitem[Potter, et al.(2002)]{potter2002} Potter, D., Mart{\'\i}n, E.~L., Cushing, M.~C., et al.\ 2002, \apj, 567, L133.
\bibitem[Reid et al.(2008)]{reid2008} Reid, I.~N., Cruz, K.~L., Kirkpatrick, J.~D., et al.\ 2008, \aj, 136, 1290 
\bibitem[Reid, et al.(2008)]{reid2008b} Reid, I.~N., Cruz, K.~L., Burgasser, A.~J., et al.\ 2008, \aj, 135, 580.
\bibitem[Reid, et al.(2006)]{reid2006} Reid, I.~N., Lewitus, E., Allen, P.~R., et al.\ 2006, \aj, 132, 891.
\bibitem[Reid, et al.(2001)]{reid2001} Reid, I.~N., Gizis, J.~E., Kirkpatrick, J.~D., et al.\ 2001, \aj, 121, 489.
\bibitem[Reid, et al.(2000)]{reid2000} Reid, I.~N., Kirkpatrick, J.~D., Gizis, J.~E., et al.\ 2000, \aj, 119, 369.
\bibitem[Reipurth \& Clarke(2001)]{reipurth2001} Reipurth, B., \& Clarke, C.\ 2001, \aj, 122, 432 
\bibitem[Ruiz et al.(1997)]{ruiz1997} Ruiz, M.~T., Leggett, S.~K., \& Allard, F.\ 1997, \apjl, 491, L107 
\bibitem[Salim, et al.(2003)]{salim2003} Salim, S., L{\'e}pine, S., Rich, R.~M., et al.\ 2003, \apj, 586, L149.
\bibitem[Salpeter(1955)]{salpeter1955} Salpeter, E.~E.\ 1955, \apj, 121, 161 
\bibitem[Saumon \& Marley(2008)]{saumon2008} Saumon, D., \& Marley, M.~S.\ 2008, \apj, 689, 1327-1344 
\bibitem[Schmidt(1968)]{schmidt1968} Schmidt, M.\ 1968, \apj, 151, 393 
\bibitem[Schmidt, et al.(2010)]{schmidt2010} Schmidt, S.~J., West, A.~A., Hawley, S.~L., et al.\ 2010, \aj, 139, 1808.
\bibitem[Schmidt, et al.(2007)]{schmidt2007} Schmidt, S.~J., Cruz, K.~L., Bongiorno, B.~J., et al.\ 2007, \aj, 133, 2258.
\bibitem[Schneider et al.(2017)]{schneider2017} Schneider, A.~C., Windsor, J., Cushing, M.~C., Kirkpatrick, J.~D., \& Shkolnik, E.~L.\ 2017, \aj, 153, 196 
\bibitem[Schneider et al.(2016)]{schneider2016} Schneider, A.~C., Greco, J., Cushing, M.~C., et al.\ 2016, \apj, 817, 112 
\bibitem[Schneider et al.(2015)]{schneider2015} Schneider, A.~C., Cushing, M.~C., Kirkpatrick, J.~D., et al.\ 2015, \apj, 804, 92
\bibitem[Schneider, et al.(2014)]{schneider2014} Schneider, A.~C., Cushing, M.~C., Kirkpatrick, J.~D., et al.\ 2014, \aj, 147, 34.
\bibitem[Scholz \& Bell(2018)]{scholz2018} Scholz, R.-D., \& Bell, C.~P.~M.\ 2018, Research Notes of the American Astronomical Society, 2, 33
\bibitem[Scholz et al.(2011)]{scholz2011} Scholz, R.-D., Bihain, G., Schnurr, O., \& Storm, J.\ 2011, \aap, 532, L5 
\bibitem[Scholz(2010)]{scholz2010a} Scholz, R.-D.\ 2010a, \aap, 510, L8 
\bibitem[Scholz(2010)]{scholz2010b} Scholz, R.-D.\ 2010b, \aap, 515, A92 
\bibitem[Scholz et al.(2003)]{scholz2003} Scholz, R.-D., McCaughrean, M.~J., Lodieu, N., \& Kuhlbrodt, B.\ 2003, \aap, 398, L29
\bibitem[Scholz \& Meusinger(2002)]{scholz2002} Scholz, R.-D. \& Meusinger, H.\ 2002, \mnras, 336, L49.
\bibitem[Skemer et al.(2016)]{skemer2016} Skemer, A.~J., Morley, C.~V., Allers, K.~N., et al.\ 2016, \apj, 826, L17.
\bibitem[Skrutskie et al.(2006)]{skrutskie2006} Skrutskie, M.~F., Cutri, R.~M., Stiening, R., et al.\ 2006, \aj, 131, 1163 
\bibitem[Smart et al.(2017)]{smart2017} Smart, R.~L., Apai, D., Kirkpatrick, J.~D., et al.\ 2017, \mnras, 468, 3764 
\bibitem[Smart et al.(2013)]{smart2013} Smart, R.~L., Tinney, C.~G., Bucciarelli, B., et al.\ 2013, \mnras, 433, 2054 
\bibitem[Smart et al.(2010)]{smart2010} Smart, R.~L., Jones, H.~R.~A., Lattanzi, M.~G., et al.\ 2010, \aap, 511, A30 
\bibitem[Smart (1977)]{smart1977} Smart, W.M. {\it Textbook on Spherical Astronomy, 6$^{th} $ Ed.} , Cambridge, UK:Cambridge Univ Press, p. 221.
\bibitem[Strauss et al.(1999)]{strauss1999} Strauss, M.~A., Fan, X., Gunn, J.~E., et al.\ 1999, \apjl, 522, L61 
\bibitem[Subasavage et al.(2009)]{subasavage2009} Subasavage, J.~P., Jao, W.-C., Henry, T.~J., et al.\ 2009, \aj, 137, 4547 
\bibitem[Sumi et al.(2011)]{sumi2011} Sumi, T., Kamiya, K., Bennett, D.~P., et al.\ 2011, \nat, 473, 349 
\bibitem[Taylor(2006)]{taylor2006} Taylor, M.~B.\ 2006, Astronomical Data Analysis Software and Systems XV, 351, 666 
\bibitem[Thies et al.(2010)]{thies2010} Thies, I., Kroupa, P., Goodwin, S.~P., Stamatellos, D., \& Whitworth, A.~P.\ 2010, \apj, 717, 577 
\bibitem[Thompson et al.(2013)]{thompson2013} Thompson, M.~A., Kirkpatrick, J.~D., Mace, G.~N., et al.\ 2013, \pasp, 125, 809
\bibitem[Thorstensen \& Kirkpatrick(2003)]{thorstensen2003} Thorstensen, J.~R. \& Kirkpatrick, J.~D.\ 2003, \pasp, 115, 1207.
\bibitem[Tinney et al.(2018)]{tinney2018} Tinney, C.~G., Kirkpatrick, J.~D., Faherty, J.~K., et al.\ 2018, \apjs, 236, 28 
\bibitem[Tinney et al.(2014)]{tinney2014} Tinney, C.~G., Faherty, J.~K., Kirkpatrick, J.~D., et al.\ 2014, \apj, 796, 39 
\bibitem[Tinney et al.(2012)]{tinney2012} Tinney, C.~G., Faherty, J.~K., Kirkpatrick, J.~D., et al.\ 2012, \apj, 759, 60
\bibitem[Tinney et al.(2005)]{tinney2005} Tinney, C.~G., Burgasser, A.~J., Kirkpatrick, J.~D., \& McElwain, M.~W.\ 2005, \aj, 130, 2326 
\bibitem[Tinney et al.(2003)]{tinney2003} Tinney, C.~G., Burgasser, A.~J., \& Kirkpatrick, J.~D.\ 2003, \aj, 126, 975 
\bibitem[Tokunaga et al.(2002)]{tokunaga2002} Tokunaga, A.~T., Simons, D.~A., \& Vacca, W.~D.\ 2002, \pasp, 114, 180 
\bibitem[Tsvetanov et al.(2000)]{tsvetanov2000} Tsvetanov, Z.~I., Golimowski, D.~A., Zheng, W., et al.\ 2000, \apjl, 531, L61 
\bibitem[van Altena et al.(1995)]{vanaltena1995} van Altena, W.~F., Lee, J.~T., \& Hoffleit, D.\ 1995, VizieR Online Data Catalog, 1174,  
\bibitem[van Leeuwen(2007)]{vanleeuwen2007} van Leeuwen, F.\ 2007, \aap, 474, 653 
\bibitem[Vigan et al.(2012)]{vigan2012} Vigan, A., Bonnefoy, M., Chauvin, G., Moutou, C., \& Montagnier, G.\ 2012, \aap, 540, A131 
\bibitem[Vrba et al.(2015)]{vrba2015} Vrba, F.~J., Munn, J.~A., Luginbuhl, C.~B., et al.\ 2015, 18th Cambridge Workshop on Cool Stars, Stellar Systems, and the Sun, 18, 945 
\bibitem[Vrba et al.(2004)]{vrba2004} Vrba, F.~J., Henden, A.~A., Luginbuhl, C.~B., et al.\ 2004, \aj, 127, 2948 
\bibitem[Warren et al.(2007a)]{warren2007} Warren, S.~J., Mortlock, D.~J., Leggett, S.~K., et al.\ 2007, \mnras, 381, 1400 
\bibitem[Warren et al.(2007b)]{warren2007b} Warren, S.~J., Hambly, N.~C., Dye, S., et al.\ 2007, \mnras, 375, 213 
\bibitem[West, et al.(2008)]{west2008} West, A.~A., Hawley, S.~L., Bochanski, J.~J., et al.\ 2008, \aj, 135, 785.
\bibitem[Whitworth \& Zinnecker(2004)]{whitworth2004} Whitworth, A.~P., \& Zinnecker, H.\ 2004, \aap, 427, 299 
\bibitem[Williams \& Berger(2015)]{williams2015} Williams, P.~K.~G. \& Berger, E.\ 2015, \apj, 808, 189.
\bibitem[Wilson, et al.(2003)]{wilson2003} Wilson, J.~C., Miller, N.~A., Gizis, J.~E., et al.\ 2003, Brown Dwarfs, 197.
\bibitem[Wright et al.(2014)]{wright2014} Wright, E.~L., Mainzer, A., Kirkpatrick, J.~D., et al.\ 2014, \aj, 148, 82 
\bibitem[Wright et al.(2013)]{wright2013} Wright, E.~L., Skrutskie, M.~F., Kirkpatrick, J.~D., et al.\ 2013, \aj, 145, 84 
\bibitem[Zacharias et al.(2012)]{zacharias2012} Zacharias, N., Finch, C.~T., Girard, T.~M., et al.\ 2012, VizieR Online Data Catalog, 1322,  
\end{thebibliography}
\end{document}